\documentclass[%
reprint,
superscriptaddress,
preprint numbers,
nofootinbib,
amsmath,amssymb,
aps,
prd
]{revtex4-2}

\usepackage{hyperref}
\usepackage{braket}
\hypersetup{colorlinks=true, citecolor=blue, urlcolor=blue, linkcolor=blue}
\usepackage[T1]{fontenc}
\usepackage{graphicx}  
\usepackage{amsmath}   
\usepackage{float}     
\usepackage{float}     
\usepackage{dcolumn}   
\usepackage{bm}        
\usepackage{graphicx}  
\usepackage{amssymb}   
\usepackage{xcolor}
\usepackage{newtxtext,newtxmath}
\usepackage[normalem]{ulem}
\usepackage{diagbox}
\usepackage{array, makecell} %

\renewcommand{\arraystretch}{1.4}
\begin{document}
\title{Modeling nonlinear scales with COLA: preparing for LSST-Y1}

\author{Jonathan Gordon}
\email{jonathan.gordon@stonybrook.edu}
\affiliation{Department of Physics \& Astronomy, Stony Brook University, Stony Brook, NY 11794, USA}
\affiliation{C. N. Yang Institute for Theoretical Physics, Stony Brook University, Stony Brook, NY, 11794, USA}
\author{Bernardo F. de Aguiar}
\affiliation{CBPF - Brazilian Center for Research in Physics, Xavier Sigaud st. 150, zip 22290-180, Rio de Janeiro, RJ, Brazil}
\author{João Rebouças}
\affiliation{Instituto de Física Teórica da Universidade Estadual Paulista, R. Dr. Bento Teobaldo Ferraz, 271, Bloco II, Barra-Funda - São Paulo/SP, Brazil}
\author{Guilherme Brando}
\affiliation{Max Planck Institute for Gravitational Physics (Albert Einstein Institute) Am M\"uhlenberg 1, 14476 Potsdam-Golm, Germany}
\author{Felipe Falciano}
\affiliation{CBPF - Brazilian Center for Research in Physics, Xavier Sigaud st. 150, zip 22290-180, Rio de Janeiro, RJ, Brazil}
\author{Vivian Miranda}
\affiliation{C. N. Yang Institute for Theoretical Physics, Stony Brook University, Stony Brook, NY, 11794, USA}
\author{Kazuya Koyama}
\affiliation{Institute of Cosmology and Gravitation, University of Portsmouth,\\ Dennis Sciama Building, Burnaby Road, Portsmouth PO1 3FX, United Kingdom}
\author{Hans A. Winther}
\affiliation{Institute of Theoretical Astrophysics, University of Oslo,\\ PO Box 1029, Blindern 0315, Oslo,
Norway}
\date{\today}

\begin{abstract}
Year 1 results of the Legacy Survey of Space and Time (LSST) will provide tighter constraints on small-scale cosmology, beyond the validity of linear perturbation theory. This heightens the demand for a computationally affordable prescription that can accurately capture nonlinearities in beyond-$\Lambda$CDM models. The COmoving Lagrangian Acceleration (COLA) method, a cost-effective \textit{N}-body technique, has been proposed as a viable alternative to high-resolution \textit{N}-body simulations for training emulators of the nonlinear matter power spectrum. In this study, we evaluate this approach by employing COLA emulators to conduct a cosmic shear analysis with LSST-Y1 simulated data across three different nonlinear scale cuts. We use the $w$CDM model, for which the \textsc{EuclidEmulator2} (\textsc{ee2}) exists as a benchmark, having been trained with high-resolution \textit{N}-body simulations. We primarily utilize COLA simulations with mass resolution $M_{\rm part}\approx 8 \times 10^{10} ~h^{-1} M_{\odot}$ and force resolution $\ell_{\rm force}=0.5 ~h^{-1}$Mpc, though we also test refined settings with $M_{\rm part}\approx 1 \times 10^{10} ~h^{-1}M_{\odot}$ and force resolution $\ell_{\rm force}=0.17 ~h^{-1}$Mpc. We find the performance of the COLA emulators is sensitive to the placement of high-resolution \textit{N}-body reference samples inside the prior, which only ensure agreement in their local vicinity. However, the COLA emulators pass stringent criteria in goodness-of-fit and parameter bias throughout the prior, when existing high-resolution $\Lambda$CDM emulators are leveraged alongside the COLA emulators to predict the respective $\Lambda$CDM parameters, suggesting a promising template for extensions to $\Lambda$CDM.

\end{abstract}

\maketitle

\section{Introduction}
In the past few decades, galaxy surveys have emerged as essential cosmological probes, with a level of precision that matches those of CMB measurements \cite{COSMOS_survey, CFHTLens, eBOSS, DEEPLENS_survey, desy1_shear, desy1_3x2, hsc_y1, hsc_y1_2pt, kids-1000, desy3_3x2, desy3_shear1, desy3_shear2, desy3_maglim, desy3_redmagic}. These surveys can break degeneracies from other probes by measuring both the background expansion and the growth of structure~\cite{mandelbaum_weak_lensing, Ruiz_2015, Zhong_2023}. As a result, they provide an important test of the $\Lambda$CDM model and help to constrain the behavior of dark energy (DE) \cite{des_y1_extensions, des_y3_extensions}. The new era of stage-IV Large Scale Structure (LSS) surveys, such as the Legacy Survey of Space and Time (LSST) \cite{lsst_book} and Euclid \cite{euclid}, will deliver additional information on smaller scales of our Universe. This poses a new challenge for the theoretical cosmology community, as novel modeling tools will be required to incorporate this information into the exploration of theories that go beyond the Standard Model of cosmology, without biasing parameter inference. 

In a weak-lensing analysis, improving small-scale modeling primarily requires refining the prediction of the matter power spectrum, $P(k,z)$. At linear order, this quantity can be quickly evaluated with accuracy using Einstein-Boltzmann codes, such as \textsc{camb}\footnote{\url{github.com/cmbant/CAMB}} and \textsc{class}\footnote{\url{github.com/lesgourg/class_public}}~\cite{Blas:2011rf} for General Relativity (GR), as well as \textsc{eftcamb}~\cite{Hu:2013twa} and \textsc{hiclass}~\cite{Zumalacarregui:2016pph} for modified gravity theories \cite{Bellini_2018}. 
However, the same is not true on small scales, where the linear theory assumption of small matter overdensities breaks down, so that alternative methods are required to accurately compute the matter power spectrum.

Recent efforts have been put forth to efficiently perform convolution integrals appearing in higher order perturbation theory~\cite{Bernardeau:2001qr}, such as the Effective Field Theory (EFT) approach~\cite{McDonald:2006mx, McDonald:2009dh, Baumann:2010tm, Carrasco:2012cv, Vlah:2015sea, Senatore:2014via, Baldauf:2015xfa, Lewandowski:2018ywf} and its implementation~\cite{Chudaykin:2020aoj, Chen:2020fxs, DAmico:2020kxu,McEwen:2016fjn, Fang:2016wcf}, as well as its generalizations for some beyond-$\Lambda$CDM models~\cite{Aviles:2021que, Noriega:2022nhf, Moretti:2023drg, Chudaykin:2020ghx, DAmico:2020kxu, DAmico:2020tty, Glanville:2022xes, Piga:2022mge}. While these perturbative templates are becoming computationally less costly, they cover a more limited range of scales, having been validated up to the order of $k \sim 0.2 \, h$Mpc$^{-1}$ at $z=0$~\cite{kmax0,kmax1,kmax2,kmax3,Bernardeau:2001qr}. However, for photometric probes, the weak lensing kernel is sensitive to sufficiently smaller scales, limiting the applicability of these approaches.

Arguably, the most accurate way of computing the matter power spectrum far into the nonlinear regime is through the use of high-resolution \textit{N}-body simulations, which are known to be time-consuming and computationally expensive\footnote{An estimate of the time taken and the computational cost of running a high-resolution simulation with stage-IV required specifications is discussed in Sec. 2.1 of~\cite{Schneider:2015yka}.}. This expense only increases when running these codes for beyond-$\Lambda$CDM models, such as in modified gravity theories, where the equations of motion for the extra degrees of freedom~\cite{Winther:2015wla} must be solved in addition to the Poisson equation. At the same time, Markov Chain Monte Carlo (MCMC) methods often require $\mathcal{O}(10^6)$ computations of the matter power spectrum throughout the parameter space in order to obtain robust constraints in standard weak-lensing analyses, necessitating predictions of the power spectrum in $\mathcal{O}(1)$ seconds. Currently, the halo model provides analytic formulas that can be fit to the results of \textit{N}-body simulations, allowing for fast predictions of the power spectrum up to scales of $k \sim 10 \, h$Mpc$^{-1}$, such as \textsc{halofit}~\cite{Smith:2002dz, Takahashi:2012em, Bird:2011rb}~\footnote{While this work was in its final stages, an improved version of \textsc{halofit} was published~\cite{new_halofit}.}, \textsc{hmcode}~\cite{Mead:2015yca, Mead:2016zqy, Mead:2020vgs} and the halo reaction model approach \textsc{react}~\cite{cataneo_react, ben_react}.

Alternatively, emulation methods have garnered attention in the literature as a means of quickly replicating the results of \textit{N}-body simulations, both in $\Lambda$CDM~\cite{Heitmann:2006hr, Habib:2007ca, Heitmann:2008eq, Heitmann:2009cu, Lawrence:2009uk, Agarwal:2012ew, Heitmann:2015xma, Lawrence:2017ost, Bocquet:2020tes, Kwan:2012nd, DeRose:2018xdj, McClintock:2018uyf, Zhai:2018plk, Nishimichi:2018etk, Kobayashi:2020zsw, Miyatake:2020uhg, Cuesta-Lazaro:2022dgr, Donald-McCann:2021nxc, Maksimova:2021ynf, Yuan:2022jqf}, and in beyond-$\Lambda$CDM models~\cite{Mauland:2023pjt, Winther:2019mus, Ramachandra:2020lue, Arnold:2021xtm, Harnois-Deraps:2022bie, Ruan:2023mgq, Fiorini:2023fjl, Sletmoen,emantis}. In particular, the \textsc{EuclidEmulator2} (\textsc{ee2})~\cite{euclidemu}, the \textsc{bacco} emulator~\cite{bacco} and the \textsc{aemulus} emulator~\cite{DeRose:2023dmk} cover a wide range of parameter space, redshifts, and scales. \textsc{ee2} was built with a training set of $127$ high-resolution \textit{N}-body simulations, run with the GPU accelerated \textsc{pkdgrav3} code~\cite{pkdgrav3}. The \textsc{bacco} emulator, utilizes the rescaling methodology developed by~\cite{rescaling_algo} to predict the nonlinear structure formation of $800$ training cosmologies from 6 high-resolution simulations, while \textsc{aemulus} was trained with 150 simulations using the \textsc{gadget-3} code. Such emulators serve as a near substitute for a high-resolution \textit{N}-body simulation in $\Lambda$CDM and minor extensions, for example \textsc{ee2} can reproduce the results of the Euclid training simulations to within $\sim 1\%$ error for $z \leq 3$ and $k \leq 1 h\mathrm{Mpc}^{-1}$.

Nevertheless, exploring the properties of dark energy requires testing a variety of models for which emulators of the nonlinear power spectrum do not yet exist. Hence, using emulators to constrain the assortment of beyond-$\Lambda$CDM models with stage-IV LSS data will require research groups to generate new training sets of $\mathcal{O}(10^2)$ simulations for each model, utilizing their available computational resources. For this reason, it was proposed in reference~\cite{Brando:2022gvg} that the COmoving Lagrangian Acceleration (COLA)~\cite{Tassev:2013pn} method could be a viable alternative to high-resolution \textit{N}-body simulations in creating training sets for beyond-$\Lambda$CDM emulators. COLA is an approximate \textit{N}-body method that allows fast generation of the matter density field with a reduced number of time-steps, and thus less computational expense, when compared to high-resolution simulations.

In this paper, we assess the performance of COLA emulators in constraining extended models by conducting LSST-Y1 simulated cosmic shear analyses. We consider the $w$CDM model, where the dark energy equation of state $w$ is a free parameter, as emulators trained on high-resolution simulations exist for this model and provide a suitable benchmark. As there are known disagreements between existing emulators in the literature on small-scales\footnote{See for instance Sec. 8 of~\cite{euclidemu2}}, we choose \textsc{ee2} as the single baseline with which we measure the performance of the COLA emulators. Consequently, our methodology is tailored to target agreement between the COLA emulators and \textsc{ee2} specifically, however we establish a general template within which one can in principle substitute a preferred emulator or other nonlinear prescription to target.

While the original proposal~\cite{Brando:2022gvg} showed sub-percent agreement between COLA and \textsc{ee2} in their nonlinear predictions down to scales of $k \sim 1~h$Mpc$^{-1}$ for some selected cosmologies, in the present work we perform stress tests of COLA's performance under large shifts throughout the $w$CDM parameter space. As beyond-$\Lambda$CDM models introduce extra parameters that will further have their own degeneracies with the $\Lambda$CDM parameters, it is ideal to have robust agreement even in extreme regions of the prior. 

One of the goals of this analysis is to determine the necessary scale cuts on LSST-Y1 cosmic shear data, such that the COLA-based constraints are equivalent to those obtained with high-resolution \textit{N}-body methods. We aim to inform future analyses of beyond-$\Lambda$CDM models, where high-resolution simulations are scarce. We also test a variety of strategies for enhancing the performance of the COLA emulators, including increasing the COLA resolution settings, and calibrating the simulations with multiple high-resolution reference samples.

This paper is organized as follows: in Sec.~\ref{sec:COLA} we discuss the COLA algorithm and the methodology for running simulations and constructing training and validation sets for the emulators. Sec.~\ref{sec:emulator} describes how we process the results from COLA simulations to build emulators and presents validation tests at the level of the power spectrum. In Sec.~\ref{sec:lsst} we provide the details of our simulated LSST-Y1 analyses that will serve as a test to the COLA emulators. Sec.~\ref{sec:lsst-results} presents cosmological parameter constraints for the LSST-Y1 simulated analysis using our COLA emulators as well as \textsc{ee2}, assessing the difference between both results. Within Sec.~\ref{sec:lsst-results}, we test two new ideas motivated by the preceding results and provide further details of the corresponding methodology in Appendix~\ref{app:inf_refs_boost}. Finally, Sec.~\ref{sec:conclusion} summarizes our conclusions. 

\section{Methodology}
\subsection{COLA Simulations}
\label{sec:COLA}

COLA~\cite{Tassev:2013pn, scola} is an approximate \textit{N}-body method that combines second-order Lagrangian Perturbation Theory (2LPT) with a Particle-Mesh (PM)~\cite{PM1988} algorithm, in which the former part of the code effectively  evolves the large and intermediate scales, while the small scale evolution is given by the latter. In COLA simulations the particles' displacements are solved on top of the LPT trajectories. In this way, we are able to reduce the number of time-steps in the simulation that would otherwise be needed to correctly capture the large scale dynamics. 

In usual \textit{N}-body simulations, the particles' positions and velocities are evaluated by:
\begin{subequations}
\begin{align}
    &\frac{\mathrm{d} \vec{x}}{\mathrm{d} t} = \vec{v}, \\
    &\frac{\mathrm{d}\vec{v}}{\mathrm{d} t} = -\vec{\nabla} \Phi.
\end{align}
\end{subequations}
In the COLA method, however, we define $\vec{x}_{\rm COLA}=\vec{x} - \vec{x}_{\rm LPT}$, and the system is then rewritten as:
\begin{subequations}
\begin{align}
    & \frac{\mathrm{d}\vec{x}}{\mathrm{d}t} = \vec{v}_{\rm COLA} + \frac{\mathrm{d} \vec{x}_{\rm LPT}}{\mathrm{d}t},  \label{eq:cola_eqn1} \\
    & \frac{\mathrm{d}\vec{v}_{\rm COLA}}{\mathrm{d} t} = - \vec{\nabla} \Phi - \frac{\mathrm{d}^{2}\vec{x}_{\rm LPT}}{\mathrm{d}t^{2}},  \label{eq:cola_eqn2}
\end{align}
\end{subequations}
so that the code evolves $\vec{x}$ and $\vec{v}_{\rm COLA}$. At the beginning of the simulation $\vec{v}_{\rm COLA}=0$, which forces the particle's trajectories to follow the LPT evolution at large scales. In this way, one can interpret the COLA method as a Particle-Mesh \textit{N}-body code in the COLA frame, where particles are bound to LPT trajectories at large scales.

Due to the approximate nature of the COLA method, previous works~\cite{Izard:2015dja, Koda:2015mca, Fiorini:2023fjl} have performed careful investigations into various simulation settings such as the initial redshift of the simulations, number of time-steps, scale factor spacing, and force resolution. COLA has passed many different benchmarking tests at the level of the power spectrum~\cite{Winther:2017jof, Fiorini:2023fjl, Brando:2022gvg, Euclid:2022qde} with respect to high-resolution $N$-body simulations. The same procedure has also been performed in modified gravity models~\cite{Valogiannis:2016ane, Winther:2017jof, Winther:2019mus, Brando:2023fzu, Mauland:2023pjt, Fiorini:2023fjl, Ramachandra:2020lue, hicola}. 

\subsubsection{COLA Simulation Settings}

\begin{table}[t]
\centering
\renewcommand{\arraystretch}{1.6}
\begin{tabular}{| c | c | c |}
\hline
 & \emph{Default-precision} & \emph{Enhanced-precision}\\
 &  (DP) & (EP)\\
\hline\hline
$N_{\rm part}$ & $1024^3$ & $1024^3$\\ \hline
$L$ [$h^{-1}$Mpc] & 1024  & 512  \\ \hline
$N_{\rm mesh}$ & $2048^{3}$ & $3072^{3}$ \\\hline
$M_{\rm part}$ [$h^{-1}M_{\odot}$] &$ 8.4 \times 10^{10}$ & $1.1 \times 10^{10}$  \\\hline
$\ell_{\rm force}$  [$h^{-1}$Mpc]  & 0.5 & 0.17  \\\hline
\end{tabular}
\caption{Important settings adopted in our COLA simulations. We run simulations with two different sets of COLA precision settings, dubbed \emph{default-precision} and \emph{enhanced-precision}. $N_\mathrm{part}$ is the total number of particles in the simulation, $L$ is the size of the simulation box, and $N_\mathrm{mesh}$ is the number of mesh grids used to calculate the force applied to each particle. $M_{\rm part}$ is the mass resolution, or the approximate mass per particle, and $\ell_\mathrm{force}$ is the size of mesh grids, called the force resolution. These two quantities are important measures of a simulation's precision.}
\label{tab:settings}
\end{table} 

In this work, we run COLA with two different precision settings, dubbed \emph{default-precision} (DP) and \emph{enhanced-precision} (EP), as outlined in Table~\ref{tab:settings}. For the \emph{default-precision} simulations, we have considered a box-size of $L=1024~ h^{-1}$Mpc, a total number of particles of $N_{\rm part} = 1024^{3}$, and a mesh grid of $N_{\rm mesh} = 2048^{3}$ cell grids, amounting to a force resolution $\ell_{\rm force} \equiv L/N_{\rm mesh}^{1/3} = 0.5~h^{-1}$Mpc. For the \emph{enhanced-precision} simulations we have decided to improve the force resolution of our simulations to $\ell_{\rm force}\approx 0.17~h^{-1}$Mpc, which is achieved by keeping the total number of particles the same as the \emph{default-precision} simulations, but reducing the box-size to $L=512~h^{-1}$Mpc and using a more refined mesh grid with $N_{\rm mesh}=3072^3$. 

All simulations were initiated at an initial redshift of $z_{\rm initial}=19$, and used a total of $51$ time-steps divided into five different redshift intervals, each with a time resolution of $\Delta a \approx 0.02$ as shown in Table~\ref{tab:timesteps}. This time resolution is motivated in~\cite{Brando:2022gvg}, where the same number of time-steps was compared with PM \textit{N}-body simulations with the same mass and force resolution as in the \emph{enhanced-precision} COLA simulations. As our simulations start at a relatively low redshift, the initial displacement fields are generated using 2LPT, which guarantees that we mitigate the effects of transients present when considering only 1LPT~\cite{Crocce:2006ve} for the initial displacements. 

Usually, the initial conditions for \textit{N}-body simulations are implemented using the back-scaling method, where one feeds the simulation the linear power spectrum at $z=0$, and then back-scales it to the initial redshift of the simulation using the $\Lambda$CDM first order growth factors. In this work, however, we implement the ``forward'' approach~\cite{Angulo_2022}, where at each time-step of the simulation we feed in the linear transfer functions that correctly capture large scale evolution. These transfer functions are computed in the \textit{N}-body gauge ~\cite{Fidler:2015npa, Fidler:2017ebh, Tram:2018znz, Brando:2020ouk} using a modified version of the Einstein-Boltzmann code \textsc{class}~\cite{class, Blas:2011rf}. The \textit{N}-body gauge is a system of coordinates in which we can introduce the effects stemming from relativistic species, radiation and neutrinos, that are only relevant on linear scales, and smoothly transition to the usual Newtonian description on small scales. 

The implementation of relativistic corrections at large scales using the \textit{N}-body gauge approach has also been followed by the Euclid collaboration in the Euclid \textit{flagship} simulations, as well as in the high-resolution \textit{N}-body simulations used to construct \textsc{ee2}. The implementation in our COLA simulations follows the procedure outlined in~\cite{Fidler:2015npa,Fidler:2017ebh}. Our COLA simulations themselves were run using the publicly available \textsc{cola-fml}\footnote{\url{https://github.com/HAWinther/FML}} code, and the power spectra from these simulations were evaluated using the on-the-fly code \textsc{ComputePowerSpectra} available in the \textsc{cola-fml} library.

Due to the finite volume of cosmological \textit{N}-body simulations, the lowest wavenumbers of the power spectra computed from the particle position are undersampled, and the effects of sample variance become more pronounced. There are a number of methods used to suppress the variance effect observed. As an example, we have the technique proposed in~\cite{Angulo:2016hjd, Villaescusa_Navarro_2018}, known as ``pair-fixing'', which consists of averaging the power spectrum between a pair of simulations with fixed amplitudes of their initial random field realizations, but opposite phases. This strategy has also been followed in the \textsc{ee2} training simulations.

Other interesting avenues to suppress cosmic variance exist, such as the control variates technique~\cite{mcbook}, which consists of reducing the variance of a random variable by considering another random variable with known mean that is highly correlated with the former. The first usage of this technique in the context of cosmology was in~\cite{Chartier:2020pmu,Chartier:2021frd,Chartier:2022kjz}, where the authors considered COLA and \textsc{fastpm} simulations as the correlated random variables with known mean to reduce the variance from high resolution \textit{N}-body simulations. Later on, the authors in~\cite{Kokron:2022iok,DeRose:2022zfu} showed that one can safely use the Zel'dovich approximation as the control variate to reduce the variance of simulations. However, in this work we have chosen to adopt pair-fixing in order to follow the strategy of our benchmark, \textsc{ee2}.
\begin{table}[t]
\centering
\renewcommand{\arraystretch}{1.6}
\begin{tabular}{| c | c |}
\hline
Redshift Range & $N_{\rm steps}$\\
\hline \hline
$19.0 \geq z > 3.0$ & $12$ \\\hline
$3.0 \geq z > 2.0 $ & $5$ \\\hline
$2.0 \geq z > 1.0 $ & $8$ \\\hline
$1.0 \geq z > 0.5 $ & $9$ \\\hline
$0.5 \geq z \geq 0.0$ & $17$ \\\hline
\end{tabular}
\caption{
Number of time-steps for numerical integration of Eqs.~\ref{eq:cola_eqn1}-\ref{eq:cola_eqn2} in different redshift intervals from $z = z_\mathrm{ini} = 19$ until $z = 0$, following~\cite{Brando:2022gvg}. In each interval, the time steps are linearly spaced in scale factor, $a$, so that for each interval, the step size is $\Delta a = (a_f - a_i)/{\rm N}_{\rm steps}$. Here,  $a_f$ and $a_i$ are the final and initial scale factors of the interval. $N_{\rm steps}$ is chosen to maintain $\Delta a \approx 0.02$ in each interval.}
\label{tab:timesteps}
\end{table}

Per design, the COLA method allows us to compute nonlinear realizations of the density field faster than a usual high-resolution \textit{N}-body simulation, typically around two orders of magnitude faster in wall-clock time~\cite{cola_mock_galaxy_catalogs}. Employing $128$ cores per simulation, the \emph{default-precision} COLA simulations take about 0.7 wall-clock hours on average, and the \emph{enhanced-precision} take about 1.5 wall-clock hours. When including a $25\%$ memory buffer and system memory, the \emph{default-precision} and \emph{enhanced-precision} simulations are able to run with 1TB and 2TB of total memory respectively, though we used 1.5TB and 3TB. All simulations were run in the SeaWulf cluster at Stony Brook University and the GridUNESP cluster at the State University of S\~ao Paulo. 

\subsubsection{Defining and Sampling the Parameter Space}
\label{sec:sampling_param}

\begin{table}[t]
\centering
\renewcommand{\arraystretch}{1.6}
\begin{tabular}{| c | c | c | c |}
\hline
Parameter & Min. & Max. & EE2 Ref.\\
\hline\hline
$\Omega_m$ & $0.24$ & $0.40$ & $0.319$\\\hline
$\Omega_b$ & $0.04$ & $0.06$ & $0.049$\\\hline
$n_s$ & $0.92$ & $1.00$ & $0.96$\\\hline
$A_s \times 10^{9}$ & $1.7 $ & $2.5 $ & $2.1$ \\\hline
$h$ & $0.61$ & $0.73$ & $0.67$ \\\hline
$w$  & $-1.3$ & $-0.7$ & $-1.0$  \\\hline
\end{tabular}
    \caption{Boundaries of the parameter space in which our emulators are designed to be valid, and our analyses are conducted. For each cosmological parameter listed, we choose minimum and maximum values following the choices of \textsc{ee2}~\cite{euclidemu2}. We populate 5\% beyond the boundaries of this space in each direction with COLA training simulations. Also listed are the values of the \textsc{ee2} reference cosmology located at the center of the space, which is used as an anchor in calibrating the output of our simulations; see Sec.~\ref{sec:boost} for details.}
\label{tab:param_space}
\end{table}

In this work, we train emulators for both the $\Lambda$CDM and $w$CDM models for comparison. The parameter space specified in Table~\ref{tab:param_space} is the intended region of validity of our emulators. Our choice is the same as \textsc{ee2}, but restricting the dark energy equation of state parameter by setting $w_a=0$ so that $w=w_0$ is a constant value (in particular $w=-1$ for $\Lambda$CDM). We also follow the \textsc{ee2} treatment of neutrinos using three degenerate massive neutrinos, but fixing their summed mass to the \textsc{ee2} reference value $\Sigma m_{\nu} = 0.058~$eV. However, in order to improve emulation near the boundary of the \textsc{ee2} parameter space, our training simulations are run over a parameter space stretched symmetrically in each parameter sampled by a total of $10\%$. Within this space, we utilize Latin Hypercube (LH) sampling to select the training points where we run simulations, ensuring a uniform distribution across each individual cosmological parameter and efficiently filling the space.

Before running COLA simulations, we determine our training cosmologies by constructing prototypes of our final emulators using \textsc{halofit}, as it allows for the quick generation of training sets of different sizes. We also generate validation sets for each model with $N_{\rm val}=100$ points chosen by LH sampling within the \textsc{ee2} boundaries in order to assess errors due to emulation. Due to the lower computational expense of COLA simulations compared to a high-resolution \textit{N}-body, we are much less constrained in the number of training points. Using the same emulation choices outlined in Sec.~\ref{sec:emulator}, we found that when using $N_{\rm train}^{\Lambda \rm CDM}=400$ training LH points for $\Lambda$CDM, we were able to safely emulate \textsc{halofit} power spectra with an error compared to the validation sets of less than $0.3\%$ for scales $10^{-2} \, h\text{Mpc}^{-1} \leq k \leq \pi \, h$Mpc$^{-1}$. For $w$CDM we obtained the same results using $N_{\rm train}^{w \rm CDM}=500$ points, increasing the number due to the extra dimension in its parameter space.


\subsection{Post-Processing COLA Simulations}

\label{sec:post_process}
In this section, we detail the methodology employed to process the raw output $P(k,z)$ from the COLA simulations and train emulators, and we evaluate the errors associated with these processes. As is common practice, we extract from the simulations only nonlinear corrections to the power spectrum, as it simplifies the data for emulation. Hence, it is these nonlinear corrections that we use to measure the disagreement directly between the COLA simulation output and \textsc{ee2}, as well as the errors in the COLA emulator predictions with respect to the COLA validation simulations.

\subsubsection{The Boost Factor: computation and validation} \label{sec:boost}

The boost factor is defined as the ratio between the nonlinear and linear power spectrum of a given cosmology, or vector of cosmological parameters, ${\boldsymbol \theta}$:
\begin{equation}\label{eq:boost}
    B_X(k, z | {\boldsymbol \theta}) \equiv \frac{P_{\rm NL}(k,z | {\boldsymbol \theta})}{P_{\rm L}(k,z| {\boldsymbol \theta})} \, .
\end{equation}
Here, $X$ denotes the nonlinear prescription used. Within the scope of this paper, this distinguishes between a COLA-based prescription or a high-resolution \textit{N}-body prescription such as \textsc{ee2}. We suppress the dependence on ${\boldsymbol \theta}$ when it is understood from context. By definition, the boost tends towards unity on large scales where linear theory is applicable, while on small scales, it primarily captures contributions due only to the nonlinear clustering of matter. We compute $B_{\rm COLA}(k,z)$ using Eq.~\ref{eq:boost} only after taking the average of the raw output $P_{\rm NL}(k,z)$ of the two paired-and-fixed simulations, and subtracting the shot noise term $L^3/N_{\rm part}$.

\begin{figure*}[t]
\centering
\includegraphics[width=\columnwidth]{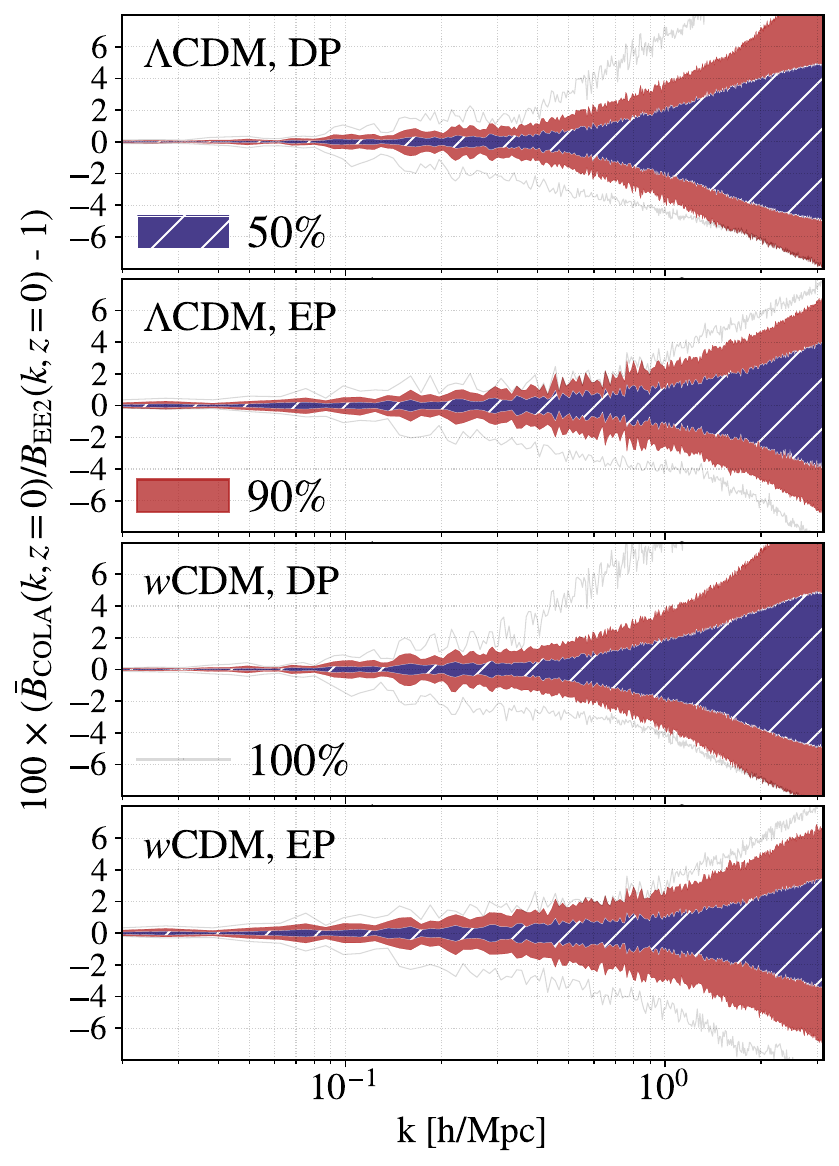}
\includegraphics[width=\columnwidth]{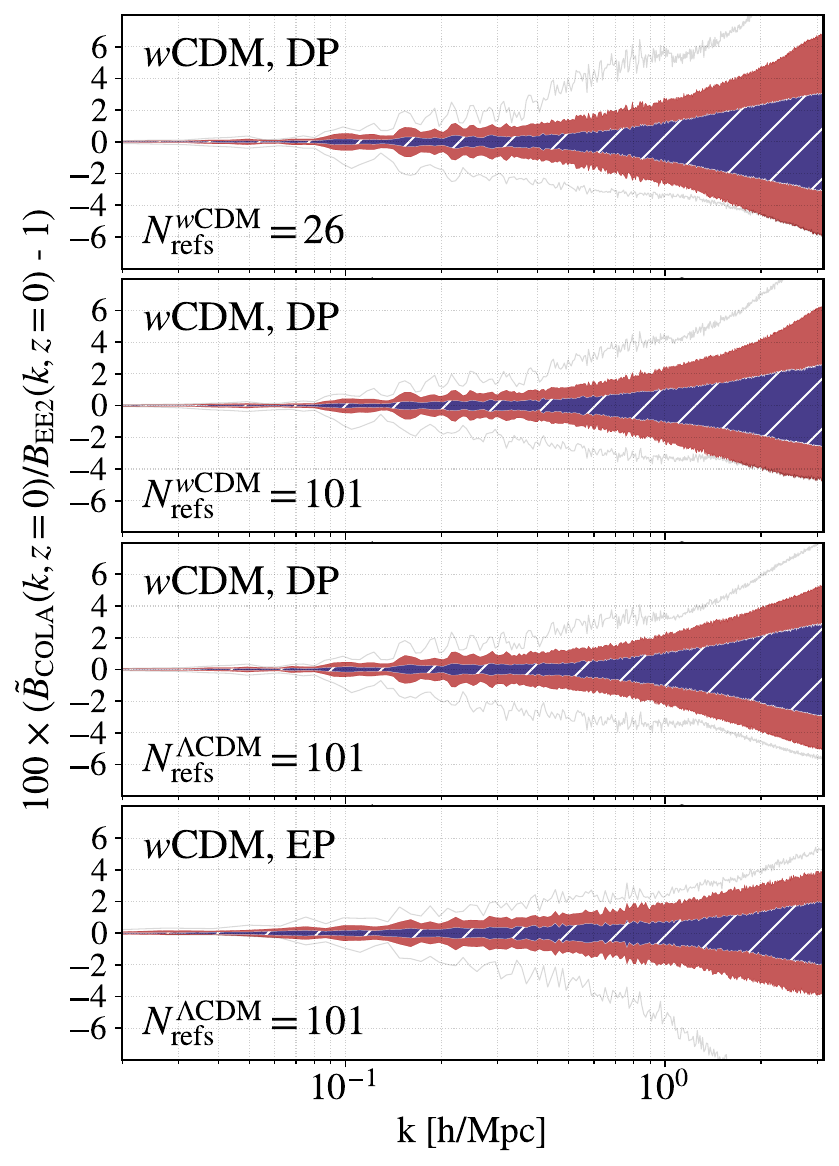}
\caption{Bounds on the fractional differences between the boost factors from our COLA training simulations with those of \textsc{ee2}, for each wavenumber at $z = 0$, for the $50\%$, $90\%$, and $100\%$ of lowest error cosmologies inside the \textsc{ee2} parameter limits. In the left panels, we use Eq.~\ref{eq:b_case} to compute the boost from our COLA simulations employing the \textsc{ee2} reference as the single reference cosmology ($N_{\rm refs}=1$). These results include both COLA precision settings defined in Table~\ref{tab:settings}, and both DE models. The right panels show the effect of using the more general Eq.~\ref{eq:mult_ref_bcase} to calibrate the $w$CDM simulations using additional reference anchors. Considering the expense of high-resolution anchors in extended models, the top right panel shows the $w$CDM \emph{default-precision} disagreement using only 25 additional references scattered throughout the $w$CDM parameter space. Alternatively, the third and fourth panels on the right display errors when the anchors were confined to the $\Lambda$CDM subspace, where we test 100 additional anchors as existing high-resolution $\Lambda$CDM emulators allow for a much greater number. The second panel on the right indicates that only a slight improvement is gained by spreading the 100 additional anchors throughout the entire $w$CDM volume compared to restricting them to $\Lambda$CDM. For similar comparisons between our COLA boosts and \textsc{ee2} at higher redshifts see Appendix~\ref{app:high_zs_ee2_comps}.} For computationally effective ways to increase $N_{\rm refs}^{\Lambda \rm CDM}$, see Sec.~\ref{sec:inf_refs} and Appendix~\ref{app:inf_refs_boost}.
\label{fig:cola_comparisons}
\end{figure*}

Following the approach in~\cite{Brando:2022gvg}, we improve the accuracy of the boost computed from the COLA simulations at a cosmology ${\boldsymbol \theta}$ by using the boost computed from a high-resolution \textit{N}-body prescription at a reference cosmology ${\boldsymbol \lambda}$:
\begin{equation}\label{eq:b_case}
    {\bar B}(k,z| {\boldsymbol \theta}, {\boldsymbol \lambda}) \equiv B_{\rm high-res}(k, z | {\boldsymbol \lambda}) \times  \bigg(\frac{B_{\rm COLA}(k,z | {\boldsymbol \theta})} {B_{\rm COLA}(k,z| {\boldsymbol \lambda})} \bigg).
\end{equation}
In this equation, cosmology-independent differences between COLA and the high-resolution method are mitigated by taking the ratio of two COLA predictions, so that COLA is used to \emph{extrapolate} the boost factor from a single "anchor" high-resolution simulation, $B_{ \rm high-res}(k, z | {\boldsymbol \lambda})$, to a wider range of cosmologies. While the reference cosmology $\boldsymbol \lambda$ may be chosen inside $\Lambda$CDM where high-resolution \textit{N}-body emulators such as \textsc{ee2} already exist, the vector ${\boldsymbol \theta}$ can nevertheless include parameters that predict new physics beyond $\Lambda$CDM. Indeed, one approach tested in this work is to use the \textsc{ee2} $\Lambda$CDM reference cosmology listed in Table~\ref{tab:param_space} as the anchor in Eq.~\ref{eq:b_case} to calibrate the $w$CDM COLA simulations. As our main interest is to benchmark our COLA emulators with \textsc{ee2}, we also use \textsc{ee2} to compute $B_{ \rm high-res}(k, z | {\boldsymbol \lambda})$. 

The left panels of Fig.~\ref{fig:cola_comparisons} compare $\bar{B}_{\rm COLA}(k, z | \boldsymbol{\theta}, \boldsymbol{\lambda})$ against $B_{\textsc{ee2}}(k, z | \boldsymbol{\theta})$ for both DE models, and both COLA precision settings listed in Table~\ref{tab:settings}. In each panel, we fix $z=0$ and plot the bands that bound the lowest error $50\%$, $90\%$, and $100\%$ of cosmologies of the respective training sets at each wavenumber $k$. Only cosmologies that lie within the \textsc{ee2} boundary are able to be compared. Therefore, the $\Lambda$CDM panels consider 240 of the $N_{\rm train}^{\Lambda \rm CDM}=400$ LH points, and the $w$CDM panels consider 279 of the $N_{\rm train}^{w \rm CDM}=500$ LH points. In either DE model, the $90\%$ error bands are within the $2\%$ range for $k \lesssim 0.5 \, h$Mpc$^{-1}$, for both COLA precision-settings. However, the improvement of the \emph{enhanced-precision} simulations appears at higher $k$ due to their refined force and mass resolutions, with the $90\%$ error bands remaining below  $3\%$ when $k \lesssim 1 \, h$Mpc$^{-1}$. More specifically, the percentage of $w$CDM cosmologies in the \emph{default-precision} runs with errors at $k = 1 \, h$Mpc$^{-1}$ outside the $2\%$, $3\%$, and $4\%$ thresholds, was $43.0\%$, $20.8\%$ and $6.81\%$, respectively. For \emph{enhanced-precision} simulations, these percentages more than halved, decreasing to $21.2\%$, $9.0\%$, and $2.9\%$.

As Eq.~\ref{eq:b_case} accounts only for cosmology-independent differences between COLA and the high-resolution \textit{N}-body method, cosmology-dependent disagreements will become more pronounced as $\boldsymbol{\theta}$ deviates more from the reference cosmology $\boldsymbol{\lambda}$. This rationale suggests incorporating multiple anchors across the parameter space in order to supplement the \textsc{ee2} reference cosmology, and offer more proximate references for every training point. We therefore generalize ${\bar B}(k,z| {\boldsymbol \theta})$ to the case of multiple reference anchors $(\boldsymbol\lambda_1, ..., \boldsymbol\lambda_{N_{\rm refs}})$ by computing a weighted sum of the boosts calibrated with each reference:
\begin{equation} \label{eq:mult_ref_bcase}
    \tilde{B}(k,z| {\boldsymbol \theta}, {\boldsymbol \lambda}_1, ..., {\boldsymbol \lambda}_{N_{\rm refs}}) \equiv \sum_{i=1}^{N_{\rm refs}} w_i  {\bar B}(k,z| {\boldsymbol \theta}, {\boldsymbol \lambda}_i) \, .
\end{equation}
The weights
\begin{equation}\label{eq:gaussian_weights}
   w_i = \frac{\exp(-d_i^2/\sigma_d^2 )}{\sum_{j=1}^{N_{\rm refs}} \exp(-d_j^2/\sigma_d^2 )} \, ,
\end{equation}
are functions of the Euclidean distance in the $D$-dimensional parameter space
\begin{equation}\label{eq:euc_dist}
    d^2_i \equiv \sum_{j=1}^{D} (\Theta_{j} - \Lambda_{ij})^2,
\end{equation}
between the normalized $\boldsymbol\theta$ and the $\boldsymbol \lambda_{i}$ vectors:
\begin{align}
    \Theta_j \equiv \frac{\theta_j - \theta_{j,\mathrm{min}}}{\theta_{j,\mathrm{max}} - \theta_{j,\mathrm{min}}} \quad 
    \Lambda_{ij} \equiv \frac{\lambda_{ij} - \lambda_{ij,\mathrm{min}}}{\lambda_{ij,\mathrm{max}} - \lambda_{ij,\mathrm{min}}} \, . 
    \label{eq:norm_param}
\end{align} 
The parameter $\sigma_d$ in the weights is chosen during the process of building the emulator, see the discussion at the end of Sec.~\ref{sec:emulator} for details.

In general, spreading high-resolution \textit{N}-body anchors $B_{\rm high-res}(k, z | {\boldsymbol \lambda_i})$ throughout the entire parameter space of an extended model will require great computational cost, substantially constraining the number of affordable references. However, using reference anchors confined only to the $\Lambda$CDM subspace, where high-resolution \textit{N}-body emulators such as \textsc{ee2} exist, reduces the cost associated with each reference to merely that of COLA, allowing for a much larger number of anchors. Indeed, the similar error profiles between $\Lambda$CDM and $w$CDM in the left panels of Fig.~\ref{fig:cola_comparisons} indicate the main source of error comes from modeling the $\Lambda$CDM parameters. Therefore, we use LH sampling to scatter 25 additional reference anchors throughout the $w$CDM space ($N_{\rm refs}^{w \rm CDM}=26$), and 100 additional reference anchors confined to the $\Lambda$CDM subspace ($N_{\rm refs}^{\Lambda \rm CDM}=101$). In the case of the fewer $N_{\rm refs}^{w \rm CDM}=26$ anchors, we reduced the parameter ranges by $15\%$ percent total from the \textsc{ee2} limits before generating the LH, in order to decrease the average Euclidean distance, $d$, between the training cosmologies inside the \textsc{ee2} boundary, and their nearest anchors.
\begin{align}
\langle d \rangle =  \frac{1}{N} \sum_{i=1}^{N} |\boldsymbol \Theta_i - \boldsymbol\Lambda_{\text{nearest to } \boldsymbol \Theta_i}| \, .
\label{eq:avg_distance_nearest_anchor}
\end{align}

The effect of using Eq.~\ref{eq:mult_ref_bcase} with multiple references is shown in the right panels of Fig.~\ref{fig:cola_comparisons} where the $w$CDM training simulations are compared to the \textsc{ee2} boost factor $B_{\textsc{ee2}}(k, z)$ at $z=0$ in 4 scenarios. The top right panel corresponds to the case of processing the $w$CDM \emph{default-precision} simulations with $N_{\rm refs}^{w \rm CDM}=26$ anchors, where we found that $21.9\%$, $7.2\%$, and $1.5\%$ of the cosmologies inside the \textsc{ee2} boundaries had errors larger than $2\%$, $3\%$ and $4\%$ at $k = 1 \, h$Mpc$^{-1}$. Therefore, the 25 reference anchors showed similar improvement at $k=1 \, h$Mpc$^{-1}$ as \emph{enhanced-precision}. 

More interestingly, in the third panel on the right of Fig.~\ref{fig:cola_comparisons} when the $w$CDM \emph{default-precision} simulations were processed with $N_{\rm refs}^{\Lambda \rm CDM}=101$ anchors confined to the $\Lambda$CDM subspace, all errors above $4\%$ at $k = 1 \, h$Mpc$^{-1}$ were eliminated, and only $13.2\%$ and $1.8\%$ of cosmologies had errors larger than $2\%$ and $3\%$. In the second panel on the right we include the same results using $N_{\rm refs}^{w \rm CDM}=101$ for visual comparison only, where $11.1\%$, $1.8\%$, and $0.4\%$ of cosmologies had errors above $2\%$, $3\%$ and $4\%$ at $k = 1 \, h$Mpc$^{-1}$, comparable to the case of $N_{\rm refs}^{\Lambda \rm CDM}=101$. These results suggest that COLA can effectively be used to model $w$CDM cosmologies by extrapolating a large number of $\Lambda$CDM anchors, without significant loss in accuracy compared to the same number of references in the entire $w$CDM space.

Finally, in the bottom right panel of Fig.~\ref{fig:cola_comparisons} we anchor the $w$CDM \emph{enhanced-precision} simulations with $N_{\rm refs}^{\Lambda \rm CDM}=101$ references, yielding $11.1\%$, $2.2\%$, and $1.8\%$ of cosmologies with errors above $2\%$, $3\%$ and $4\%$ at $k = 1 h$Mpc$^{-1}$. The comparison between the first and third panels on the right may suggest that using a large number of $\Lambda$CDM anchors is the best approach for further reducing disagreement. However, Eq.~\ref{eq:b_case} shows that every additional anchor requires a new COLA simulation to compute $B_{\rm COLA}(k,z| {\boldsymbol \lambda})$. Running beyond $N_{\rm refs}^{\Lambda \text{CDM}} = 101$ COLA simulations approaches the cost of the training set itself, therefore one might consider avoiding the expense by building a dedicated emulator for $B_{\rm COLA}(k,z| {\boldsymbol \lambda})$ from the training simulations, provided this emulation does not introduce large errors. This possibility is discussed in Sec.~\ref{sec:inf_refs}, where we also test using the $B_{\rm COLA}(k,z| {\boldsymbol \lambda})$ emulator to calibrate the COLA predictions post-emulation rather than calibrating the training set.

\begin{figure}[t]
\centering
\includegraphics[width=0.96\columnwidth]{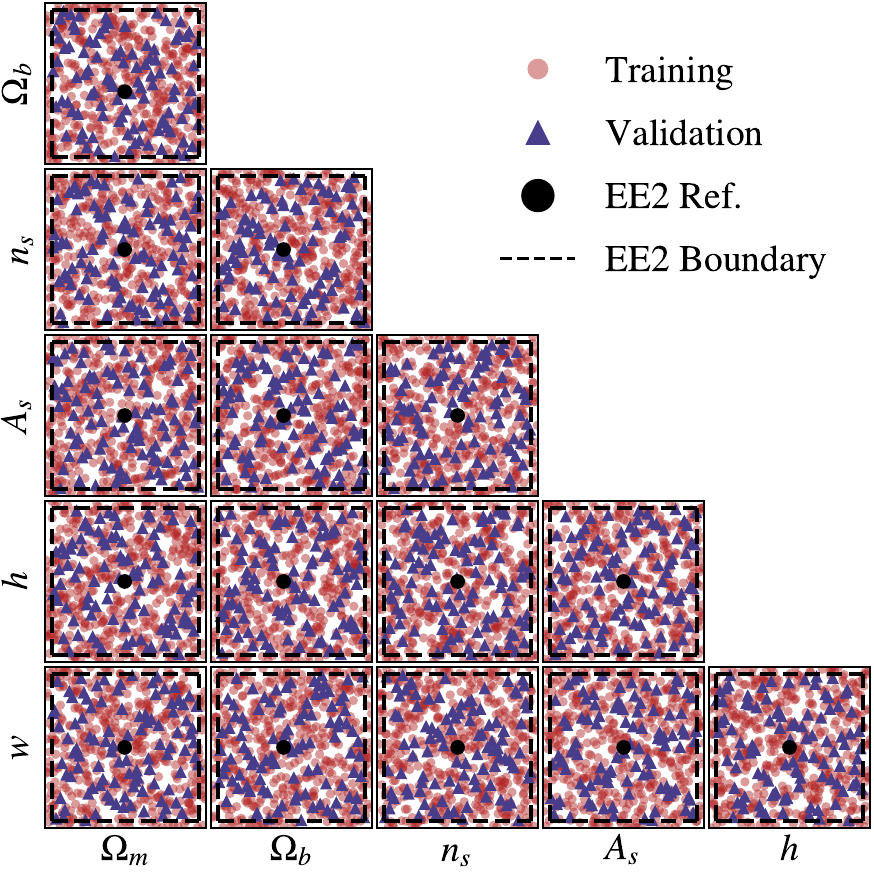}
\caption{2D projections of the cosmologies where our $w$CDM COLA simulations are run. We use 500 training points (red circles) and 100 validation points (blue triangles), drawing both using LH sampling. As the desired region of validity of the emulators is within the \textsc{ee2} parameter boundaries presented in Table~\ref{tab:param_space}, we confine the validation set to within these bounds (dashed black line). However, we train in a region expanded by $5\%$ for each parameter in each direction, to ensure accurate emulation near the \textsc{ee2} boundary. We also mark the \textsc{ee2} reference cosmology near the center of the space.}
\label{fig:lhs}
\end{figure}

\begin{figure*}[t] 
    \centering
    \begin{minipage}[t]{0.33\textwidth}
        \centering
        \includegraphics[width=\linewidth]{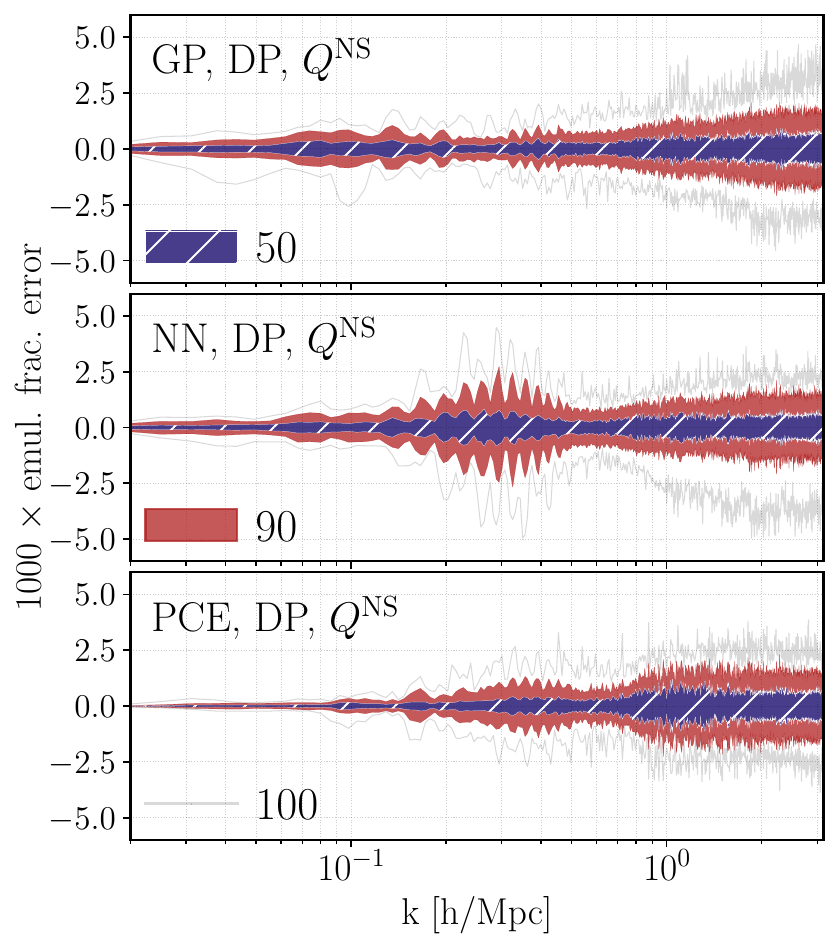}
    \end{minipage}%
    \begin{minipage}[t]{0.33\textwidth}
        \centering
        \includegraphics[width=\linewidth]{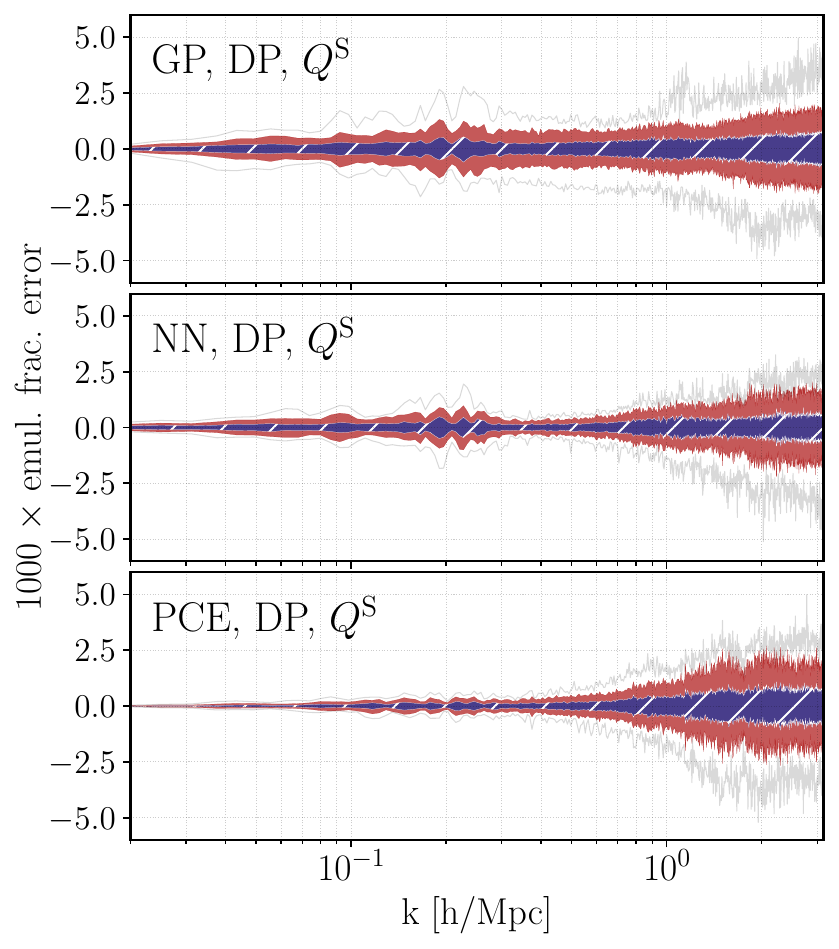}
    \end{minipage}
        \begin{minipage}[t]{0.33\textwidth}
        \centering
        \includegraphics[width=\linewidth]{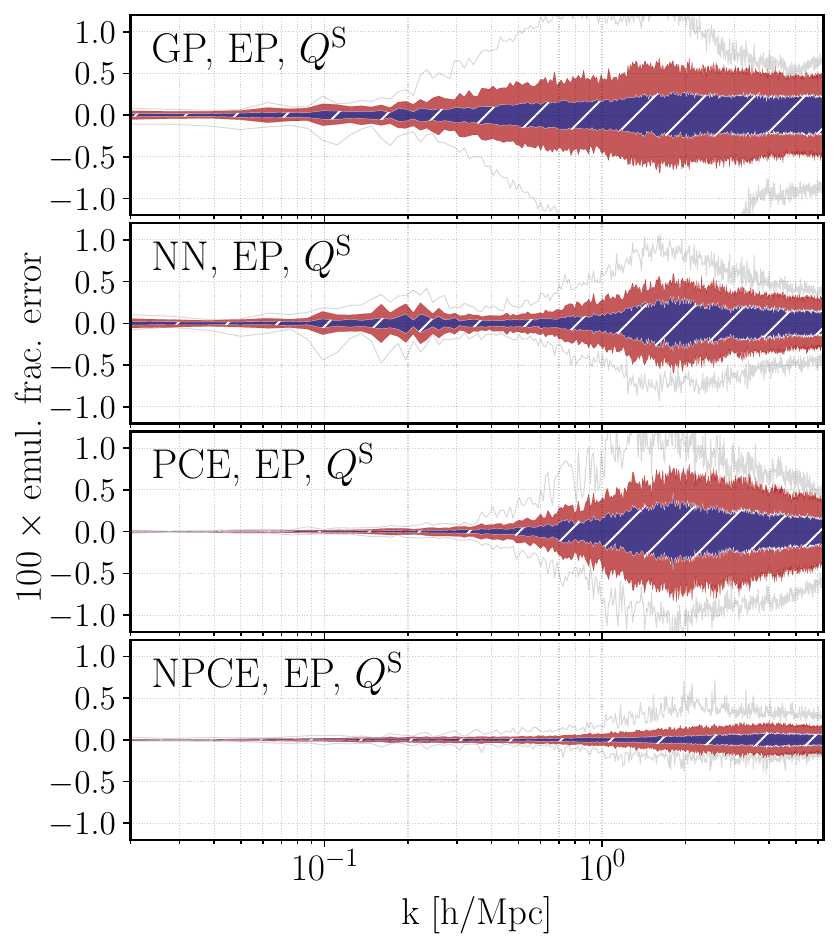}
    \end{minipage}
    \caption{Bounds on the fractional differences between the boost functions at $z=0$ from various COLA $w$CDM emulators, and those from the 100 validation simulations, for the best $50\%$, $90\%$, and $100\%$ of cosmologies. The middle panel shows the slight improvement in the \emph{default-precision} emulation errors when incorporating the smearing procedure utilized by \textsc{bacco} via the emulation variable $Q^{\rm S}(k,z)$, compared to the analogous case when the emulation variable is $Q^{\rm NS}(k,z)$ following \textsc{ee2}. The emulated boost is recovered from the respective emulation variables via Eqs.~\ref{eq:pk_from_logb}-\ref{eq:pk_from_qbacco}, while the boost from the simulations is computed directly via Eq.~\ref{eq:mult_ref_bcase}. While all \emph{default-precision} emulators are equivalent to the COLA simulations within a $\sim 0.5\%$ margin for $k \leq \pi$ $h$Mpc$^{-1}$, only the NPCE method is able to achieve comparable errors for \emph{enhanced-precision}, which it does for scales up to $k\leq 2\pi$ $h$Mpc$^{-1}$. For these comparisons we use only the \textsc{ee2} reference cosmology to calibrate simulations ($N_{\rm refs}=1$).}
    \label{fig:combined}
\end{figure*}

\subsubsection{Emulating the Boost Factor}
\label{sec:emulator}
In this section, we will describe the steps taken to construct emulators for the boost factor $\tilde{B}(k,z | {\boldsymbol \theta}, {\boldsymbol \lambda}_1, ..., {\boldsymbol \lambda}_{N_{\rm refs}})$ defined in Eq.~\ref{eq:mult_ref_bcase}, so that it may be inferred for arbitrary choices of the 6 cosmological parameters listed in Tab.~\ref{tab:param_space} $\boldsymbol \theta=(\Omega_m,\Omega_b,n_s,A_s,h,w)$ using the training set. We tested emulators using three methods found in the literature: Gaussian Process Regression (GP), neural networks (NN), and Polynomial Chaos Expansion (PCE); and combined PCE with an NN to form a fourth method which we denote "Neural Polynomial Chaos Expansion" (NPCE). Their implementation is detailed in Appendices~\ref{app:grp},~\ref{app:nn_emul},~\ref{app:pce}, and~\ref{app:pce_nn}. 

In order to improve the emulation, \textsc{ee2} processes the training data by computing the logarithm of the boost factor. We test this approach using the boost computed from our simulations $\tilde{B}(k,z)$:
\begin{multline}  
    Q^{\rm NS}(k,z|\boldsymbol{\theta},\boldsymbol{\lambda}_1,...,\boldsymbol{\lambda}_{N_{\rm refs}}) \equiv \\ \log[\tilde{B}(k,z)|\boldsymbol{\theta},\boldsymbol{\lambda}_1,...,\boldsymbol{\lambda}_{N_{\rm refs}}] .
\end{multline}
Here the superscript denotes that a "smearing" of the BAO oscillations has not been applied, which differs from the approach of the \textsc{bacco} emulator~\citep{bacco}. The BAO smearing procedure used by \textsc{bacco} dampens oscillations in the boost which are due to large scale bulk flow differences around the BAO peaks between linear and nonlinear prescriptions. This smearing reduces the features in the quantity to be emulated. In our approach, much of this mismatch of the BAO is already mitigated by using Eq.~\ref{eq:b_case}. Nevertheless, we implement this procedure by using as an emulation variable the following quantity:
\begin{multline} \label{eq:qbacco}
    Q^{\rm S}(k,z | {\boldsymbol \theta}, {\boldsymbol \lambda}_1, ..., {\boldsymbol \lambda}_{N_{\rm refs}})  \equiv \\ \log \bigg[\tilde{B}(k,z| {\boldsymbol \theta}, {\boldsymbol \lambda}_1, ..., {\boldsymbol \lambda}_{N_{\rm refs}}) \times \bigg(\frac{P_\mathrm{L}(k,z| {\boldsymbol \theta})}{P^{\text{S}}_\mathrm{L}(k,z| {\boldsymbol \theta})}\bigg) \bigg] \, ,
\end{multline}
where $P^{\rm{S}}_{\rm L}(k,z | {\boldsymbol \theta})$ is a combination of the linear matter spectrum and a version where its BAO signal is removed. Specifically, this is defined as:
\begin{multline}\label{eq:smearing}
    P^{\text{S}}_\mathrm{L}(k,z | {\boldsymbol \theta}) \equiv P_\mathrm{L}(k,z| {\boldsymbol \theta}) G(k,z) \, + \, \\ P^{\text{NBAO}}_\mathrm{L}(k,z| {\boldsymbol \theta}) (1-G(k,z)), \, \,
\end{multline}
where
 \begin{equation}
     G(k,z) \equiv \exp (-0.5k^2/k^2_{*}(z)),
 \end{equation} 
 and 
 \begin{equation}
     k_{*}^{-2}(z) \equiv (3\pi^2)^{-1} \int dk P_\mathrm{L}(k,z| {\boldsymbol \theta}) .
 \end{equation}
To obtain the linear power spectrum devoid of the BAO signal, $P^{\text{NBAO}}_\mathrm{L}(k,z)$, we modify the publicly available code from \textsc{bacco}\footnote{\url{https://bitbucket.org/rangulo/baccoemu/src/master/baccoemu/matter_powerspectrum.py}}. The details of the algorithm are described in Appendix C of~\cite {baumann_removing_bao}, and our modifications are discussed in Appendix~\ref{app:bao-smearing}. When employing this method, the smearing procedure must be applied both when training the emulator, and in the subsequent recovery of the nonlinear matter power spectrum. 

We obtain the nonlinear power spectrum from the emulator depending on the emulation variable, using the relevant equation:
\begin{align}
 \label{eq:pk_from_logb}
    P_{\rm NL}(k,z| {\boldsymbol \theta}) &=  \exp[Q^{\rm NS}(k,z| {\boldsymbol \theta})] \times P_{\rm L}(k,z  | {\boldsymbol \theta}) , \\
    P_{\rm NL}(k,z| {\boldsymbol \theta}) &=  \exp[Q^{\rm S}(k,z| {\boldsymbol \theta})] \times P^\mathrm{S}_{\rm L}(k,z  | {\boldsymbol \theta}) .
    \label{eq:pk_from_qbacco}
\end{align}
Here and throughout, we suppress the explicit dependence on the anchors $({\boldsymbol \lambda}_1, ..., {\boldsymbol \lambda}_{N_{\rm refs}})$ for simplicity. In this work, $Q^{\rm S}(k,z)$ serves as the default emulation variable, used for the vast majority of our emulators. Exceptions, where we utilize $Q^{\rm NS}(k,z)$ instead, are explicitly noted. The following methodology outlined in this section remains applicable in either case.

We then normalize the input and output of the emulators to keep all parameters on a common scale. The emulators were implemented with slightly different normalization procedures without noticeable differences in the final result. All emulators normalized the input cosmological parameters via Eq.~\ref{eq:norm_param}, however the GP and NN emulators used the \textsc{ee2} limits from Tab.~\ref{tab:param_space} as the maximum and minimum values, while the PCE emulator used the maximum and minimum values of each parameter that were sampled in the training set. In addition, the PCE and NN emulators normalize the output two-dimensional array $Q(k_i,z_j)$, with $k_i$ and $z_j$ being the sampled wave vector and redshift output from the simulations. The PCE emulator does this by finding for every wavenumber and redshift the minimum and maximum values $Q_\mathrm{min}(k_i,z_j)$ and $Q_\mathrm{max}(k_i,z_j)$ over all cosmologies, and then computing:
\begin{equation}
    \hat Q({k_i,z_j | \boldsymbol \theta}) \equiv \frac{Q(k_i,z_j | {\boldsymbol \theta}) - Q_\mathrm{min}(k_i,z_j)}{Q_\mathrm{max}(k_i,z_j) - Q_\mathrm{min}(k_i,z_j)},
\end{equation}
In the neural network, we normalized using the minimum and maximum per redshift instead: $Q_\mathrm{min}(z_j) = \min_{i} Q_\mathrm{min}(k_i,z_j)$ and $Q_\mathrm{max}(z_j) = \max_{i} Q_\mathrm{max}(k_i,z_j)$.

Next, we decompose $\hat Q(k_i,z_j)$ into a linear combination of basis vectors called principal components (PCs) in order to decrease the dimensionality of the emulator's output. The principal components capture the maximum variance of the data, therefore we are able to decrease the number of parameters we emulate per redshift from $N_{k} =  512$, to a much smaller number $N_{\rm PC} \sim 10$ of coefficients, without significant loss of information. To find the principal components at each redshift $z_j$, we first calculate the sample average across the training set
\begin{align}
\langle \hat Q \rangle(k_i,z_j)= \frac{1}{N_{\rm cosmo}} \sum_{{\boldsymbol \theta}} Q(k_i,z_j | {\boldsymbol \theta} ) \, ,
\end{align}
as well as the sample covariance $\mathcal{M}_{j}$
\begin{multline}
(\mathcal{M}_{j})_{ip} = \frac{1}{N_{\rm cosmo}-1} \sum_{{\boldsymbol \theta}}  \Big[ \big(\hat Q(k_i,z_j | {\boldsymbol \theta} )-\langle \hat Q \rangle(k_i,z_j) \big) \times \\ \big(\hat Q(k_p,z_j | {\boldsymbol \theta} ) -\langle \hat Q \rangle(k_p,z_j)\big)\Big] \,.
\end{multline}

Finally, we diagonalize each covariance matrix to obtain the $N_\mathrm{PC}$ eigenvectors $\mathcal{E}_{\ell j}(k_i)$ with the largest eigenvalues, and project $\hat{Q}(k_i,z_j)$ onto the subspace spanned by these PCs. Mathematically, this is expressed by:
\begin{equation}
    \hat Q(k_i,z_j | {\boldsymbol \theta} ) = \langle \hat Q \rangle(k_i,z_j) + \sum_{\ell=1}^{N_\mathrm{PC}} \alpha_{\ell j}({\boldsymbol \theta}) \mathcal{E}_{\ell j}(k_i) \, .
    \label{eq:pca}
\end{equation}
Here, the coefficients $\alpha_{\ell j}({\boldsymbol \theta})$ are functions of the cosmological parameters that constitute the direct output of our emulators. For each emulation method, we determine $N_\mathrm{PC}$ by considering the trade off between computational speed and accuracy. For the \emph{default-precision} emulators, we set $N_\mathrm{PC} = (10, 11, 13)$ for the GP, NN, and PCE, respectively. For \emph{enhanced-precision}, we selected $N_\mathrm{PC} = (13, 11, 15, 15)$ for GP, NN, PCE, and NPCE.

In Fig.~\ref{fig:combined} we display the fractional difference in the $w$CDM validation set at $z=0$ between the emulated boost, obtained by dividing Eqs.~\ref{eq:pk_from_logb}-\ref{eq:pk_from_qbacco} by $P_{\rm L}(k,z | \boldsymbol{\theta})$, and those computed directly from the simulations using Eq.~\ref{eq:mult_ref_bcase}. For the \emph{default-precision} simulations, we show our emulation errors which use $Q^{\rm S}(k,z)$ as the emulation variable in the middle panel, but also include for visual comparison in the left panel the analogous errors using $Q^{\rm NS}(k,z)$. Comparing the range $0.1 h \text{Mpc}^{-1}< k < 0.5h \text{Mpc}^{-1}$ between the middle and left panels shows the modest effect of the smearing algorithm on emulation errors. Therefore, we stress that the smearing procedure is not a paramount step that needs to be performed when following our emulation strategy. The GP, NN, and PCE \emph{default-precision} emulators all perform similarly with emulation errors below $\sim 0.5\%$ up to the chosen cutoff for \emph{default-precision}, $k_{\rm max} = \pi \, h$Mpc$^{-1}$\footnote{For \emph{default-precision} emulators, we applied $k_{\rm max} = \pi \, h$Mpc$^{-1}$ for $z < 2$ and $k_{\rm max} = \frac{\pi}{2} \, h$Mpc$^{-1}$ for $z>2$. Conversely, for the \emph{enhanced-precision} emulators $k_{\rm max} = 2 \pi$ $h$Mpc$^{-1}$ for $z < 2$ and $k_{\rm max} = \pi$ $h$Mpc$^{-1}$ for $z>2$. Shot noise can dominate the power spectrum output from the simulations on small scales. Because power dissipates on larger scales for higher redshifts, we adjust $k_{\rm max}$ for $z >2$, to prevent shot noise subtraction from producing spurious negative power spectrum values at these redshifts.}.

The \emph{enhanced-precision} emulation errors are shown in the right panel of Fig.~\ref{fig:combined} where we see that GP, NN, and PCE all exhibited inaccuracies above $1\%$, using a higher $k_{\rm max} = 2 \pi$ $h$Mpc$^{-1}$ due to the increased force-resolution. Therefore, for \emph{enhanced-precision} we developed the NPCE method, discussed in Appendix~\ref{app:pce_nn}, which restored emulation errors to below the $0.5\%$ level. However, comparing Fig.~\ref{fig:cola_comparisons} and Fig.~\ref{fig:combined} shows that emulation errors are subdominant on all scales when compared to COLA inaccuracies, so emulation errors are not expected to heavily bias our analysis. Of course, it is possible that a more pronounced difference in performance would appear between the emulation methods in the presence of other emulation choices, such as a lower number of training points, a different $N_{\rm PC}$, or different dimensionality of the input. In Appendix~\ref{app:emulator_equiv}, we perform a test to confirm that the GP, NN, and PCE \emph{default-precision} emulators agree at the level of parameter inference.

As a final note, the emulation errors shown in Fig.~\ref{fig:combined} were computed using COLA simulations calibrated to only a single reference anchor $N_{\rm refs}=1$. For multiple anchors, the emulation errors are sensitive to the choice of the parameter $\sigma_d$ in Eq.~\ref{eq:gaussian_weights}. While low values of $\sigma_d$ suppress the sensitivity of $\tilde{B}(k,z)$ to anchors at larger distances $d_i$, they can also cause inconsistent modeling between different regions of the parameter space which manifest in the emulation errors. This is because the training points nearest to a given test point, and the test point itself, may heavily weight different anchors, causing large disagreements between the emulator and the test simulations. Indeed, avoiding this inconsistency is the motivation for using a weighted sum in Eq.~\ref{eq:mult_ref_bcase} as opposed to using solely the nearest anchor. Therefore, we choose $\sigma_d$ for every set of references by computing the emulation errors of the validation set for different values of $\sigma_d$ and using the lowest value possible before the emulation errors significantly increase beyond those in Fig.~\ref{fig:combined}. We found that regardless of the DE model, $\sigma_d=0.5$ for $N_{\rm refs}=26$, and $\sigma_d=0.3$ for $N_{\rm refs}=101$, did not significantly increase emulation errors above those of the single-anchor cases in Fig.~\ref{fig:combined}.

\begin{table}[t]
	\centering
	\begin{tabular}{| lcc|} 
        \hline
        \textbf{Parameter} & \textbf{Fiducial} & \textbf{Prior}\\ \hline
        \textbf{Survey specifications} & & \\
        Area & $12300 \; \mathrm{deg}^2$ & --\\
        Shape noise per component & $0.26$ & --\\
        $n_\mathrm{eff}^\mathrm{sources}$ & $11.2 \; \mathrm{arcmin}^{-2}$ & -- \\
        \hline
        \textbf{Photometric redshift offsets} & & \\
        $\Delta z_{\mathrm{source}}^{i}$ & 0 & $\mathcal{N}$[0, 0.02]\\
        \hline
        \textbf{Intrinsic alignment (NLA)}& &\\
        $a_{1}$ & 0.7 & $\mathcal{U}$[-5, 5]\\
        $\eta_{1}$ & -1.7 &  $\mathcal{U}$[-5, 5]\\
        \hline
        \textbf{Shear calibration} & & \\
        $m^i$ & 0 & $\mathcal{N}$[0, 0.005]\\
        \hline
	\end{tabular}
    \caption{Survey specifications on which the simulated analysis is based, and priors for LSST nuisance parameters sampled in the MCMCs. $\mathcal{U}[a,b]$ represents an uniform distribution with edges $[a,b]$, while $\mathcal{N}[a,b]$ represents a Gaussian distribution with mean $a$ and standard deviation $b$. $i$ represents the index of the galaxy bin, and all our priors are the same for all bins.}
    \label{tab:lsst_specifications}
\end{table}

\subsection{Analysis of LSST-Y1 Simulated Data}
In this Section, we describe the simulated cosmic shear dataset we analyze with MCMCs, how the data vector and covariance matrix are generated, and how we quantify deviations between COLA and \textsc{ee2}.

\subsubsection{Simulating Cosmic Shear Data}
\label{sec:lsst}
The weak gravitational lensing of light in the large-scale structure of the Universe distorts galaxies' shapes, imposing measurable correlations that depend on the matter power spectrum (see \cite{weak_lensing_review} for a detailed review). Our cosmic shear analysis is comprised of a simulated survey based on LSST-Y1, as in \cite{accelerated-inference-emulator-lsst}.  The fiducial data vectors and covariance matrix used in the analysis are computed following the methodology introduced in~\cite{cosmocov}. The specific survey specifications, such as footprint area and effective number of galaxies per solid angle $n_\mathrm{eff}$ are based on the LSST DESC Science Requirement Document~\citep{lsst-survey-specifications}, and are shown in Table~\ref{tab:lsst_specifications}. The source galaxy redshifts are drawn from a Smail distribution, $n(z) \propto z^2 \exp[ -(z/z_0)^\alpha ]$, normalized by $n_\mathrm{eff}$, where $(z_0, \alpha) = (0.191, 0.870)$. The drawn samples are divided into 5 tomographic redshift bins with equal number of galaxies. The bins are then convolved with a Gaussian redshift uncertainty of $0.02(1+z)$.
The cosmic shear angular two-point correlation function is calculated from theory in the following way. First, the lensing efficiency $q_\kappa^i$ is defined as:
\begin{equation}
    q_\kappa^i(\chi) = \frac{3\Omega_mH_0^2}{2} \times \int_\chi^{\chi_H} d\chi' \left(\frac{\chi' - \chi}{\chi}\right) n^i_\mathrm{s}(z(\chi')) \frac{dz}{d\chi'},
\end{equation}
where $\chi$ is the comoving distance. The two-point correlations in Fourier space can be calculated via the Limber approximation:
\begin{equation}\label{eq:cosmic_shear_harm}
    C^{ij}_{\kappa\kappa}(\ell) = \int \frac{d\chi}{\chi^2} q_\kappa^i(\chi) q_\kappa^j(\chi)P_{\mathrm{NL}}\left(k = \frac{\ell + 1\big/2}{\chi}, z(\chi)\right).
\end{equation}
We convert to real-space correlations by transforming the Fourier-space correlations:
\begin{equation}
    \xi^{ij}_{\pm}(\theta) = \sum_\ell \frac{2\ell + 1}{4\pi} \frac{2(G^+_{\ell,2}(x) \pm G^-_{\ell,2}(x))}{\ell^2(\ell+1)^2}C^{ij}_{\kappa\kappa}(\ell),
    \label{eq:xi}
\end{equation}
where $x = \cos\theta$ and $G^{\pm}$ are analytic functions described in Appendix A of \cite{desy3cov}. The $\xi^{ij}_{\pm}(\theta)$ correlation functions are computed in 26 logarithmically-spaced angular bins between $2.5$ and $900$ arcmin. To compute correlation functions over the whole angular bin, the term
\begin{equation}
G^+_{\ell,2}(x) \pm G^-_{\ell,2}(x),
\end{equation}
is swapped by the averaged function 
\begin{equation}
\overline{G^+_{\ell,2}(x) \pm G^-_{\ell,2}(x)} = \frac{\int_{\theta_\mathrm{min}}^{\theta_\mathrm{max}}(G^+_{\ell,2}(\cos\theta) \pm G^-_{\ell,2}(\cos\theta))d\theta }{\theta_\mathrm{max} - \theta_\mathrm{min}},
\end{equation}
for which analytic expressions are known \cite{desy3_cov}.

Finally, the data vector must be corrected for several systematic effects:
\begin{itemize}
    \item \textbf{Photometric redshift uncertainties:} to account for systematic biases in the photometric redshift measurements, we introduce an offset parameter for the redshift of each tomographic bin, such that the galaxy number densities are computed with $n^i(z + \Delta z^i)$. For generating the fiducial data vector, we assume no redshift uncertainties.
    \item \textbf{Multiplicative shear calibration factor:} to account for systematic biases in the galaxy shape measurements, we introduce a multiplicative factor $m^i$ for each tomographic bin, and the cosmic shear two-point correlation functions are corrected as $\xi^{ij}_\pm \rightarrow (1+m^i)(1+m^j)\xi^{ij}_\pm$.
    \item \textbf{Intrinsic Alignment (IA):} tidal gravitational fields can align galaxy shapes, accounting for correlations that are not caused by weak lensing. To model these correlations, we use the "nonlinear linear alignment" (NLA) model, where we use a linear IA model sourced by the nonlinear matter power spectrum\footnote{The same nonlinear power spectrum used to calculate the data vector.}. The model characterized by an amplitude $a$ and exponent $\eta$ (see, \textit{e.g.} \cite{ia-overview, ia-review}). In this model, the intrinsic shape field $\gamma_{ij}^\mathrm{IA}$ is given by:
    \begin{equation}
        \gamma_{ij}^\mathrm{IA} = -a_1 \bar{C}_1\frac{\rho_\mathrm{c}\Omega_m}{D(z)}\left(\frac{1 + z}{1 + z_0}\right)^{\eta_1}\left(k_ik_j - \frac{1}{3}k^2\delta_{ij}\right)\delta_m(k),
    \end{equation}
    where $\bar{C}_1 = 5 \times 10^{-14}\mathrm{M}^{-1}_\odot h^{-2}\mathrm{Mpc}^3$ is a normalization constant, $\rho_\mathrm{c} = 3H_0^2/8\pi G$ is the critical density, $D(z)$ is the growth factor, $z_0 = 0.62$ is a pivot redshift and $a_1$ and $\eta_1$ are model parameters.
\end{itemize}
The priors and fiducial values for the nuisance parameters are shown in Table~\ref{tab:lsst_specifications}.

We perform the MCMC analysis using \textsc{Cocoa}, the \textsc{Cobaya}-\textsc{Cosmolike} Architecture\footnote{\url{https://github.com/CosmoLike/cocoa}} . \textsc{Cocoa} is a modified version of \textsc{CosmoLike}~\cite{Krause:2016jvl} multi-probe analysis software incorporated into the \textsc{Cobaya} framework~\cite{Lewis:2013hha,Torrado:2020dgo}. The $\xi_{\pm}$ angular correlation functions are calculated in \textsc{CosmoLike}, which takes the linear matter power spectrum from \textsc{CAMB}~\cite{Lewis:1999bs,Howlett:2012mh} corrected for nonlinearities by one of the emulators. We establish convergence of the chains using the Gelman-Rubin criterion when $|R-1| < 0.03$.

The fiducial data vectors of our analyses were generated at different cosmologies in the prior, using \textsc{ee2} as the nonlinear prescription. One fiducial data vector is generated at the \textsc{ee2} reference cosmology, with four more being taken from a $2\times 2$ grid in the $\Omega_m-A_s$ plane, and similarly another four from the $A_s-n_s$ plane, keeping other cosmological parameters at the \textsc{ee2} reference values. These 8 fiducial grid points are chosen for each parameter such that there is a "low" ($\downarrow$) and a "high" ($\uparrow$) value, corresponding to normalized values of $\Theta \in \{0.25,0.75\}$, leading to $\Omega_m \in \{0.28, 0.36\}$, $A_s\times 10^{9} \in \{1.9, 2.3\}$ and $n_s \in \{0.94, 0.98\}$. 

For the four $\Omega_m^{\uparrow}$ cosmologies, we also compute data vectors outside of the $\Lambda$CDM subspace, choosing $w \in \{ -1.1, -0.9\}$. In total, we generate 17 fiducial data vectors. The values of the shifted cosmological parameters in the fiducial data vectors are summarized in Table~\ref{tab:fiducials}. The covariance matrix was computed in \textsc{CosmoCov}, using an analytical approach. Finally, a Gaussian noise realization was generated from the covariance matrix and added to the fiducial data vectors.

\begin{table}[t]
\centering
\renewcommand{\arraystretch}{1.6}
\begin{tabular}{| c | c | c |}
\hline
  $\theta$ & $\theta^{\downarrow}$ & $\theta^{\uparrow}$ \\
\hline \hline
$\Omega_m$ & 0.28 & 0.36 \\\hline
$A_s \times 10^9$ & 1.9 & 2.3 \\\hline
$n_s$ & 0.94 & 0.98 \\\hline
$w$ & -1.1 & -0.9 \\\hline
\end{tabular}
\caption{"Low" ($\downarrow$) and "high" ($\uparrow$) cosmological parameter values used to create the fiducial cosmologies of our analyses.  In this notation, fiducial cosmologies are labeled in the text by the parameters shifted from the \textsc{ee2} reference values listed in Table~\ref{tab:param_space}.}
\label{tab:fiducials}
\end{table}

As we expect more significant discrepancies between \textsc{ee2} and COLA at smaller scales, we test different scale cuts to the data vector. We exclude data points in increasingly smaller angles based on an angular cutoff $\theta_\mathrm{min}$. Since $\xi_-$ at a given angle maps to smaller scales than $\xi_+$, its cutoff is more stringent. We test three angular cutoffs for the cosmic shear correlation functions $\xi_+$ and $\xi_-$:
\begin{itemize}
    \item Cutoff 1 (C1): $\theta_\mathrm{min} = 22.0'$ for $\xi_+$ and $\theta_\mathrm{min} = 69.6'$ for $\xi_-$. This leaves intact 405 out of 780 elements in the data vector.
    \item Cutoff 2 (C2): $\theta_\mathrm{min} = 11.0'$ for $\xi_+$ and $\theta_\mathrm{min} = 34.8'$ for $\xi_-$. This leaves intact 495 out of 780 elements in the data vector.
    \item Cutoff 3 (C3): $\theta_\mathrm{min} = 5.5'$ for $\xi_+$ and $\theta_\mathrm{min} = 17.4'$ for $\xi_-$. This leaves intact 600 out of 780 elements in the data vector.
\end{itemize}

To compute the cosmic shear integral in Eq.~\ref{eq:cosmic_shear_harm}, we must provide the nonlinear matter power spectrum up to $k$ values beyond the emulator limits. We perform a linear extrapolation of $\exp(Q(k))$ vs. $\log(k)$ for the COLA emulators, and of $B(k)$ vs. $\log(k)$ for \textsc{ee2} from its maximum value of $k_{\rm max}^{\textsc{ee2}} \approx 9.41  \, h \text{Mpc}^{-1}$. A Savitzky-Golay filter of order one was used to filter the last $k$-bins of $Q(k)$ from the COLA emulators prior to performing the extrapolation to avoid noise errors. 

\begin{table}[t]
\centering
\begin{tabular}{| c | c | c | c | c | c |}
\hline
\rule{0pt}{15pt} \backslashbox{Cutoff}{$\left<z\right>$}  &  $0.33$ & $0.54$ & $0.74$ & $1.01$ & $1.62$ \\
\hline \hline
Cutoff 1 & 1.4 & 1.1 & 0.9 & 0.9 & 0.8 \\\hline
Cutoff 2 & 2.9 & 2.2 & 1.9 & 1.7 & 1.6 \\\hline
Cutoff 3 & 5.7 & 4.3 & 3.8 & 3.4 & 3.3 \\\hline
\end{tabular}
\caption{Approximate scales in wavenumber $k$, measured in $h$Mpc$^{-1}$, for each galaxy source bin that correspond to the angular cutoffs in $\xi_+$  tested in our LSST-like cosmic shear analysis. The scales are approximated by computing the angular diameter distance to the mean redshift of each bin and converting to a wavenumber.}
\label{tab:k_cuts}
\end{table}

\subsubsection{Quantifying tensions between analyses choices}
One of the main goals of this analysis is to assess the scale cuts that need to be adopted so that COLA provides unbiased constraints to cosmological parameters when compared to analyses based on the \textsc{ee2} emulator. To quantify deviations between parameter constraints from the LSST-Y1 simulated analysis, we first assess the 1D biases in the cosmological parameters $\Omega_m$, $S_8$ and $w$, defined as:
\begin{equation}
    \frac{\Delta \theta}{\sigma_\theta} = \frac{\braket{\theta}_\mathrm{COLA} - \braket{\theta}_\textsc{ee2}}{\sqrt{\sigma_{\theta,\textsc{ee2}}^2 + \sigma_{\theta,\mathrm{COLA}}^2}},
    \label{eq:1d_fob}
\end{equation}
where $\theta$ denotes one of the parameters $\Omega_m$, $S_8$ and $w$, and $\braket{\theta}$ and $\sigma^2_\theta$ are the sample mean and sample variance of the parameter $\theta$ in the chains. We take the difference between COLA and \textsc{ee2} means, rather than the difference between means and fiducial parameters, to avoid projection effects. 

The one-dimensional bias does not capture correlations between cosmological parameters. Therefore, we also employ the Figure of Bias (FoB) metric, which generalizes the 1D bias for multidimensional distributions, defined by:
\begin{equation}
    \mathrm{FoB}(\boldsymbol{\theta}) = [\Delta\braket{\boldsymbol{\theta}}^\mathrm{T} \cdot (C_\mathrm{COLA} + C_\textsc{ee2})^{-1} \cdot \Delta\braket{\boldsymbol{\theta}}]^{1/2},
    \label{eq:fob}
\end{equation}
Here $\boldsymbol{\theta}$ denotes the vector cosmological parameters, $\cdot$ denotes matrix multiplication, $\Delta\braket{\boldsymbol{\theta}} = \braket{\boldsymbol{\theta}}_\mathrm{COLA} - \braket{\boldsymbol{\theta}}_\textsc{ee2}$ is the difference in parameter sample means in the chains and $C_\mathrm{COLA}$ and $C_\textsc{ee2}$ are the sample covariance matrices of the cosmological parameters obtained in the MCMCs. In this analysis, $\boldsymbol{\theta}$ contains the 6 cosmological parameters listed in Table~\ref{tab:param_space}, thus we refer to it as the 6D bias.

As a goodness-of-fit test, we compute the difference in $\chi^2$ between the COLA emulators and \textsc{ee2}, $\Delta \chi^2 \equiv \chi^2_{\rm COLA} - \chi^2_{\textsc{ee2}}$, by performing importance sampling of the \textsc{ee2} chains and recalculating the $\chi^2$ for the same points using the COLA emulators as the nonlinear prescription.


\section{Results for LSST-Y1 Simulated Analysis}
\label{sec:lsst-results}

\begin{figure}[t]
    \centering
    \includegraphics[width=0.9\columnwidth]{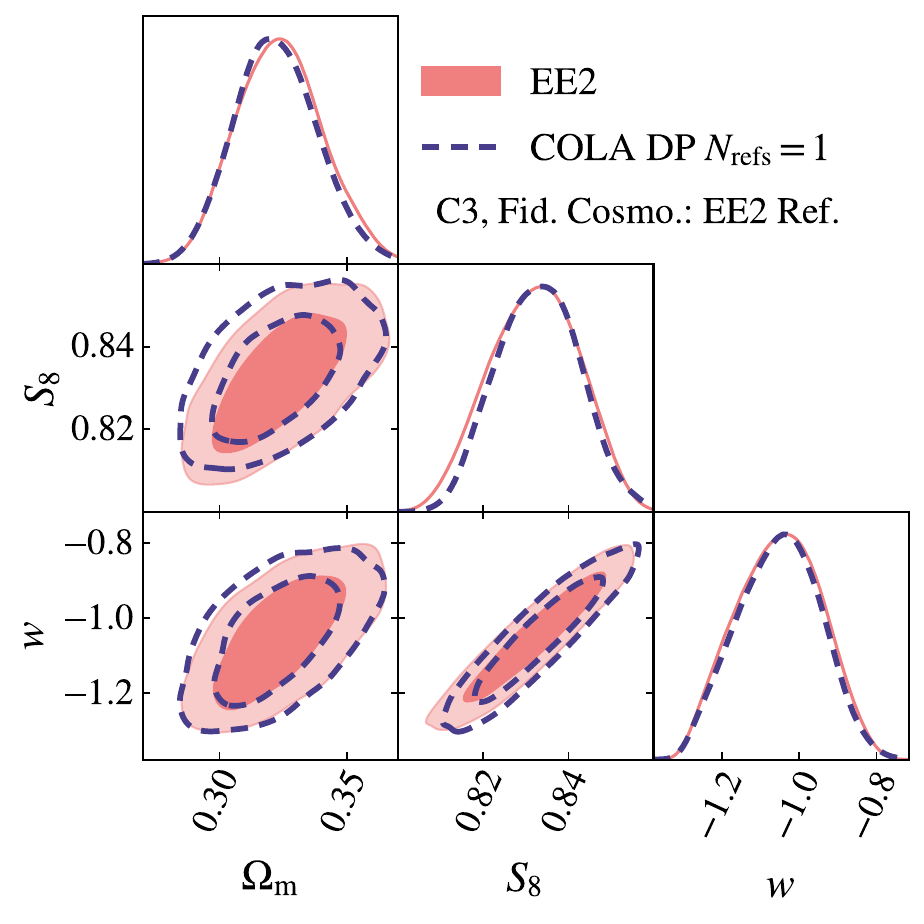}
    \includegraphics[width=0.9\columnwidth]{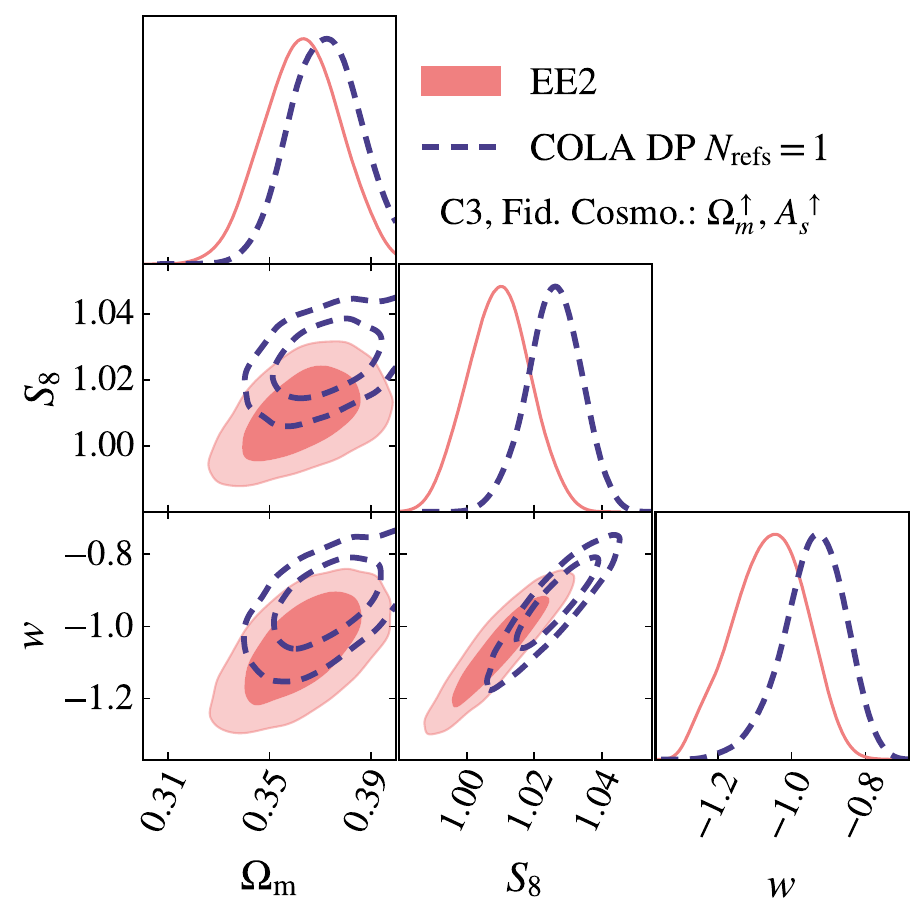}
    \caption{Confidence contours (68\% and 95\%) of the parameters $\Omega_m$, $S_8$, and $w$ when the nonlinear matter power spectrum is computed using either \textsc{ee2} (pink filled contours) or the baseline $w$CDM COLA emulator (dashed blue lines), all using the most aggressive Cutoff 3. In the top panel we place the fiducial cosmology at the reference cosmology of the emulator. This yields a low $\rm{FoB}~=0.11$, indicating that only the baseline emulator is needed when the reference is optimally positioned inside the posterior. However, the bottom panel shows that this same emulator produces large biases in regions far from the reference, yielding an $\rm{FoB}~=5.37$. These biases correlate strongly with $\sigma_8$, with the fiducial value used in the bottom panel ($\sigma_8 = 0.93$, $S_8=1.02$) being significantly greater than that of the \textsc{ee2} reference cosmology ($\sigma_8 = 0.82$, $S_8=0.84$).}
\label{fig:triangle_plots_cola_vs_ee2}
\vspace{-0.5cm}
\end{figure}

In this section, we present results from our $w$CDM analysis of the LSST-Y1 simulated data using COLA emulators. In Appendix~\ref{app:emulator_equiv}, it can be seen that the GP, NN, and PCE \emph{default-precision} emulators show strong agreement in parameter estimation. Therefore, we simply employ the PCE method for all \emph{default-precision} COLA emulators in the subsequent tests, unless otherwise stated. However, the NPCE method was used for all \emph{enhanced-precision} emulators, due to its improved accuracy shown in Fig.~\ref{fig:combined}.

We run MCMCs for a variety of fiducial cosmologies and angular scale cuts, using both \textsc{ee2} and our COLA emulators to provide the nonlinear correction for the matter power spectrum. We test a baseline COLA emulator, which employs \emph{default-precision} simulations and a single anchor at the \textsc{ee2} reference cosmology ($N_{\rm refs}=1$). We also test modifications to this baseline approach, including the use of \emph{enhanced-precision} COLA settings, $N_{\rm refs}^{w \rm CDM}=26$ reference anchors, and $N_{\rm refs}^{\Lambda \rm CDM}=101$ reference anchors. Following those results, we introduce and investigate two more methods which we label $N_{\rm refs}^{\Lambda \rm CDM}=500$ and $N_{\rm refs}^{\Lambda\rm CDM}=\infty$, discussed in Sec.~\ref{sec:inf_refs}. We will use as a guideline the criteria adopted in the DES-Y3 analysis~\cite{krause2021dark}, seeking a 6D $\rm{FoB}<0.3$ (Eq.~\ref{eq:fob}) in the parameters $\{ H_0, \Omega_m, \Omega_b, A_s, n_s, w\}$ between \textsc{ee2} and COLA constraints, and a difference $|\Delta\chi^2|<1$ for importance-sampled points. 

\subsection{Default-Precision Settings Using \texorpdfstring{$N_\mathrm{refs}=1$}{}}

We begin by testing our baseline COLA emulator in the most idealized scenario when the fiducial cosmology is exactly equal to the reference cosmology of the emulator. In this case, the emulator will be calibrated by Eq.~\ref{eq:b_case} to agree with \textsc{ee2} well for points in the high-likelihood region of the posterior, thus one expects optimal agreement between the two prescriptions. While we perform this test using the baseline emulator, we also apply the most aggressive mask Cutoff 3. The results are shown in the triangle plot in the top panel of Fig.~\ref{fig:triangle_plots_cola_vs_ee2}, where we see strong agreement in 2D contours between the COLA emulator and \textsc{ee2}. Specifically, the 6D FoB given by Eq.~\ref{eq:fob} was $0.11$. With regards to goodness-of-fit, we obtain in this case $66.5\%$ of points with $|\Delta\chi^2|<1$. Therefore, we have the immediate result that even our baseline COLA approach has the potential for strong agreement with a high-resolution \textit{N}-body emulator on the most nonlinear scales tested in this paper, as long as the reference cosmology is optimally located inside the high-likelihood region. This suggests that performing exploratory analyses, such as initial chains using linear scale cuts, could be a useful approach to identify ideal locations to place anchors before the final chain. 

\begin{figure}[t]
    \centering
    \includegraphics[width=0.95\columnwidth]{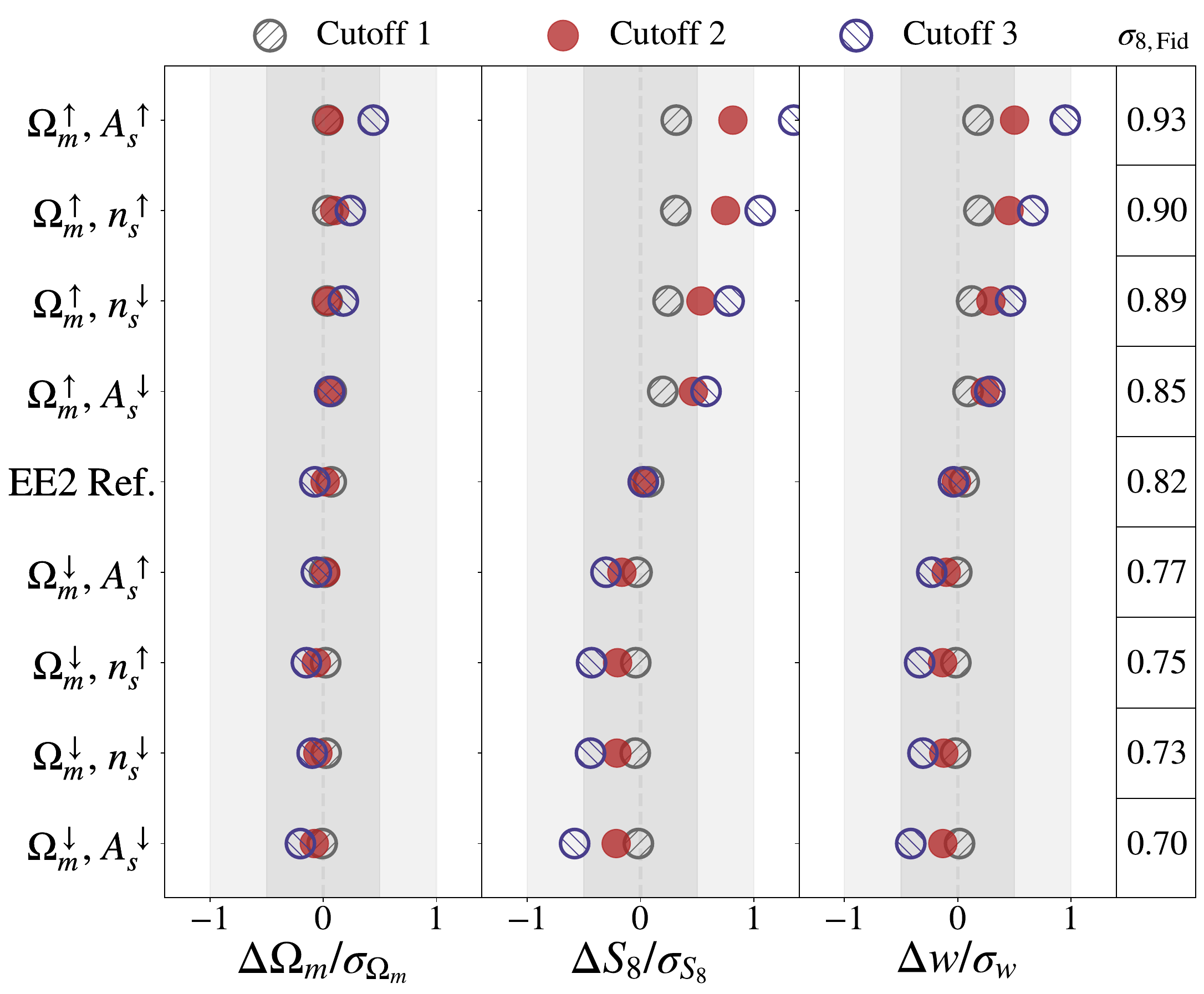}
    \caption{The 1D measure of bias defined in Eq.~\ref{eq:1d_fob}, between the \emph{default-precision} COLA emulator and \textsc{ee2}, when only the single \textsc{ee2} reference anchor is used. In addition to the \textsc{ee2} reference, we test this baseline emulator on 8 fiducial cosmologies far from the reference, applying three different scale cuts described in Sec.~\ref{sec:lsst}. The 8 shifted fiducial cosmologies on the vertical axis are obtained by starting from the \textsc{ee2} reference values shown in Table~\ref{tab:param_space}, and shifting pairs of values to those listed in Tab.~\ref{tab:fiducials}. The fiducial cosmologies on the vertical axis are listed in decreasing order of their associated $\sigma_8$ values, showing a strong correlation with the biases. We use the same color scheme to distinguish the scale cuts in plots of 1D bias throughout the paper.}
    \label{fig:1d_diff}
\end{figure}

\begin{figure}[t]
    \centering
    \includegraphics[width=1.0\columnwidth]{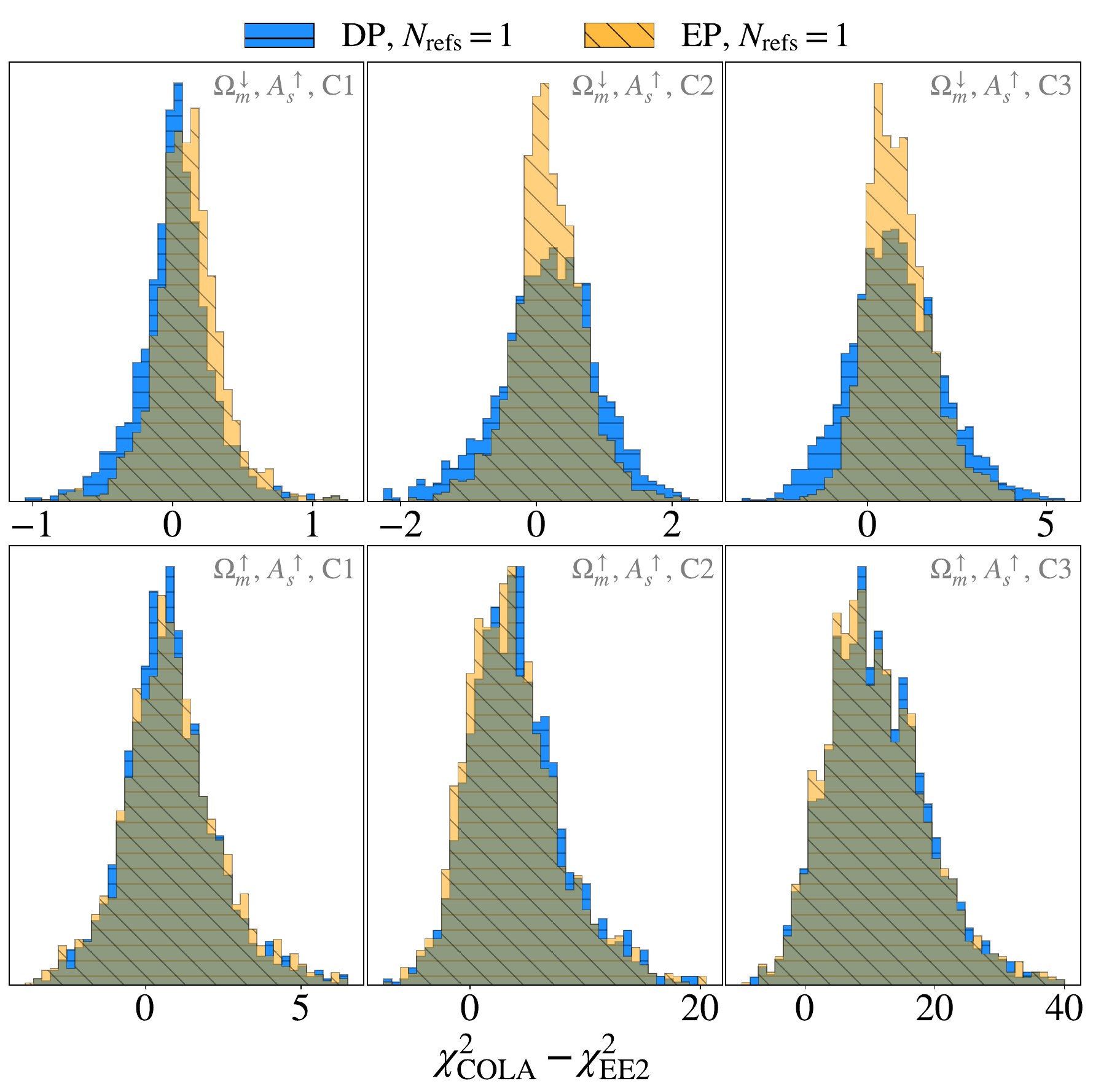}
    \caption{Histograms of $\chi^2$ differences between COLA and \textsc{ee2} at two fiducial cosmologies: ($\Omega_m^{\uparrow}$, $A_s^{\uparrow}$) and ($\Omega_m^{\downarrow}$, $A_s^{\uparrow}$), and the three angular scale cuts tested. We burn-in and thin the \textsc{ee2} chains, and calculate the $\chi^2$ for these cosmologies when using COLA as the nonlinear model, in order to compute the difference. Here we use COLA emulators calibrated only with the \textsc{ee2} reference cosmology, testing both COLA precision settings: \emph{default-precision} (blue with horizontal stripes) and \emph{enhanced-precision} (yellow with diagonal stripes).}
    \label{fig:chi2_precision}
\end{figure}

Nevertheless, given that in general the high-likelihood region is not known a priori, it is desirable to have an emulator that obtains accurate constraints throughout the whole parameter space. Hence, we perform MCMCs using fiducial cosmologies in which two cosmological parameters at a time are significantly shifted from the \textsc{ee2} reference values, in order to reassess the agreement between the baseline COLA emulators and \textsc{ee2} in these regions of the prior. We begin these tests by considering all 4 combinations of high and low $\Omega_m$ and $A_s$ values, and all 4 combinations of high and low $\Omega_m$ and $n_s$ values, according to Table~\ref{tab:fiducials}. We then repeat the previous analysis using the baseline COLA emulator with Cutoff 3 with these 8 fiducial cosmologies. The 2D contours corresponding to the fiducial cosmology $(\Omega_m^\uparrow, A_s^\uparrow)$, the most biased of these 8 cases, are shown in the bottom panel of Fig. ~\ref{fig:triangle_plots_cola_vs_ee2} where we now see a strong disagreement between \textsc{ee2} and COLA constraints. Specifically, the full $\rm{FoB}=5.37$ when the fiducial was shifted to $(\Omega_m^\uparrow, A_s^\uparrow)$. However, in a test where the emulator's single reference was set to the fiducial values of $(\Omega_m^\uparrow, A_s^\uparrow)$, the FoB was restored to a low value of $0.40$, near the target of 0.3.

More comprehensively, the 1D bias metric of Eq.~\ref{eq:1d_fob} between the baseline COLA emulator and \textsc{ee2} for all 8 of the aforementioned fiducial cosmologies, and all three angular scale cuts, is shown in Fig.~\ref{fig:1d_diff} for the parameters $\Omega_m$, $S_8$ and $w$. The fiducial cosmologies are listed in decreasing order of $\sigma_8$ as it displays a clear correlation with all parameter biases, especially for fiducial $\sigma_8$ values larger than the reference value $\sigma_{8,\rm ref}=0.82$. The largest 1D biases in $S_8$ between COLA and \textsc{ee2} are for the fiducial cosmology $(\Omega_m^\uparrow, A_s^\uparrow)$, which has $\sigma_8=0.93$. We also notice that $S_8$ tends to be the most biased parameter, as it is the most constrained by cosmic shear surveys. For the $(\Omega_m^\uparrow, A_s^\uparrow)$ cosmology, the baseline emulator produces 1D biases in $S_8$ of $0.32\sigma_{S_8}$, $0.82\sigma_{S_8}$ and $1.35\sigma_{S_8}$ using Cutoffs 1, 2, and 3, respectively. The 6D FoB in these cases is $1.12$, $3.18$ and $5.37$. However, the maximum FoB for Cutoff 1 occurred for the  $(\Omega_m^\uparrow, n_s^\uparrow)$ cosmology, which has the second largest $\sigma_8=0.90$, producing an FoB of 1.27. 

\begin{figure}[t]
    \centering
    \includegraphics[width=\columnwidth]{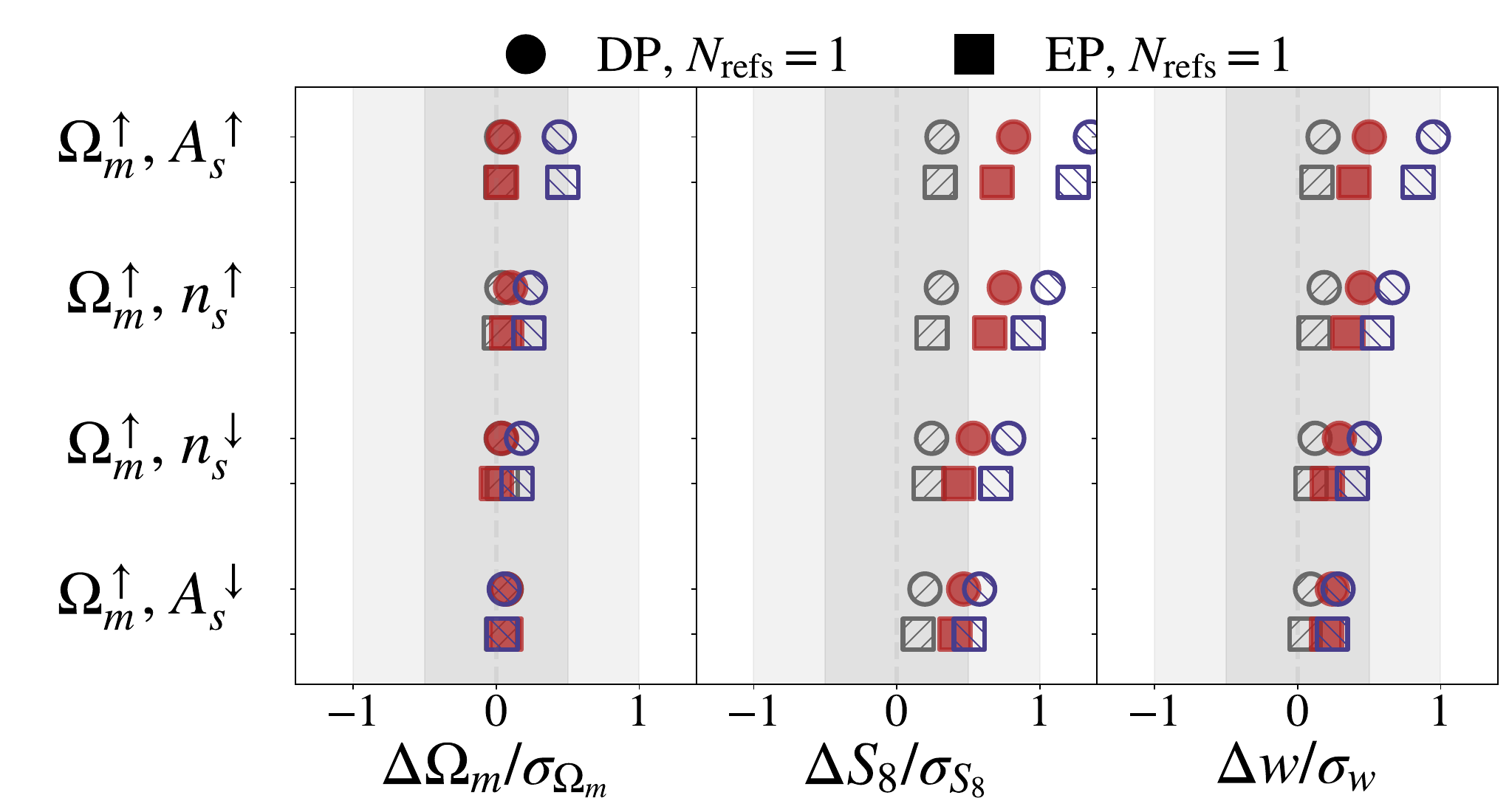}
    \includegraphics[width=\columnwidth]{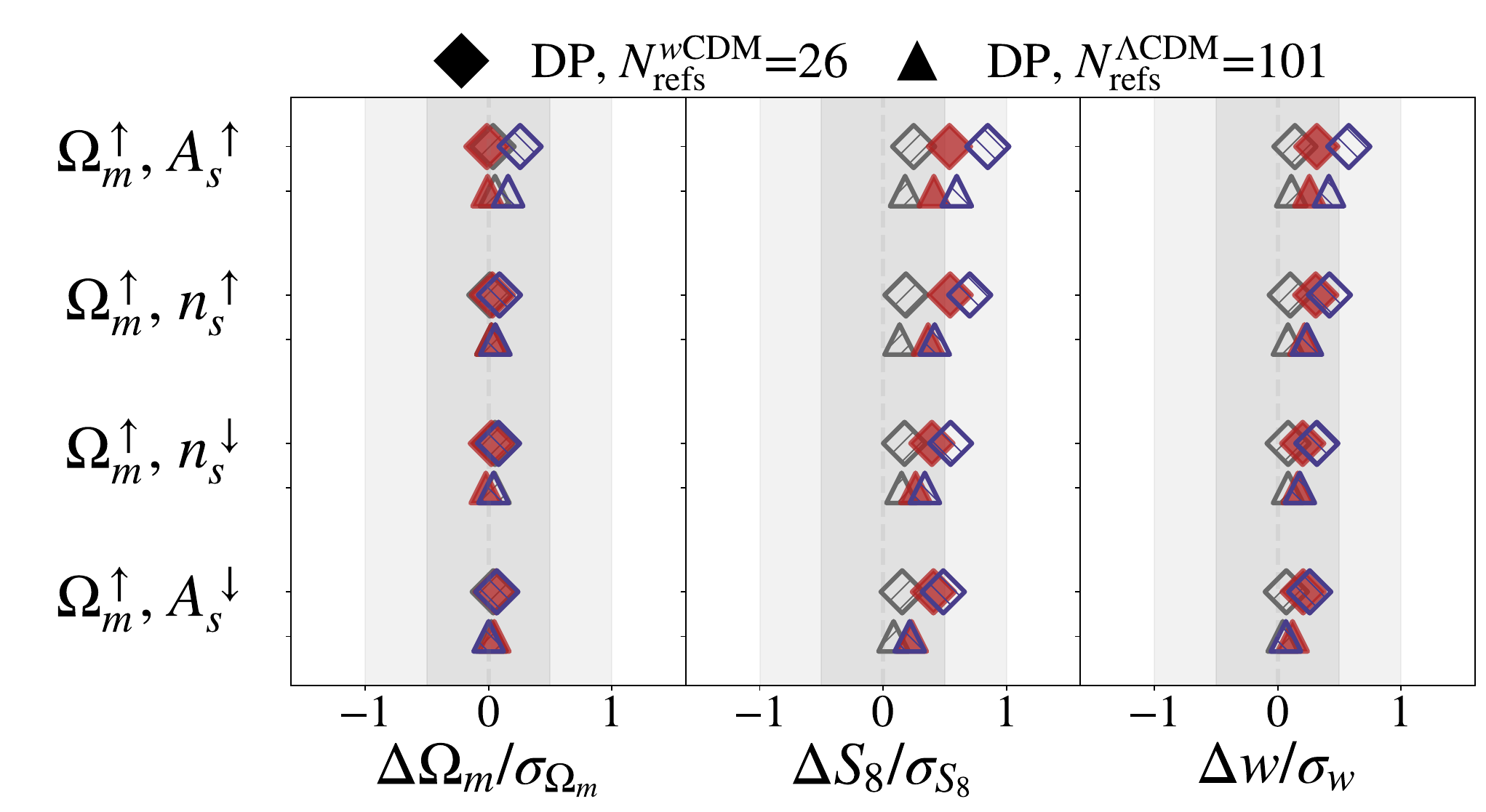}
    \caption{The 1D measures of bias against \textsc{ee2}, now including three additional COLA emulators, using the high $\Omega_m$ fiducial cosmologies of Fig. \ref{fig:1d_diff}. The top panel shows the improvement in bias due to using \emph{enhanced-precision} simulations for the single-reference emulators. Alternatively, the bottom panel compares two approaches of increasing the number of reference anchors: using $N_{\rm refs}^{w \rm CDM}=26$ references, vs. $N_{\rm refs}^{\Lambda \rm CDM}=101$. 
    }
    \label{fig:1d_colas}
\end{figure}

Additionally, in Fig.~\ref{fig:chi2_precision} we show the $\Delta \chi^2$ distribution for cosmologies from the \textsc{ee2} chains for a fiducial cosmology with low bias $(\Omega_m^\downarrow, A_s^\uparrow)$, and with high bias $(\Omega_m^\uparrow, A_s^\uparrow)$. For the latter region, the $\Delta \chi^2$ disagreement reaches as high as $5$, $20$, and $40$ for the three respective scale cuts. Therefore, while the baseline approach achieves low bias near the reference, in other regions of the prior it can impose non-negligible bias and $\Delta \chi^2$ compared to \textsc{ee2}, for all three scale cuts.

\subsection{Enhanced-Precision Settings Using \texorpdfstring{$N_{\rm refs}=1$}{}}
With the objective of creating an emulator with low bias throughout the prior, we now investigate the impact of using our \emph{enhanced-precision} COLA emulator with $N_{\rm refs}=1$, and compare it to the baseline emulator. In the top panel of Fig.~\ref{fig:1d_colas} we repeat the tests of the 1D bias from Fig.~\ref{fig:1d_diff} for the \emph{enhanced-precision} emulator, but only for the high $\Omega_m$ fiducial cosmologies, as they all displayed higher bias than their low $\Omega_m$ counterparts. The observed trend indicates that the \emph{enhanced-precision} emulator did consistently reduce biases, though minimally. Considering the most biased fiducial cosmology of the \emph{enhanced-precision} emulator $(\Omega_m^\uparrow, A_s^\uparrow)$, we compare the tension to the \emph{default-precision} case:

\begin{itemize}
    \item DP $N_\mathrm{refs} = 1$: $\Delta S_8 / \sigma_{S_8} = 0.32$, FoB $=1.12$ (C1);
    \item EP $N_\mathrm{refs} = 1$: $\Delta S_8 / \sigma_{S_8} = 0.30$, FoB $=0.98$ (C1);
    \item DP $N_\mathrm{refs} = 1$: $\Delta S_8 / \sigma_{S_8} = 0.82$, FoB $=3.18$ (C2);
    \item EP $N_\mathrm{refs} = 1$: $\Delta S_8 / \sigma_{S_8} = 0.70$, FoB $=2.52$ (C2).
\end{itemize}

\begin{figure}[t]
    \centering
    \includegraphics[width=1.0\columnwidth]{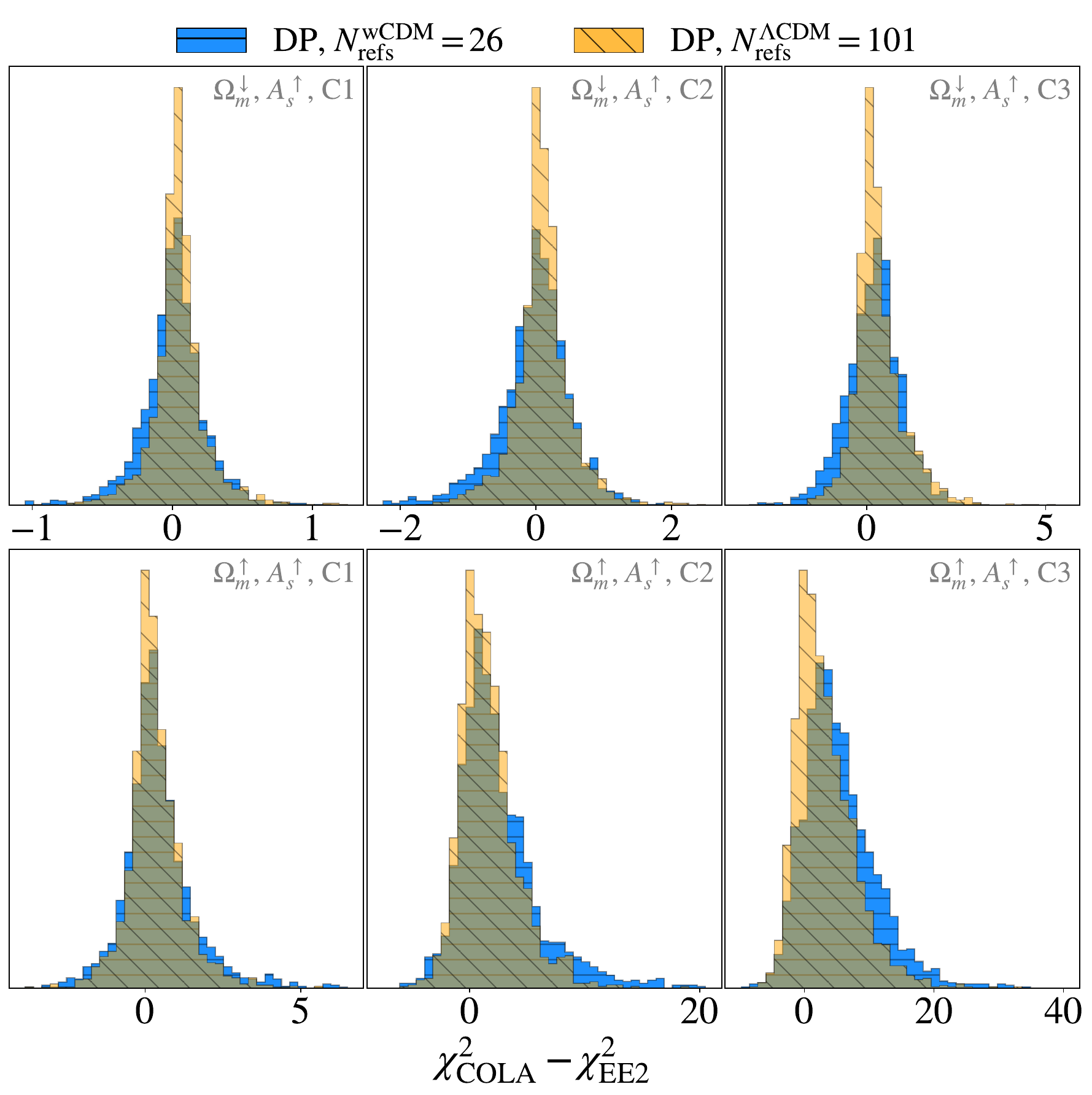}
    \caption{$\Delta\chi^2$ histograms between the COLA emulators and \textsc{ee2}, as in Fig.~\ref{fig:chi2_precision}, now for \emph{default-precision} emulators calibrated to multiple references. We assess the $N_{\rm refs}^{w\rm CDM}=26$ emulator (blue with horizontal stripes) and the $N_{\rm refs}^{\Lambda\rm CDM}=101$ emulator (yellow with diagonal stripes).}
    \label{fig:chi2_references}
\end{figure}

Fig.~\ref{fig:chi2_precision} shows the difference in $\chi^2$ between the $N_{\rm refs}=1$ emulators and \textsc{ee2} at two of the fiducial cosmologies used in Fig.~\ref{fig:1d_diff}: $(\Omega_m^\downarrow, A_s^\uparrow)$ and $(\Omega_m^\uparrow, A_s^\uparrow)$. The lower bias fiducial cosmology with $\Omega_m=\Omega_m^\downarrow$ also shows a better agreement in $\chi^2$ with \textsc{ee2}, with $98.7\%$ of the points having $|\Delta\chi^2| < 1$ for the \emph{default-precision} emulator using Cutoff 1, increasing to $98.9\%$ for the \emph{enhanced-precision} emulator. More significant disagreements begin to appear at Cutoff 2, with $80.3\%$ of the points having $|\Delta\chi^2| < 1$ for \emph{default-precision} and $92.3\%$ for \emph{enhanced-precision}. For the $(\Omega_m^\uparrow, A_s^\uparrow)$ fiducial cosmology, even Cutoff 1 shows a significant disagreement between COLA and \textsc{ee2}, with only $51.7\%$ of the points having $|\Delta\chi^2| < 1$ with \emph{default-precision}, decreasing to $48.1\%$ using \emph{enhanced-precision}. More details are provided in Table~\ref{tab:fraction_chi2}.

These results indicate that the \emph{enhanced-precision} emulator only slightly reduced bias against \textsc{ee2} compared to the \emph{default-precision} case, managing for Cutoff 1 to keep the largest $S_8$ bias to below 0.3$\sigma_{S_8}$ and FoB $<1$. However, both single-reference emulators produced large biases and fail to achieve the desired FoB $<0.3$ throughout the prior, particularly in regions with $\sigma_8$ far from the reference value.

\subsection{Default-Precision Settings Using \texorpdfstring{$N_{\rm refs}^{w \rm CDM}=26$}{} and \texorpdfstring{$N_{\rm refs}^{\Lambda \rm CDM}=101$}{}}
Alternatively, we now consider two approaches for calibrating the \emph{default-precision} COLA simulations with multiple reference anchors according to Eq.~\ref{eq:mult_ref_bcase}. We compare the approaches of using a low number of anchors in the extended model $N_{\rm refs}^{w \rm CDM}=26$, and using a high number confined to the $\Lambda$CDM subspace $N_{\rm refs}^{\Lambda \rm CDM}=101$. In the bottom panel of Fig.~\ref{fig:1d_colas}, as in the \emph{enhanced-precision} tests, we present the 1D biases of the multiple-reference emulators for the high $\Omega_m$ fiducial cosmologies. A comparison to the top panel suggests that either approach of using additional anchors to calibrate the COLA simulations is more effective at mitigating bias than using the \emph{enhanced-precision} settings with a single anchor.

\begin{table}[t]
\centering
\setlength{\tabcolsep}{6pt}
\renewcommand{\arraystretch}{1.6}
\begin{tabular}{ |c |c|c|c|c| }
\hline
\rule{0pt}{15pt} \backslashbox{Cosmo.}{Emul.} & DP, 1 & EP, 1 &  \makecell{DP, 26 \\ (wCDM)}  &  \makecell{DP, 101 \\ ($\Lambda$CDM)}  \\\hline\hline
$\Omega_m^\downarrow$, $A_s^\uparrow$, C1   & $98.7\%$      & $98.9\%$    & $99.3\%$    & $99.2\%$   \\ \hline
$\Omega_m^\downarrow$, $A_s^\uparrow$, C2   & $80.3\%$      & $92.3\%$     & $92.4\%$    & $96.0\%$  \\ \hline
$\Omega_m^\downarrow$, $A_s^\uparrow$, C3   & $50.4\%$     & $58.9\%$     & $80.2\%$     & $83.7\%$   \\ \hline\hline
$\Omega_m^\uparrow$, $A_s^\uparrow$, C1 & $51.4\%$            & $48.2\%$     & $71.0\%$     & $77.8\%$   \\ \hline
$\Omega_m^\uparrow$, $A_s^\uparrow$, C2 & $14.2\%$             & $17.0\%$   & $28.5\%$    & $35.8\%$      \\ \hline
$\Omega_m^\uparrow$, $A_s^\uparrow$, C3 & $4.1\%$                & $4.3\%$    & $10.6\%$    & $22.6\%$      \\ \hline
\end{tabular}
\caption{Percentages of MCMC accepted points with $|\chi^2_\mathrm{COLA} - \chi^2_\textsc{ee2}| < 1$ at two representative cosmologies. We include four different emulators, labeled by the precision of the COLA simulations used, and the number of \textsc{ee2} reference anchors used. The results overall show the best agreement in goodness-of-fit for the $N_{\rm refs}^{\Lambda \rm CDM}=101$ emulator relative to \textsc{ee2}. We consistently see that the COLA errors depend significantly on the value of $\Omega_m$, which is correlated with $\sigma_8$, with better agreement when using fiducial cosmologies with low $\Omega_m$ values compared to high $\Omega_m$.}
\label{tab:fraction_chi2}
\end{table}

\begin{figure}[t]
    \centering
    \includegraphics[width=\columnwidth]{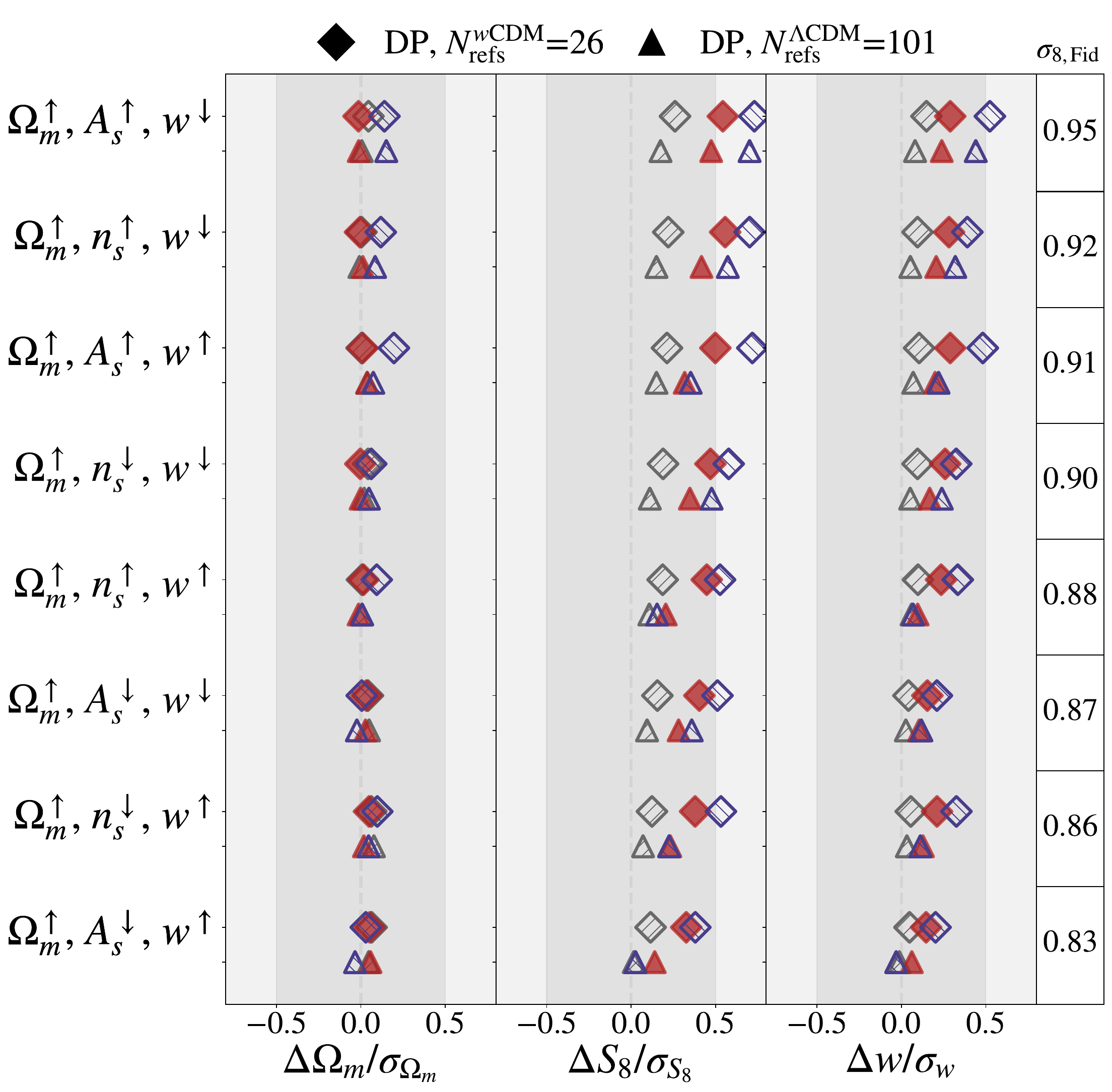}
    \caption{1D biases between two COLA \emph{default-precision} emulators and \textsc{ee2} for fiducial cosmologies outside of the $\Lambda$CDM region, for the two most aggressive scale cuts. While Fig.~\ref{fig:1d_colas} showed less bias for the $N_{\rm refs}^{\Lambda \rm CDM}=101$ emulator compared to the $N_{\rm refs}^{w \rm CDM}=26$ emulator for fiducial cosmologies within $\Lambda$CDM, we consider whether this holds when the same fiducial cosmologies are shifted in $w$.}
    \label{fig:w_fiducials_1d}
\end{figure}

Additionally, the bottom panel of Fig.~\ref{fig:1d_colas} shows that for these $\Lambda$CDM fiducial cosmologies, the strategy of using a large number of $\Lambda$CDM anchors consistently produced less bias against \textsc{ee2}, than did a low number of $w$CDM anchors. The largest biases continue to occur at the fiducial cosmology $(\Omega_m^\uparrow, A_s^\uparrow)$. For this cosmology, we now see improvements in the tension for Cutoff 1 and Cutoff 2:
\begin{itemize}
    \item  DP $N_\mathrm{refs}^{w\mathrm{CDM}} = 26$: $\Delta S_8 / \sigma_{S_8} = 0.25$, FoB $=0.91$ (C1);
    \item  DP $N_\mathrm{refs}^{\Lambda\mathrm{CDM}} = 101$: $\Delta S_8 / \sigma_{S_8} = 0.18$, FoB $=0.69$ (C1);
    \item  DP $N_\mathrm{refs}^{w\mathrm{CDM}} = 26$: $\Delta S_8 / \sigma_{S_8} = 0.54$, FoB $=2.17$ (C2);
    \item  DP $N_\mathrm{refs}^{\Lambda\mathrm{CDM}} = 101$: $\Delta S_8 / \sigma_{S_8} =0.41$, FoB $=1.61$ (C2);
\end{itemize}

Furthermore, Fig.~\ref{fig:chi2_references} indicates that using multiple anchors improved the goodness-of-fit of COLA, with the emulator calibrated to a high number of $\Lambda$CDM anchors showing the smallest differences in $\chi^2$ compared to \textsc{ee2}. Whereas the baseline emulator had $80.3\%$ of the points with $|\Delta\chi^2| < 1$ compared to \textsc{ee2} using Cutoff 2 and the $(\Omega_m^{\downarrow},A_s^{\uparrow})$ fiducial cosmology, the $N_\mathrm{refs}^{w\mathrm{CDM}} = 26$ and $N_\mathrm{refs}^{\Lambda\mathrm{CDM}} = 101$ emulators managed to achieve $92.4\%$ and $96.0\%$ respectively. In addition, for the $(\Omega_m^{\downarrow},A_s^{\uparrow})$ fiducial cosmology using Cutoff 1, the $N_\mathrm{refs}^{w\mathrm{CDM}} = 26$ and $N_\mathrm{refs}^{\Lambda\mathrm{CDM}} = 101$ emulators had $71.0\%$ and $77.8\%$ of points with $|\Delta\chi^2| < 1$, compared to the $51.4\%$ of the baseline emulator. A more thorough summary is provided in Table~\ref{tab:fraction_chi2}.

As a final comparison between these two approaches of using multiple anchors, we test whether the $N_{\rm refs}^{w\mathrm{CDM}}=26$ emulator has an advantage in regions of the prior outside of the $\Lambda$CDM space, by varying the fiducial value of $w$. We start from the 4 fiducial cosmologies with high $\Omega_m$, but shift the value of $w$ to $w^\uparrow = -0.9$ and $w^\downarrow = -1.1$, creating 8 new fiducial cosmologies. The 1D biases at those fiducial cosmologies are shown in Fig.~\ref{fig:w_fiducials_1d}. The result appears to hold that the $N_\mathrm{refs}^{\Lambda\mathrm{CDM}} = 101$ emulator consistently mitigates large biases more effectively than the $N_\mathrm{refs}^{w\mathrm{CDM}} = 26$ emulator. We perform our remaining tests using fiducial cosmologies with $w \neq -1$, and present more results regarding the $N_\mathrm{refs}^{\Lambda\mathrm{CDM}} = 101$ approach in the next section.

We emphasize that we are investigating the COLA method in the context of extended models for which there exist at best only computationally expensive prescriptions, hence this expense extends to computing reference anchors that fill the entire parameter space. Because high-resolution \textit{N}-body emulators already exist for $\Lambda$CDM, the only significant computational expense to adding $\Lambda$CDM anchors under the present approach is the associated COLA simulations, though we test eliminating this expense as well in Sec.~\ref{sec:inf_refs}. Therefore, using a large number of $\Lambda$CDM references to calibrate COLA emulators in extended models is likely more computationally efficient than even a modest number of anchors that do not leverage existing emulators. Hence, we perform our remaining tests using a large number of $\Lambda$CDM references.

\subsection{Enhanced-Precision Settings Using \texorpdfstring{$N_{\rm refs}^{\Lambda \rm CDM}=101$}{}}

\begin{figure}[t]
    \centering
    \includegraphics[width=\columnwidth]{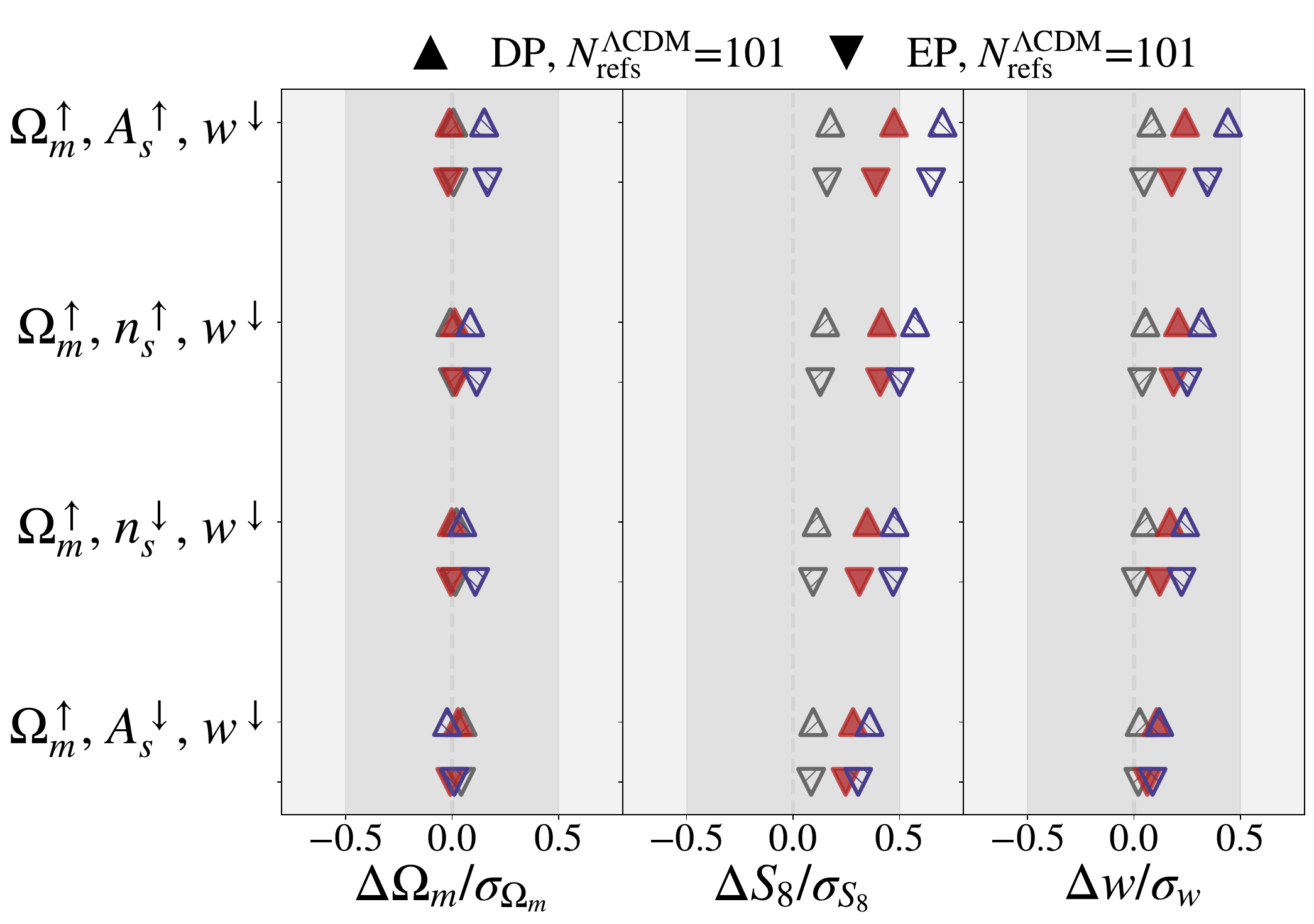}
    \caption{1D measures of bias between COLA emulators and \textsc{ee2} for fiducial cosmologies outside of the $\Lambda$CDM region, for the all three scale cuts. We use fiducial cosmologies with $w=w^{\downarrow}$ as Fig.~\ref{fig:w_fiducials_1d} shows larger bias in these cases compared to the analogous $w=w^{\uparrow}$ cases. We test \emph{default-precision} and \emph{enhanced-precision} emulators with number of references $N_{\rm refs}^{\Lambda \rm CDM}=101$.}
    \label{fig:w_fiducials_1d_precision}
\end{figure}

Due to the reduction of bias of the COLA emulators using the $N_{\rm refs}^{\Lambda \rm CDM}=101$ anchors approach, and the improved goodness-of-fit, we now train an $N_\mathrm{refs}^{\Lambda\mathrm{CDM}} = 101$ emulator with \emph{enhanced-precision} simulation settings to assess whether the combination of the increased precision settings and multiple anchors techniques can further reduce bias across the prior. We test this emulator for the most biased fiducial cosmologies from Fig.~\ref{fig:w_fiducials_1d}, those with $w=w^{\downarrow}$, displaying in Fig.~\ref{fig:w_fiducials_1d_precision} the 1D biases and comparing them to the \emph{default-precision} case. As in the case of $N_{\rm refs}=1$, we see only moderate improvement in the bias due to the use of the \emph{enhanced-precision} simulations over \emph{default-precision}. The largest tensions occur for the fiducial cosmology ($\Omega_m^{\uparrow}$, $A_s^{\uparrow}$, $w^{\downarrow}$) which has a high $\sigma_8=0.95$:
\begin{itemize}
    \item DP $N_\mathrm{refs}^{\Lambda\mathrm{CDM}} = 101$: $\Delta S_8 / \sigma_{S_8} =0.17$, FoB $=0.54$ (C1);
    \item EP $N_\mathrm{refs}^{\Lambda\mathrm{CDM}} = 101$: $\Delta S_8 / \sigma_{S_8} =0.16$, FoB $=0.40$ (C1);
    \item DP $N_\mathrm{refs}^{\Lambda\mathrm{CDM}} = 101$: $\Delta S_8 / \sigma_{S_8} =0.47$, FoB $=1.54$ (C2);
    \item EP $N_\mathrm{refs}^{\Lambda\mathrm{CDM}} = 101$: $\Delta S_8 / \sigma_{S_8} =0.39$, FoB $=1.10$ (C2);
    \item DP $N_\mathrm{refs}^{\Lambda\mathrm{CDM}} = 101$: $\Delta S_8 / \sigma_{S_8} =0.70$, FoB $=2.61$ (C3);
    \item EP $N_\mathrm{refs}^{\Lambda\mathrm{CDM}} = 101$: $\Delta S_8 / \sigma_{S_8} =0.65$, FoB $=2.20$ (C3).
\end{itemize}
Thus, even in the most extreme cases tested, both of these emulators maintain low bias for Cutoff 1, with below $0.2\sigma_{S_8}$ tension in $S_8$. However, neither emulator achieves the desired $\rm{FoB}<0.3$. Additionally, for Cutoff 2 both emulators were pushed above $0.3\sigma_{S_8}$ tension in $S_8$ and an $\rm{FoB}>1$.

\begin{table}[t]
\centering
\setlength{\tabcolsep}{6pt}
\renewcommand{\arraystretch}{1.6}
\begin{tabular}{ |c |c|c| }
\hline
\rule{0pt}{15pt} \backslashbox{Cosmo.}{Emul.} &  \makecell{DP, 101 \\ ($\Lambda$CDM)} & \makecell{EP, 101 \\ ($\Lambda$CDM)}  \\\hline\hline
$\Omega_m^\uparrow$, $A_s^\uparrow$, $w^\downarrow$, C1   & $76.8\%$  & $73.1\%$    \\ \hline
$\Omega_m^\uparrow$, $A_s^\uparrow$, $w^\downarrow$, C2   & $32.7\%$      & $36.6\%$     \\ \hline\hline
$\Omega_m^\uparrow$, $A_s^\downarrow$, $w^\downarrow$, C1   & $83.6\%$     & $81.7\%$      \\ \hline
$\Omega_m^\uparrow$, $A_s^\downarrow$, $w^\downarrow$, C2   & $52.6\%$      & $52.9\%$       \\ \hline\hline
$\Omega_m^\uparrow$, $n_s^\uparrow$, $w^\downarrow$, C1   & $76.9\%$      & $71.4\%$    \\ \hline
$\Omega_m^\uparrow$, $n_s^\uparrow$, $w^\downarrow$, C2   & $37.3\%$      & $38.5\%$     \\ \hline\hline
$\Omega_m^\uparrow$, $n_s^\downarrow$, $w^\downarrow$, C1   & $82.3\%$     & $81.2\%$      \\ \hline
$\Omega_m^\uparrow$, $n_s^\downarrow$, $w^\downarrow$, C2   & $44.1\%$      & $45.5\%$       \\ \hline
\end{tabular}
\caption{Percentages of points within the desired range of $|\chi^2_\mathrm{COLA} - \chi^2_\textsc{ee2}| < 1$ using $N_{\rm refs}^{\Lambda\rm CDM}=101$ references with different sets of COLA resolution settings. Both emulators show similar goodness-of-fit, falling outside of the desired criteria at Cutoff 2.
}
\label{tab:fraction_chi2_100refs}
\end{table}

Table~\ref{tab:fraction_chi2_100refs} shows the fraction of points with $|\Delta \chi^2| < 1$ for both $N_{\rm refs}^{\Lambda \rm CDM}=101$ emulators for the two most biased fiducial cosmologies. Across both fiducial cosmologies and three scale cuts we see very comparable goodness-of-fit. We conclude that, even with additional references, the \emph{enhanced-precision} settings can only slightly help mitigating the COLA emulator biases. Given the improvement of using $N_{\rm refs}^{\Lambda \rm CDM}=101$, in Appendix~\ref{app:pca} we test whether marginalizing over differences in the COLA and \textsc{ee2} data vectors can reduce the remaining biases.

\subsection{\texorpdfstring{Using $N_{\rm refs}^{\Lambda \rm CDM} > 101$}{}: Emulating \texorpdfstring{$B_{\rm COLA}(k, z)$}{}}
\label{sec:inf_refs}

Our results thus far have indicated that the most effective way to improve the agreement between COLA emulators and \textsc{ee2} is using a large number of anchors inside the $\Lambda$CDM subspace, proving effective even in regions where $w \neq -1$. Due to the computational expense of running the necessary COLA simulation for every anchor to compute Eq.~\ref{eq:b_case}, our previous tests limited the number of $\Lambda$CDM anchors to $N_{\rm refs}^{\Lambda\rm CDM}=101$. However, the $N_{\rm refs}^{\Lambda\rm CDM}=101$ emulators did not stringently satisfy the DES-inspired accuracy requirement of ${\rm FoB} < 0.3$ and $|\Delta \chi^2 |< 1$ throughout the prior, for the scale cuts tested. Nevertheless, the demonstrated improvement of this strategy motivates initial testing to assess whether the uncalibrated boost $B_{\rm COLA}(k,z)$ can effectively be emulated, eliminating the need to run extra COLA simulations in order to increase the number of $\Lambda$CDM anchors. Here, we perform such a preliminary study using the \emph{default-precision} simulations, as we have shown the \emph{enhanced-precision} simulations provide only slight improvements.

\begin{figure}[t]
    \centering
    \includegraphics[width=\columnwidth]{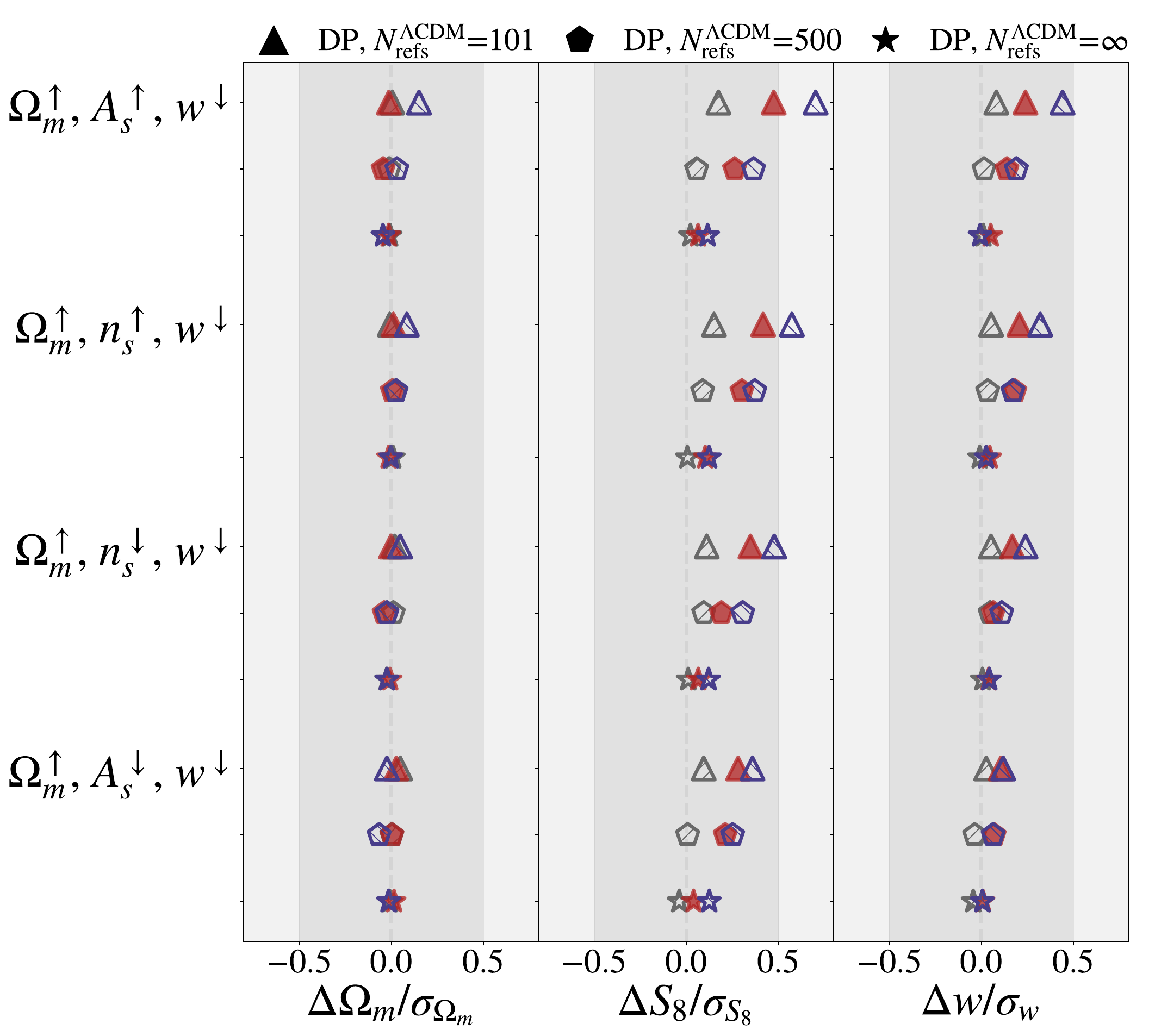}
    \caption{1D biases for both approaches of using $N_{\rm refs}^{\Lambda \rm CDM} > 101$. We test these approaches on the low $w$ fiducial cosmologies of Fig.~\ref{fig:w_fiducials_1d}, located outside of $\Lambda$CDM. While both the $N_{\rm refs}^{\Lambda \rm CDM}=500$ and $N_{\rm refs}^{\Lambda \rm CDM}=\infty$ approaches meet the bias requirements for Cutoff 1 well outside of the $\Lambda$CDM region, only the latter is able to meet or border the requirements for Cutoffs 2 and 3. }
\label{fig:inf_refs}
\end{figure}

When using Eq.~\ref{eq:mult_ref_bcase} to calibrate the training simulations, the most heavily weighted boost, $\bar{B}(k,z)$, is at the reference cosmology with the smallest distance, $d_i$. For each $w$CDM training point $\boldsymbol\theta$, the nearest possible location for a $\Lambda \rm CDM$ reference is at the $\Lambda$CDM-projected point $\boldsymbol{\theta}_{\Lambda \rm CDM}$, in which $w$ is set to $-1$ and the other parameters are unchanged. Under this rationale, the optimal calibration of the training set then requires computing $B_{\rm COLA}(k,z|\boldsymbol{\theta}_{\Lambda \rm CDM})$ for each of the $N_{\rm train}^{w \rm CDM}=500$ $w$CDM training points,\footnote{Because we stretch the training space beyond the \textsc{ee2} boundary, some $w$CDM training points have a $\Lambda$CDM-projection outside of \textsc{ee2} limits. In these cases we compute $\boldsymbol{\theta}_{\Lambda \rm CDM}$ by setting the relevant parameters to their values at the nearest boundary.} $\boldsymbol{\theta}$. Therefore, we use the $B_{\rm COLA}(k,z)$ emulator to generate these 500 predictions before computing $\tilde{B}(k,z)$ with Eq.~\ref{eq:mult_ref_bcase}, and label this approach $N_{\rm refs}^{\Lambda \rm CDM}=500$. Details regarding our $B_{\rm COLA}(k,z)$ emulator are provided in Appendix~\ref{app:inf_refs_boost}.

Notwithstanding this idea, emulating $B_{\rm COLA}(k,z)$ opens another avenue for modeling the boost. Building a $B_{\rm COLA}(k,z)$ emulator provides all the components needed to compute Eq.~\ref{eq:b_case} using an arbitrary $\Lambda$CDM cosmology as the reference in $\mathcal{O}(1)$ seconds. This allows one to optimize the choice of reference cosmology for every point sampled in an MCMC. Therefore, we use the $\Lambda$CDM-projection of the point being \emph{predicted}, as opposed to training points as the reference. We restate Eq.~\ref{eq:b_case} in the present context, with all three boosts being produced separately by an emulator:
\begin{multline}
\label{eq:b_case_inf}
    \bar{B}(k,z| {\boldsymbol \theta}) = B_{\textsc{ee2}}(k, z | \boldsymbol{\theta}_{\Lambda \rm CDM})  \, \times \\ \, \bigg(\frac{B_{\rm COLA}(k,z | {\boldsymbol \theta})} {B_{\rm COLA}(k,z| \boldsymbol{\theta}_{\Lambda \rm CDM})} \bigg), 
\end{multline}
By computing a high-resolution prediction for the $\Lambda$CDM-projection of each sampled point, the COLA emulators serve to correct the $\Lambda \rm CDM$ prediction for the extended model parameter $w$. While in this work we use \textsc{ee2}, $B_{\textsc{ee2}}(k, z | \boldsymbol{\theta}_{\Lambda \rm CDM})$ can be substituted in Eq.~\ref{eq:b_case_inf} using any fast high-resolution $\Lambda$CDM prescription. Hence, compartmentalizing the calculation of $\bar{B}(k,z)$ with emulators allows us to be flexible in our choice of anchor. We refer to this as the infinite references approach, $N_{\rm refs}^{\Lambda \rm CDM}=\infty$.  Comparisons at the boost level between COLA and \textsc{ee2} using both approaches are provided in Appendix~\ref{app:inf_refs_boost}.

\begin{table}[t]
\centering
\setlength{\tabcolsep}{6pt}
\renewcommand{\arraystretch}{1.6}
\begin{tabular}{ |c |c|c| }
\hline
\rule{0pt}{15pt} \backslashbox{Cosmo.}{Emul.} &  \makecell{DP, 500 \\ ($\Lambda$CDM)} & \makecell{DP, $\infty$ \\ ($\Lambda$CDM)}  \\\hline\hline
$\Omega_m^\uparrow$, $A_s^\uparrow$, $w^\downarrow$, C1   & $89.1\%$  & $97.2\%$    \\ \hline
$\Omega_m^\uparrow$, $A_s^\uparrow$, $w^\downarrow$, C2   & $54.4\%$      & $87.8\%$     \\ \hline
$\Omega_m^\uparrow$, $A_s^\uparrow$, $w^\downarrow$, C3   & $35.9\%$      & $60.8\%$     \\ \hline\hline
$\Omega_m^\uparrow$, $A_s^\downarrow$, $w^\downarrow$, C1   & $88.4\%$     & $96.5\%$      \\ \hline
$\Omega_m^\uparrow$, $A_s^\downarrow$, $w^\downarrow$, C2   & $58.4\%$      & $92.5\%$       \\ \hline
$\Omega_m^\uparrow$, $A_s^\downarrow$, $w^\downarrow$, C3   & $43.6\%$      & $63.7\%$       \\ \hline\hline
$\Omega_m^\uparrow$, $n_s^\uparrow$, $w^\downarrow$, C1   & $86.5\%$      & $96.1\%$    \\ \hline
$\Omega_m^\uparrow$, $n_s^\uparrow$, $w^\downarrow$, C2   & $50.2\%$      & $87.1\%$     \\ \hline
$\Omega_m^\uparrow$, $n_s^\uparrow$, $w^\downarrow$, C3   & $32.5\%$      & $58.3\%$     \\ \hline\hline
$\Omega_m^\uparrow$, $n_s^\downarrow$, $w^\downarrow$, C1   & $90.4\%$     & $98.2\%$      \\ \hline
$\Omega_m^\uparrow$, $n_s^\downarrow$, $w^\downarrow$, C2   & $62.4\%$      & $91.1\%$       \\ \hline
$\Omega_m^\uparrow$, $n_s^\downarrow$, $w^\downarrow$, C3   & $43.0\%$      & $64.9\%$       \\ \hline
\end{tabular}
\caption{Percentages of points satisfying the desired goodness-of-fit criteria $|\chi^2_\mathrm{COLA} - \chi^2_\textsc{ee2}| < 1$, for the $N_{\rm refs}^{\Lambda \rm CDM}=500$ and $N_{\rm refs}^{\Lambda \rm CDM}=\infty$ approaches. Both methods show improved agreement in goodness-of-fit with \textsc{ee2} throughout the prior, particularly in the $N_{\rm refs}^{\Lambda \rm CDM}=\infty$ case, which produces a majority of points within the desired $\Delta\chi^2$ range using all scale cuts.
}
\label{tab:fraction_chi2_infrefs}
\end{table}

Fig.~\ref{fig:inf_refs} shows the impact of both approaches on the 1D biases for the $w=w^{\downarrow}$ fiducial cosmologies from Fig.~\ref{fig:w_fiducials_1d}. Both approaches show a large reduction in bias compared to previous emulators. The improvement is particularly pronounced for the $N_{\rm refs}^{\Lambda \rm CDM}=\infty$ case, where we also no longer see a strong correlation between bias and $\sigma_8$. We list the largest biases occurring for each prescription at each scale cut, now occurring at various fiducial cosmologies: 

\begin{itemize}
    \item DP $N_\mathrm{refs}^{\Lambda\mathrm{CDM}} = 500$: $\Delta S_8 / \sigma_{S_8} =0.09$, FoB $=0.31$ (C1);
    \item DP $N_\mathrm{refs}^{\Lambda\mathrm{CDM}} = \infty$: $\Delta S_8 / \sigma_{S_8} =0.04$, FoB $=0.29$ (C1);
    \item DP $N_\mathrm{refs}^{\Lambda\mathrm{CDM}} = 500$: $\Delta S_8 / \sigma_{S_8} =0.26$, FoB $=1.08$ (C2);
    \item DP $N_\mathrm{refs}^{\Lambda\mathrm{CDM}} = \infty$: $\Delta S_8 / \sigma_{S_8} =0.06$, FoB $=0.29$ (C2);
    \item DP $N_\mathrm{refs}^{\Lambda\mathrm{CDM}} = 500$: $\Delta S_8 / \sigma_{S_8} =0.36$, FoB $=1.26$ (C3);
    \item DP $N_\mathrm{refs}^{\Lambda\mathrm{CDM}} = \infty$: $\Delta S_8 / \sigma_{S_8} =0.12$, FoB $=0.33$ (C3).
\end{itemize}

While the $N_{\rm refs}^{\Lambda \rm CDM}=500$ emulator only comes close to $\rm{FoB}<0.3$ on Cutoff 1, having in the worst case tested an $\rm{FoB}=0.31$, it exhibits more significant bias on the two more aggressive scale cuts. Meanwhile, the $N_{\rm refs}^{\Lambda \rm CDM}=\infty$ approach maintains a bias below or very near the desired $\rm{FoB}<0.3$ on all three scale cuts for these fiducial cosmologies, only reaching an $\rm{FoB}=0.33$ in the worst case for Cutoff 3. Furthermore, the goodness-of-fit of these approaches can be gauged from the $\Delta\chi^2$ data presented in Table~\ref{tab:fraction_chi2_infrefs}. The $N_{\rm refs}^{\Lambda \rm CDM}=500$ emulator surpasses all previous emulators in goodness-of-fit for chains with high $\Omega_m$ fiducial cosmologies, having more than $86.5\%$ of points within $|\Delta\chi^2|<1$ with \textsc{ee2} on Cutoff 1. However, the $N_{\rm refs}^{\Lambda \rm CDM}=\infty$ approach is more successful than that even using Cutoff 2, where it maintains at least $87.1\%$ of points within $|\Delta\chi^2|<1$, increasing to $96.1\%$ for Cutoff 1.

Therefore, emulating $B_{\rm COLA}(k,z)$ makes using an arbitrarily high number of $\Lambda$CDM references affordable, which improves the agreement between COLA-based emulators and those trained with high-resolution simulations. Both $N_{\rm refs}^{\Lambda \rm CDM}=500$ and $N_{\rm refs}^{\Lambda \rm CDM}=\infty$ approaches achieve a significant improvement over any other emulator tested, reducing bias with \textsc{ee2} and matching its goodness-of-fit. The $N_{\rm refs}^{\Lambda \rm CDM}=500$ approach represents the logical extent of our previous methodology of attempting to refine the training data, producing an optimally positioned reference for each training point. However, the results of the $N_{\rm refs}^{\Lambda \rm CDM}=\infty$ method suggest that it is more effective not to calibrate the COLA training data beforehand, instead using the COLA emulator in tandem with a high-resolution $\Lambda$CDM emulator to individually calibrate each prediction during the MCMC analysis.

\section{Conclusion}
\label{sec:conclusion}
One of the greatest challenges for constraining beyond-$\Lambda$CDM models with upcoming Stage-IV surveys such as LSST and Euclid will be developing methods to quickly model the matter power spectrum on nonlinear scales. Without this, beyond-$\Lambda$CDM analyses will have to resort to linear scale cuts, discarding an abundance of data and wasting much of the constraining power provided by these surveys. While emulators trained with high-resolution \textit{N}-body simulations are effective tools for this task, such simulations require a formidable computational investment, which would have to be made for every beyond-$\Lambda$CDM model of interest. 

In this work, we have tested using COLA-based emulators as a computationally affordable alternative for modelling the nonlinear power spectrum. We achieved this by comparing their performance to the high-resolution \textsc{EuclidEmulator2} in a $w$CDM cosmic shear analysis of LSST-Y1 simulated data, employing three nonlinear scale cuts. In order to improve the performance of the COLA emulators, we tested two general techniques: incorporating high-resolution reference samples from \textsc{ee2} and refining the resolution settings of the COLA simulations. We emphasize that the choice to use reference predictions from \textsc{ee2} was made solely because it was the benchmark by which we were measuring accuracy of our COLA emulators, and we encourage future exploration into using other prescriptions as references. We found that optimally placing a single reference prediction ($N_{\rm refs}=1$) is enough to calibrate COLA training simulations in a desired region of the prior for even the most aggressive scale cut, leading to low bias. However, this approach suffers from large bias in distant regions of the prior, especially with significantly different $\sigma_8$ values from that of the reference cosmology. 

Our choice of performing analyses using fiducial cosmologies with significant shifts in the parameters was motivated by Stage-IV galaxy survey requirements for investigations in beyond-$\Lambda$CDM models. The fiducial cosmologies employed include values of $\sigma_{8}$ which lie outside the typical range for $\Lambda$CDM investigations, and are challenging to model for any nonlinear prescription. The \textsc{bacco} emulator for instance, would not be able to perform a similar analysis, as some of our fiducial cosmologies themselves lie outside of its parameter space. However, when working with extensions to the Standard Model, it is imperative to develop nonlinear prescriptions for the power spectrum that do not degrade in recovering the model data vector in any region of the parameter space. Extensions to $\Lambda$CDM normally introduce extra parameters with degeneracies with the six vanilla parameters, therefore, it should be expected that unusual regions of the original parameter space will be explored.

Neither increasing to \emph{enhanced-precision} settings, nor $N_{\rm refs}=26$ reference samples spread throughout the entire space were able to eliminate bias across the prior, though offering some alleviation. For beyond-$\Lambda$CDM models which do not contain $\Lambda$CDM within their parameter space, or another sub-model with existing high-resolution emulators, our methodology therefore suggests that careful placement of reference anchors is crucial to avoid large bias. This can be done by performing initial explorations of the posterior in advance of the final chain, such as using linear scale cuts, in order to identify ideal reference cosmologies. 

However, for extensions to $\Lambda$CDM, leveraging existing high-resolution $\Lambda$CDM emulators proved a computationally affordable strategy to consistently eliminate significant bias in otherwise difficult regions of the prior. This was achievable with our less computationally demanding \emph{default-precision} COLA simulations, using a mass resolution $M_{\rm part}\approx 8 \times 10^{10}h^{-1}M_{\odot}$ and force resolution $\ell_{\rm force}=0.5~h^{-1}$Mpc. While the initial methodology explored in the paper aimed to refine the training input to the emulators, so that one self-contained emulator could be incorporated into an MCMC analysis, our $N_{\rm refs}^{\Lambda\rm CDM}=\infty$ results suggest reference samples are best incorporated post-emulation. By training the emulator for the raw boost from the COLA simulations, $B_{\rm COLA}(k,z)$, we are able to calibrate the prediction of the boost during the MCMC to its nearest $\Lambda$CDM point by using \textsc{ee2}. This approach proved to consistently eliminate bias in all regions of the prior and scale cuts on which it was tested, either meeting the desired criteria or bordering the threshold. 

It must be acknowledged that the $w$CDM model is only a minor extension to $\Lambda$CDM, introducing only one additional parameter. For more exotic models with several additional parameters, any approach relying upon $\Lambda$CDM references may degrade somewhat in performance. Nevertheless, we expect that these results hold for beyond-$\Lambda$CDM models that modify the expansion history of our Universe, such as dynamical dark energy theories, as well as theories that modify the linear order growth of structure in a scale-independent way~\cite{growth_fac_paper, Nguyen_2023, Wen_2023}.

Additionally, the COLA input itself can also be further improved beyond refining the force and mass resolutions. It is known that Particle-Mesh based \textit{N}-body codes suffer from a loss of power at small scales due to their inability of resolving the internal structure of halos. To bypass this, different alternatives have been proposed in the literature, such as the Potential Gradient Descent methods~\cite{pgd_simulations} and neural network corrections to simulations~\cite{hybrid_physical_neural_nbody}. Another possible avenue would be to use field-level emulation techniques such as the ones presented in~\cite{NECOLA, renan, Schmidt_2021, nguyen2024information}. However, as our main proposal is to use COLA as a viable alternative to model nonlinear scales in beyond-$\Lambda$CDM models, an obvious setback in the implementation of field-level methods, is that they require us to run a decent amount of high-resolution \textit{N}-body simulations of a given model to act as a true model. Therefore, while field level techniques are an extremely appealing alternative, they lack the needed predictability for investigations of alternatives to the Standard Model.

\section*{Acknowledgements}
The authors would like to thank Stony Brook Research Computing and Cyberinfrastructure and the Institute for Advanced Computational Science at Stony Brook University for access to the high-performance SeaWulf computing system, which was made possible by \$1.85M in grants from the National Science Foundation (awards 1531492 and 2215987) and matching funds from the the Empire State Development’s Division of Science, Technology and Innovation (NYSTAR) program (contract C210148). This research was also supported by resources supplied by the Center for Scientific Computing (NCC/GridUNESP) of the São Paulo State University (UNESP). The authors acknowledge the National Laboratory for Scientific Computing (LNCC/MCTI, Brazil) for providing HPC resources of the SDumont supercomputer, which have contributed to the research results reported within this paper. 
GB is supported by the Alexander von Humboldt Foundation. FTF acknowledges financial support from the National Scientific and Technological Research Council (CNPq, Brazil). JR acknowledges the financial support from FAPESP under grant 2020/03756-2, São Paulo Research Foundation (FAPESP) through ICTP-SAIFR. KK is supported by STFC grant ST/W001225/1. HAW acknowledges support from the Research Council of Norway under grant 287772.

\bibliography{nls_short}

\appendix

\section{Emulation Implementation Details}
\label{app:emulators}

\subsection{Separating the BAO Signal from Linear Power Spectrum}
\label{app:bao-smearing}
In this Appendix, we explain in detail our algorithm to remove the BAO signal from the linear power spectrum, a step required in emulating the quantity $Q^{\rm S}(k,z)$ (Eq.~\ref{eq:qbacco}). We follow the \textsc{bacco} emulator and employ the methodology outlined in Appendix C of \cite{baumann_removing_bao}, with slight modifications. The process can be summarized as follows: 

\begin{enumerate}
    \item \textbf{Standardize $\mathbf{k}$-range:} Ensure $P_{\rm L}(k)$ covers the range of $10^{-4}~h\mathrm{Mpc}^{-1} \leq k \leq 5~h\mathrm{Mpc}^{-1}$. If necessary, linearly extrapolate the power spectrum.
    
    \item \textbf{Linear Log-Log Interpolation:} Interpolate the power spectrum $P_{\rm L}(k)$ using cubic spline at $2^n$ logarithmically-spaced points in $k$, ensuring data is evenly spaced in log-log space for further analysis.
    
    \item \textbf{Fast Sine Transform:} Generate an array of $\log[kP_{\rm L}(k)]$ evaluated at the $2^n$ $k$-bins and apply an orthonormalized type-II fast sine transform to that array.
    
    \item \textbf{Separating Even and Odd Entries:} Denoting the index of the sine-transformed array by $i$, divide the array into two separate arrays of size $2^{n-1}$, corresponding to components with even $i$ and odd $i$. Save the even and odd indices into another two arrays.
    
    \item \textbf{Interpolation and Differentiation:} Interpolate both even and odd sine-transformed arrays using the respective set of indices. Then, calculate the second derivative by either differentiating the cubic spline or computing the finite differences to identify the locations of baryonic features. 
    
    \item \textbf{Noise Reduction:} Average the second derivative points with their immediate neighbors to reduce noise.
    
    \item \textbf{Identifying Bumps:} Locate the BAO bumps by finding the minimum and maximum values of the second derivatives for both even and odd arrays, locating the bump to be removed.
    
    \item \textbf{Removing Bumps:} Remove the region between the minimum and maximum second derivatives from both even and odd sine-transformed arrays, filling the gaps by interpolating the remaining data, scaled by $(i+1)^2$, with cubic splines. 
    
    \item \textbf{Inverse Fast Sine Transform:} Recombine the even and odd sine-transformed arrays, dividing the $(i+1)^2$ scaling, and apply an inverse fast sine transform to reconstruct $\log[k P_{\rm L}^{\mathrm{NBAO}}(k)]$.

\end{enumerate}

The \textsc{scipy}\footnote{\url{https://scipy.org/}} \cite{2020SciPy-NMeth} library was used to perform the sine transforms on the power spectra, GSL was used for interpolation. Our implementation is given in this repository\footnote{\url{https://github.com/bernardo7crf/Cython_Filter_Files}}. We remark that the BAO signal separation is not unique but rather depends on the specific methodology employed. Some examples, as discussed in \cite{Vlah_2016}, are the Savitzky–Golay or Gaussian filters, fitting models such as B-Splines, employing a power law model with the form $k^n$, or leveraging the output from Boltzmann codes with no baryons. We found that the method of \cite{baumann_removing_bao} was sufficient to our needs, and we have not tested other methods.

\subsection{Gaussian Process Regression}
\label{app:grp}
The first method tested to perform emulation is Gaussian Process regression. GP regression is a non-parametric Bayesian method which assumes any set of function values has a joint multivariate Gaussian distribution. We denote an arbitrary principal component amplitude from Eq.~\ref{eq:pca} at an arbitrary redshift as $\alpha$, which we treat as a function of a $D$-dimensional vector of normalized cosmological parameters $\boldsymbol{\theta}$. Our training set provides a vector $\boldsymbol{\alpha}_{\rm obs}$ of $N_{\rm train}$ observations of the function at the training points $\boldsymbol{\theta}_i$, which generally contain noise:
\begin{equation}
    \alpha_{\mathrm{obs}, i} = \alpha(\boldsymbol{\theta}_i) + \epsilon_i .
\end{equation}
For simplicity, we assume the function values have been centered to have mean 0. We take the noise to be Gaussian, such that $\epsilon_i \sim \mathcal{N}(0,\sigma_{\epsilon}^2)$, where $\sigma_{\epsilon}^2$ is called the noise variance. 

We use a Radial Basis Function (RBF) kernel to model the covariance between values of the function:
\begin{equation}
     k(\boldsymbol{\theta},\boldsymbol{\theta}') \equiv \sigma^2 \exp \Bigg[ -\frac{1}{2} \sum^{D}_{j=1} \frac{(\theta_{j} - \theta'_{j})^2}{l_j^2} \Bigg].
\end{equation}
where we have the hyperparameters $\sigma^2$, called the signal variance, and the $\{ l_j \}$, called the correlation lengths. We can form a covariance matrix, $\boldsymbol{K}$, from the training points defined by $K_{pq} \equiv k(\boldsymbol{\theta}_p,\boldsymbol{\theta}_q)$. However, we modify this matrix to account for noise $\hat{\boldsymbol{K}} \equiv \boldsymbol{K} + \sigma_{\epsilon}^2 \boldsymbol{I}$, where $\boldsymbol{I}$ is the identity matrix. The hyperparameters $\sigma^2$, $\{ l_j \}$, and $\sigma_{\epsilon}^2$, are determined from the training data by maximizing the marginal likelihood:
\begin{equation}
    p(\boldsymbol{\alpha}_{\rm obs}|K) = \frac{1}{\sqrt{(2\pi)^{N_{\rm train}} \left| \hat{\boldsymbol{K}} \right| }}\exp \Big[-\frac{1}{2} \boldsymbol{\alpha}_{\rm obs}^T \hat{\boldsymbol{K}}^{-1} \boldsymbol{\alpha}_{\rm obs} \Big].
\end{equation}

Aside from the training points, we consider a test point $\boldsymbol{\theta}^*$ at which we wish to infer the value of the function $\alpha(\boldsymbol{\theta}^*)$. The GP assumption states that the value of the function at the test point is normally distributed along with the observed values at the training points:
\begin{equation}
    \left[
    \begin{array}{c}
    \boldsymbol{\alpha}_{\rm obs} \\
    \alpha(\boldsymbol{\theta}^*)
    \end{array}
    \right] \sim \mathcal{N} 
    \left( \left[
    \begin{array}{c}
    \boldsymbol{0}\\
    0
    \end{array}
    \right], 
    \left[
    \begin{array}{cc}
    \hat{\boldsymbol{K}} & \boldsymbol{K}^*\\
    \boldsymbol{K}^{*T} & K^{**}
    \end{array} 
    \right]
    \right)
\end{equation}
The vector $\boldsymbol{K}^*$ contains the covariances between the function values at the test point and training points, $\boldsymbol{K}^*_p \equiv k(\boldsymbol{\theta}^*,\boldsymbol{\theta}_p)$, and we define $K^{**} \equiv k(\boldsymbol{\theta}^*,\boldsymbol{\theta}^*)$. In order to predict the value of the function at $\boldsymbol{\theta}^*$, and the prediction uncertainty, we take the mean and variance of the conditional distribution $p(\alpha(\boldsymbol{\theta}^*)| \{ \boldsymbol{\theta}_i \} , \boldsymbol{\alpha_}{\rm obs} )$ respectively:
\begin{align}
    \alpha_{\rm pred}(\boldsymbol{\theta}^*) &= \boldsymbol{K}^{*T} \hat{\boldsymbol{K}}^{-1} \boldsymbol{\alpha}_{\rm obs} \\
    \sigma^2_{\rm pred} &= K^{**} - \boldsymbol{K}^{*T} \hat{\boldsymbol{K}}^{-1} \boldsymbol{K}^{*} .
\end{align}
Here $\alpha_{\rm pred}(\boldsymbol{\theta}^*)$ serves as our (centered) prediction of the principal component amplitude at the test point, with $\sigma^2_{\rm pred}$ providing the uncertainty in the prediction. We perform a different regression for every principal component, for every redshift, using the \textsc{gpy}\footnote{\url{https://gpy.readthedocs.io/en/deploy/}} library.

\subsection{Neural Network}
\label{app:nn_emul}
Another emulation strategy we have employed is through the use of neural networks. We use a fully-connected, feed-forward network consisting of the input layer of normalized cosmological parameters, three hidden layers with $512$ neurons each, and the output layer of principal components $\alpha_{\ell j}$ according to Eq.~\ref{eq:pca}. We train one NN model for each redshift, and thus the index $j$ will be omitted in the following. The value of the neurons in each subsequent layer are calculated via linear transformations (\textit{i.e.} the application of weights), the addition of biases to the previous layer, and the application of an activation function used by \cite{cosmopower, speculator}, which has the form:
\begin{equation}
    y^{m+1}_n = \left[\gamma^m_n + (1 - \gamma^m_n)\frac{1}{1 + e^{-\beta^m_n y^m_n}}\right]\tilde{y}^m_n,
    \label{eq:nn_activation}
\end{equation}
where $y^{m+1}_n$ is the value of the $n$-th neuron from the $(m+1)$-th layer, $\tilde{y}^m_n$ is the is the value of the $n$-th neuron from the $(m+1)$-th layer after the application of weights and biases, and $\gamma^m_n$ and $\beta^m_n$ are parameters of the activation function that can be backpropagated during the training. We have tested several activation functions, and their emulation errors at $z=0$ and $k \lesssim 1 h/\mathrm{Mpc}$ were within approximately $0.8\%$ (Rectified Linear Unit), $0.5\%$ (sigmoid) and $0.2\%$ (Eq.~\ref{eq:nn_activation}). We have also tested different numbers of layers, neurons and residual layers, finding no improvement on the emulation performance.

In order to train the neural network emulators, we adjust the weights and biases to minimize the $L_1$ loss function, or the mean absolute error:
\begin{equation}
    L_1 = \sum_{\mathrm{training}}\sum_{\ell=1}^{N_\mathrm{PC}} |\alpha_{\mathrm{truth},\ell} - \alpha_{\mathrm{pred},\ell}|,
\end{equation}
where $\alpha_{\mathrm{truth},\ell}$ are the principal component weights from the training sample, $\alpha_{\mathrm{pred},\ell}$ are the neural network predictions for the training sample, and the outermost sum goes through all samples in the training set.

In order to prevent overfitting, we perform a $L_1-L_2$ regularization, also known as Elastic Net regularization, of the weights and biases. This encourages the training process to keep weights and biases to a low absolute value, limiting the complexity of the model. We keep $10\%$ of the training cosmologies as a validation set for early stopping: a technique that prevents overfitting by ensuring that the validation loss does not grow during the training. The training process is halted after 2500 epochs, with an initial learning rate of $10^{-3}$ which, starting from epoch $1500$, halves every $200$ epochs. Our neural network is implemented using the \textsc{keras}\footnote{\url{https://keras.io}} library, and the loss function is minimized using the \textsc{adam} optimizer \cite{adam_opt}. The training process for all redshifts takes 2 hours without GPU acceleration, and the full emulator takes around 250MB of disk space.

\subsection{Polynomial Chaos Expansion}
\label{app:pce}

Polynomial Chaos Expansion is another surrogate modelling technique, which expands the model's output in orthogonal polynomials \cite{Blatman2010,Blatman2011,lüthen2021automatic, Marelli2018,Torre2019,Kaintura2018}. These polynomials capture the functional relationship between inputs and outputs. The choice of polynomial basis in PCE corresponds to the input variables' probability distribution, and if the inputs do not follow a specific distribution then PCE reduces to a simple polynomial expansion.

To perform the PCE, we start with the principal components $\alpha_{\ell j}$ of Eq.~\ref{eq:pca}, calculated for the training set. At fixed redshift, this represents a $N_\mathrm{train} \times N_\mathrm{PC}$ matrix. We decompose this matrix as a product of two matrices $X \times W$, where $W$ will contain the polynomial coefficients to be determined, and $X$ is the polynomial basis matrix representing the orthogonal polynomial expansions. To construct the matrix $X$, we calculate its elements $X_{ij}$ using:

\[
X_{ij} = \prod^{N_\mathrm{PC}}_{k=1} L_{\beta_{jk}}(x_{ik}).
\]
Here, $L_{\beta_{jk}}$ are the orthogonal polynomial basis functions, $x_{ik}$ are powers of the input variables and $\boldsymbol{\beta}$ is a multi-index matrix. The multi-index $\boldsymbol{\beta}_j = (\beta_{j1}, \beta_{j2}, \ldots, \beta_{jD})$ specifies the polynomial orders for each input dimension in the $j$th column of $X$. Each $\beta_{jk}$ is a non-negative integer representing the degree of the $k$th input parameter in the $j$th polynomial term. For example, consider a 2-dimensional input $(x_1, x_2)$ expanded up to the 2nd order. The $\boldsymbol{\beta}$ matrix would be:

\[
\boldsymbol{\beta} =
\begin{pmatrix}
0 & 0 \\
1 & 0 \\
0 & 1 \\
2 & 0 \\
1 & 1 \\
0 & 2
\end{pmatrix}
\]
Each row in $\boldsymbol{\beta}$ represents a different combination of polynomial terms, with values indicating the powers of $x_1$ and $x_2$ for that combination. We define the candidate set $\mathcal{A}_{\text{cand}}$ of multi-indices $\boldsymbol{\beta}$ to include in the expansion, based on constraints like the maximum total polynomial degree $p$, maximum interaction order $r$, and an optional $q$-norm penalty to induce sparsity:

\[
\mathcal{A}_{\text{cand}} = \left\{ \boldsymbol{\beta}_i : \left( \sum_{k=1}^{N_\mathrm{PC}} \beta_{ik}^q \right)^{\frac{1}{q}} \leq p, \sum_{\substack{k=1 \\ \beta_{ik} \neq 0}}^{N_\mathrm{PC}} 1 \leq r \right\}
\]
We then use Elastic-Net regression to determine the optimal PCE coefficients $W$ that balance the mean-squared error between the PCE model and training data with $L_1$ (lasso) and $L_2$ (ridge) regularization terms to help prevent overfitting.

After solving the regression problem to obtain the optimal PCE coefficient matrix $W$, we can evaluate the principal component amplitudes $\alpha$ for a new cosmology $\boldsymbol{\theta}^*$ as:

\[
\alpha_{\rm pred}(\boldsymbol{\theta}^*) = X_{\boldsymbol{\theta}^*} \times W
\]
where $X_{\boldsymbol{\theta}^*}$ is the row of the polynomial basis matrix corresponding to $\boldsymbol{\theta}^*$, evaluated using the selected multi-indices in $\mathcal{A}$. 

\subsection{Neural PCE}
\label{app:pce_nn}
 
During the emulation of the \emph{enhanced-precision} simulations, we observed performance issues characterized by increased prediction errors. To address this problem, we developed a hybrid model, combining PCE with neural networks, which we have termed the Neural PCE (NPCE). In NPCE, PCE is used for initial prediction, and the NN performs a correction phase, similar to the iterative refinement process in ODE solvers.

A standalone PCE emulator was trained as described in \ref{app:pce}, and its outputs served as inputs to a neural network model. The neural network was designed with input and output layers sized to match the PCE output, incorporating 2 to 7 hidden layers based on problem complexity, with each layer containing 32 to 128 neurons. The activation function specified in \ref{eq:nn_activation} was used alongside a residual connection between the input and output layers. The network's weights were initialized randomly and trained using the PCE predictions as inputs against the data, aiming to minimize mean squared error via the L-BFGS algorithm.

For predictions at new cosmologies, the NPCE model refined initial PCE emulator predictions through the neural network, yielding improved accuracy. The residual connection in the neural network allowed it to learn corrections to the PCE predictions. It is important to note that we trained the PCE and neural network components separately, rather than as an integrated model. By adding the neural network, we increased the model parameters by approximately $10\%$ compared to using PCE alone, without the need to modify the PCE training process itself.

The effectiveness of this model is demonstrated through a comparison of the prediction errors between the existing emulators and NPCE in Figure~\ref{fig:combined}. The results show a reduction in error rates using the new model, with the error being under $\sim 0.25\%$ for most cosmologies on all  scales, indicating a successful improvement in performance. 

\begin{figure}[t]
    \centering
    \includegraphics[width=\columnwidth]{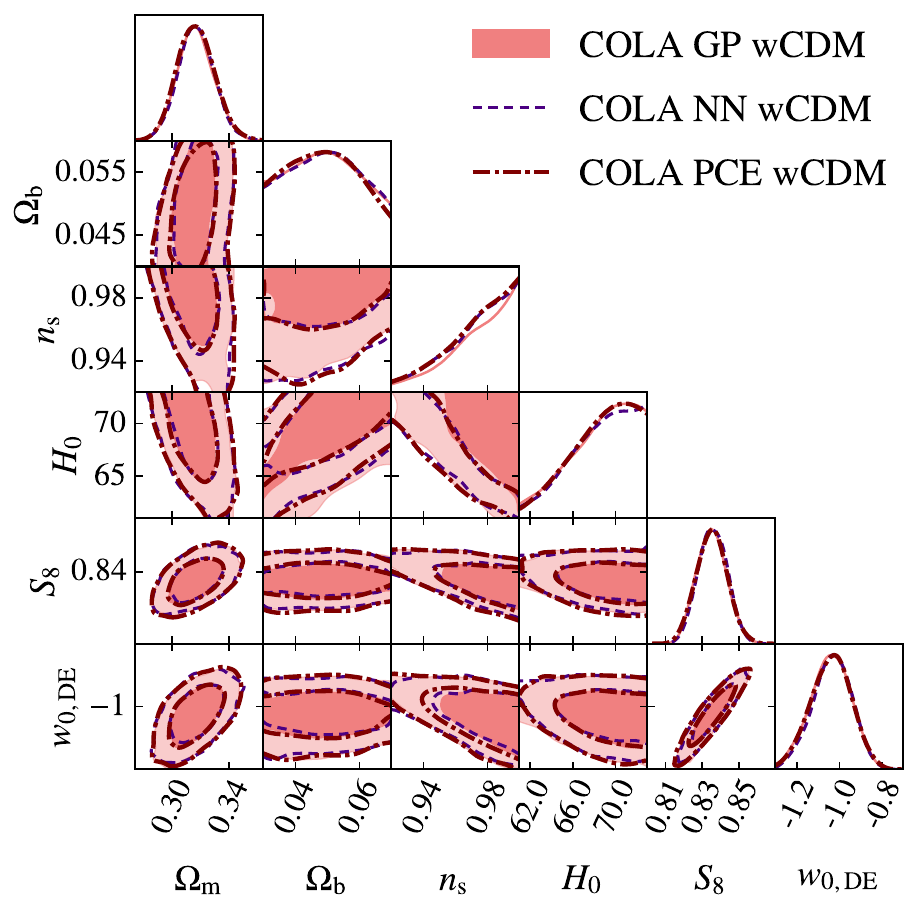}
    \caption{68\% and 95\% confidence contours of several cosmological parameters from MCMCs run with three emulation techniques: Gaussian Process regression (pink filled contours), neural network (blue dashed lines) and Polynomial Chaos Expansion (maroon dotted-dashed lines). In both panels, the COLA emulators are trained with \emph{default-precision} simulations calibrated to the \textsc{ee2} reference cosmology, which also serves as the fiducial cosmology. We use a scale cut even more aggressive than Cutoff 3 for this test, with $\theta_\mathrm{min} = 2.7'$ for $\xi_+$ and $\theta_\mathrm{min} = 8.7'$ for $\xi_-$.}
    \label{fig:emus-equiv}
\end{figure}

\subsection{Testing Emulator Equivalence}
\label{app:emulator_equiv}

We now test whether the three emulator techniques, namely Gaussian Process regression, neural networks, and Polynomial Chaos Expansion, are equivalent at the level of parameter inference. For this test, we ran MCMCs using the LSST-Y1 simulated cosmic shear data. We utilized the three emulated methods, where each emulator was trained with \emph{default-precision} simulations and $N_{\rm refs}=1$. We use the \textsc{ee2} reference cosmology as the fiducial, and employ a scale cut even more aggressive than Cutoff 3, with $\theta_\mathrm{min} = 2.7'$ for $\xi_+$ and $\theta_\mathrm{min} = 8.7'$ for $\xi_-$. Fig.~\ref{fig:emus-equiv} shows the confidence contours and marginalized posteriors of several cosmological parameters obtained with the three emulators. We see strong agreement in the 1D distributions, and we conclude that any of the three emulation techniques can be used interchangeably. However, we note that there are differences in the training time of the various emulators. Considering all principal components and redshifts, and using only CPUs, the GP emulators took $\sim 10$ minutes to train on 1 core, the NN took $\sim 2$ hours to train on 1 core, the PCE emulators took $\sim 10$ hours to train on 16 cores, and the NPCE emulators took an extra $\sim 6$ hours to train compared to the PCE.

\section{\texorpdfstring{$N_{\rm refs}^{\Lambda\rm CDM}=500$}{} and \texorpdfstring{$N_{\rm refs}^{\Lambda\rm CDM}=\infty$}{}, Comparisons at the Boost Level}
\label{app:inf_refs_boost}

\begin{figure}[t]
    \centering
    \includegraphics[width=\columnwidth]{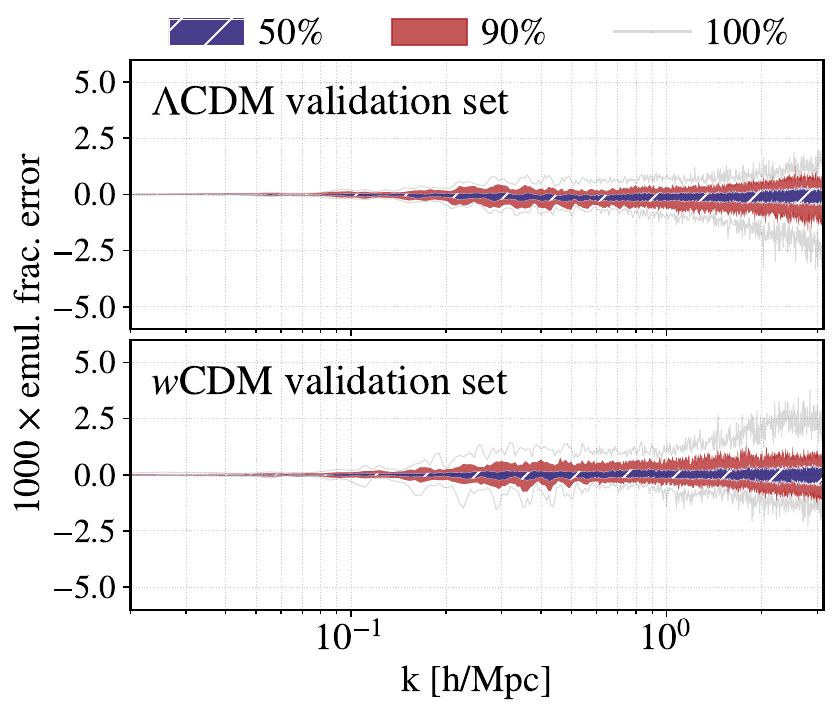}
    \caption{Bounds on the best $50\%$, $90\%$, and $100\%$ of emulation errors at $z=0$ for our NPCE $w$CDM emulator of the uncalibrated $B_{\rm COLA}(k,z)$ used for the $N_{\rm refs}^{\Lambda \rm CDM}=500$ and $N_{\rm refs}^{\Lambda \rm CDM}=\infty$ approaches from Sec.~\ref{sec:inf_refs}. The emulation errors in the $\Lambda$CDM validation set quantify the error introduced in emulating the raw COLA prediction $B_{\rm COLA}(k,z)$ at the $\Lambda$CDM reference cosmologies, which is involved in both approaches. The errors in the $w$CDM validation set do similarly for the $w$CDM prediction of $B_{\rm COLA}(k,z)$ in the $N_{\rm refs}^{\Lambda \rm CDM}=\infty$ approach.}
\label{fig:bcola_emu}
\end{figure}

\begin{figure}[t]
    \centering
    \includegraphics[width=\columnwidth]{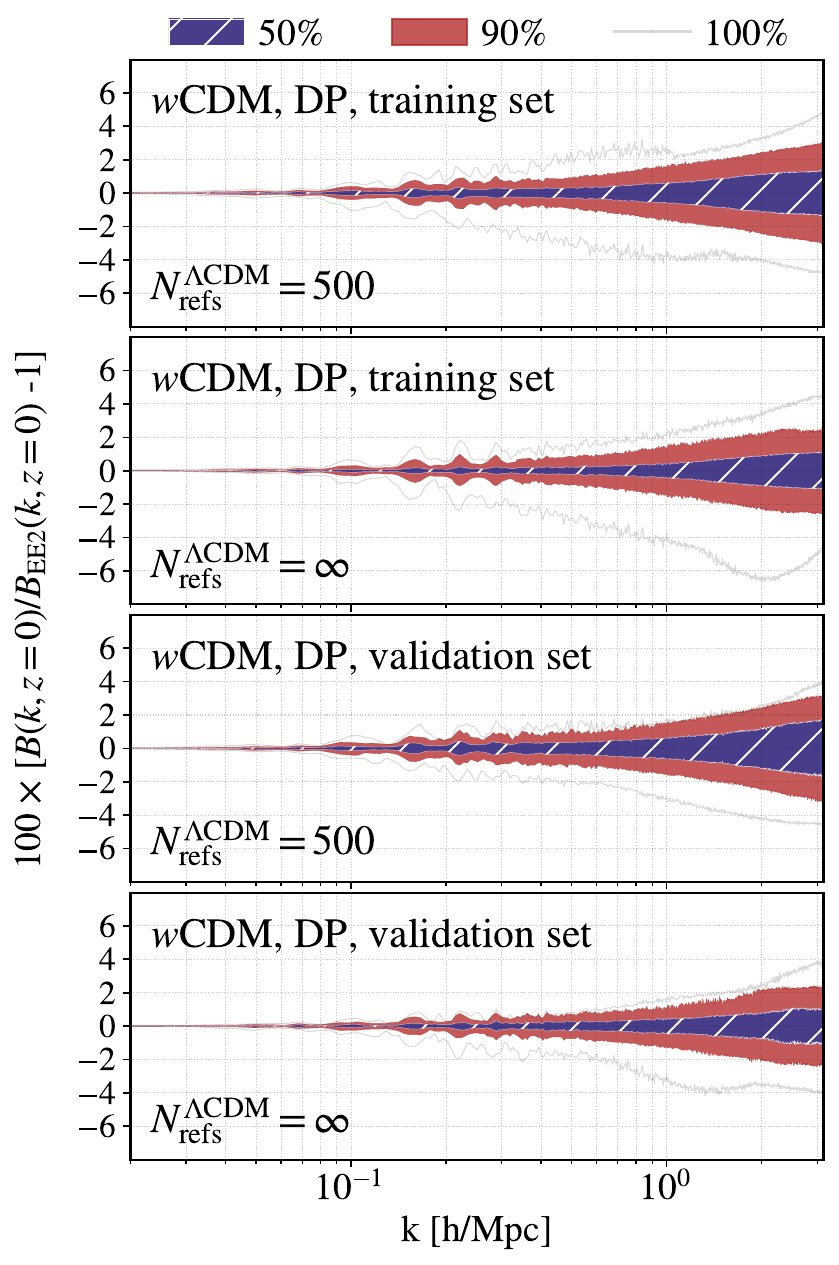}
    \caption{Following Fig.~\ref{fig:cola_comparisons}, the comparisons of the boost from COLA to \textsc{ee2}, for the training set and validation sets, using the $N_{\rm refs}^{\Lambda\rm CDM}=500$ and $N_{\rm refs}^{\Lambda\rm CDM}=\infty$ methodology. For $N_{\rm refs}^{\Lambda\rm CDM}=500$ references we compute the boost using $B(k,z)=\tilde{B}(k,z)$ as in Eq.~\ref{eq:mult_ref_bcase}, while for $N_{\rm refs}^{\Lambda\rm CDM}=\infty$ references we use $B(k,z)=\bar{B}(k,z)$ as in Eq.~\ref{eq:b_case_inf}. For these comparisons, we evaluate the respective equations for the boost by computing $B_{\rm COLA}(k,z)$ directly from simulations for the $w$CDM predictions, while using the NPCE emulator for the $\Lambda$CDM-projected predictions as we lack the requisite $\Lambda$CDM simulations. For comparisons at higher redshift see Appendix~\ref{app:high_zs_ee2_comps}.}
\label{fig:inf_refs_boost_ee2}
\end{figure}
In order to implement the $N_{\rm refs}^{\Lambda\rm CDM}=500$ and $N_{\rm refs}^{\Lambda\rm CDM}=\infty$ approaches, the same emulator for the uncalibrated boost $B_{\rm COLA}(k,z)$ was used, trained on the $N_{\rm train}^{w\rm CDM}=500$ $w$CDM cosmologies. In both cases emulation now enters the pipeline in two places. In the $N_{\rm refs}^{\Lambda\rm CDM}=500$ approach, the $B_{\rm COLA}(k,z)$ emulator serves to compute the COLA prediction at the reference cosmologies in Eq.~\ref{eq:b_case}, before the $\tilde{B}(k,z)$ emulator is then subsequently trained. Meanwhile, in the $N_{\rm refs}^{\Lambda\rm CDM}=\infty$ approach the $B_{\rm COLA}(k,z)$ emulator is used in Eq.~\ref{eq:b_case_inf} both to evaluate the COLA prediction at the $w$CDM point and again at the $\Lambda$CDM-projected point. Therefore, we use the more effective NPCE method, and a larger $N_{\rm PC}=25$, in order to prevent a compounding of emulation errors in our analysis. With the higher number of principal components, there was little improvement observed in implementing the BAO-smearing procedure, therefore we use $Q^{\rm NS}(k,z)$ as the emulation variable for the $B_{\rm COLA}(k,z)$ emulator. This is advantageous in the $N_{\rm refs}^{\Lambda\rm CDM}=\infty$ approach as one would otherwise need to compute $P_{\rm L}(k,z|\boldsymbol{\theta}_{\Lambda \rm CDM})$ in addition to $P_{\rm L}(k,z|\boldsymbol{\theta})$ merely to recover the boosts from the emulation variable.

The NPCE $B_{\rm COLA}(k,z)$ emulator utilizes a polynomial expansion up to the 12th order in the cosmological parameters. Additionally, the neural network responsible for the correction phase consists of 3 layers and 128 neurons, employing the activation function presented in Eq.~\ref{eq:nn_activation}. Fig.~\ref{fig:bcola_emu} shows the emulation errors in $B_{\rm COLA}(k,z)$ at $z=0$. As the reference COLA predictions in both $N_{\rm refs}^{\Lambda\rm CDM}=500$ and $N_{\rm refs}^{\Lambda\rm CDM}=\infty$ approaches are confined to $\Lambda$CDM, we compute the emulation errors in the $\Lambda$CDM validation set, in addition to the $w$CDM validation set. We see that the emulator of the uncalibrated $B_{\rm COLA}(k,z)$ achieves an error of less than 0.3\% on all scales used $k \leq \pi h$Mpc$^{-1}$ for either validation set. In the $N_{\rm refs}^{\Lambda\rm CDM}=500$ approach, we opted to use the NPCE method for the $\tilde{B}(k,z)$ emulator as well following the choices detailed in Sec.~\ref{sec:emulator}, using $N_{\rm PC}=25$. We set the parameter $\sigma_d=0.2$ in Eq.~\ref{eq:gaussian_weights}, which yields emulation errors similar to the PCE errors displayed in the middle panel of Fig.~\ref{fig:combined}. To improve predictive performance we increased the number of layers in the neural network step to 7. 

In Fig.~\ref{fig:inf_refs_boost_ee2} we present the disagreement between the COLA boosts and \textsc{ee2} at $z=0$, when using $\tilde{B}(k,z)$ with $N_{\rm refs}^{\Lambda\rm CDM}=500$ and $\bar{B}(k,z)$ for the $N_{\rm refs}^{\Lambda\rm CDM}=\infty$ method. Here, we use the simulation output directly to compute $B_{\rm COLA}(k,z)$ for $w$CDM cosmologies in the calculation of the respective boosts, and use the NPCE $B_{\rm COLA}(k,z)$ emulator to obtain the COLA predictions at $\Lambda$CDM references. We perform this comparison for both the 279 $w$CDM training cosmologies inside the \textsc{ee2} parameter space as was done for the right panels in Fig.~\ref{fig:cola_comparisons}, as well as for the $w$CDM validation set. This is because both approaches are similar in their treatment of the training set, calibrating the training simulations, either exclusively or most heavily, to their $\Lambda$CDM-projected cosmologies. However, for predictions of the validation points, the $N_{\rm refs}^{\Lambda\rm CDM}=\infty$ strategy is able to calibrate these predictions to their respective $\Lambda$CDM-projected cosmologies, unlike the $N_{\rm refs}^{\Lambda\rm CDM}=500$ case. 

For the training cosmologies, the $N_{\rm refs}^{\Lambda\rm CDM}=500$ approach produced errors at $k=1h$Mpc$^{-1}$ above $1\%$, $1.5\%$, and $2\%$ for only $26.5\%$, $10.4\%$, and $2.5\%$ of cosmologies respectively. Meanwhile the $N_{\rm refs}^{\Lambda\rm CDM}=\infty$ approach had $20.8\%$, $9.0\%$, and $4.3\%$ of these cosmologies outside the $1\%$, $1.5\%$, and $2\%$ thresholds. However, in the validation set, the $N_{\rm refs}^{\Lambda\rm CDM}=500$ references yielded $31.0\%$, $12.0\%$, and $1.0\%$ of cosmologies outside of those respective criteria, while the $N_{\rm refs}^{\Lambda\rm CDM}=\infty$ method only yielded $17.0\%$, $3.0\%$, and $1.0\%$. Therefore, excluding the $\mathcal{O}(0.1\%)$ emulation errors shown in Fig.~\ref{fig:bcola_emu}, the $N_{\rm refs}^{\Lambda\rm CDM}=\infty$ methodology manages to reproduce the \textsc{ee2} boost predictions at $k=1h$Mpc$^{-1}$ at sub-percent error for $\sim 80\%$ of cosmologies throughout the $w$CDM parameter space.

\section{Boost Comparisons to \textsc{ee2} at Higher Redshift}
\label{app:high_zs_ee2_comps}
\begin{figure*}[t] 
    \centering
        \begin{minipage}[t]{\textwidth}
        \centering
        \includegraphics[width=\linewidth]{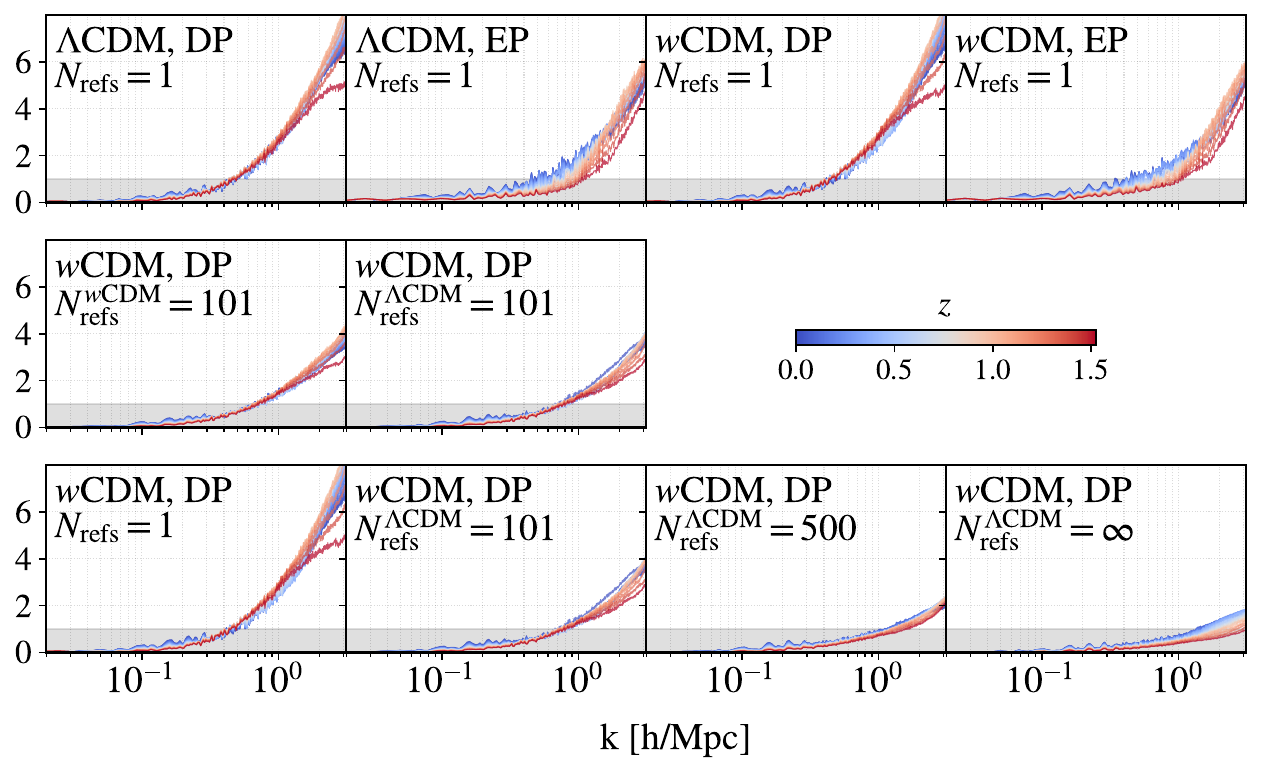}
    \end{minipage}
    \caption{In this figure we show the 68th percentile error between the post-processed boost from our COLA training simulations and those from \textsc{ee2}, at each wavenumber and redshift, using many of the different approaches tested in this paper. The error is computed as the absolute value in the percentage difference in the boost $100\times |B/B_{\textsc{ee2}} -1 |$, where $B$ is replaced by $\bar{B}_{\rm COLA}$ or $\tilde{B}_{\rm COLA}$ via Eqs.~\ref{eq:b_case}, \ref{eq:mult_ref_bcase}, or ~\ref{eq:b_case_inf} in the respective cases. The labels identify the model within which the training simulations are spread, the precision settings of the simulations, and the number and model of the references used. While the top row compares results for $N_{\rm refs}=1$, the middle row compares the $N_{\rm refs}=101$ cases showing the lack of degradation when confining the references to $\Lambda$CDM, and the bottom row shows the improvement due to increasing $N_{\rm refs}^{\Lambda \rm CDM}$.} 
    \label{fig:high_zs}
\end{figure*}

In this section we include comparisons of the post-processed boosts from our COLA simulations to those computed by \textsc{ee2}, at higher redshifts that more significantly impact the LSST simulated analyses performed in the paper (see Tab.~\ref{tab:k_cuts} for the mean redshift of each galaxy source bin used). These comparisons are presented in Fig.~\ref{fig:high_zs} where we show the 68th percentile error in the boost at each wavenumber and redshift. From the figure we see that in all cases the disagreement at higher redshifts is either roughly consistent with the error at $z=0$ or noticeably decreased.

\section{Generating Fiducial Data with the \textsc{bacco} Emulator}
~\label{app:bacco_fid}

In the analysis presented in~\ref{sec:lsst-results}, \textsc{ee2} was used both in generating the fiducial data, as well as in modelling the data vector at the sampled points in the chains. This is true not only for chains in which \textsc{ee2} was used directly to model the nonlinear power spectrum, but also for COLA emulator chains since \textsc{ee2} was used to compute the reference predictions. While our original analysis only compared the MCMC constraints between the COLA emulators and \textsc{ee2}, it remains possible in theory that generating the fiducial data with another prescription could cause the COLA emulator and \textsc{ee2} constraints to diverge more. This is an important consideration as real data from LSST will not be generated with \textsc{ee2}.

We briefly explore whether our results depend strongly on the choice of nonlinear prescription used in generating the fiducial data, by repeating previous tests with fiducial data generated with the \textsc{bacco} emulator. As it was determined in Sec.~\ref{sec:lsst-results} that the $N_{\rm refs}^{\Lambda \rm CDM}=500$ and $N_{\rm refs}^{\Lambda \rm CDM}=\infty$ approaches were the most accurate, we repeat only tests similar to those presented in Fig.~\ref{fig:inf_refs}. However, since \textsc{bacco} emulator only covers values of $\sigma_8 < 0.9$, we generated fiducial data vectors at $\Omega_m = 0.33$ while maintaining the other cosmological parameters. We also used \textsc{ee2} to generate fiducial data for $z>1.5$ as this is outside of the \textsc{bacco} emulator's range.

\begin{figure}[h]
    \centering
    \includegraphics[width=\columnwidth]{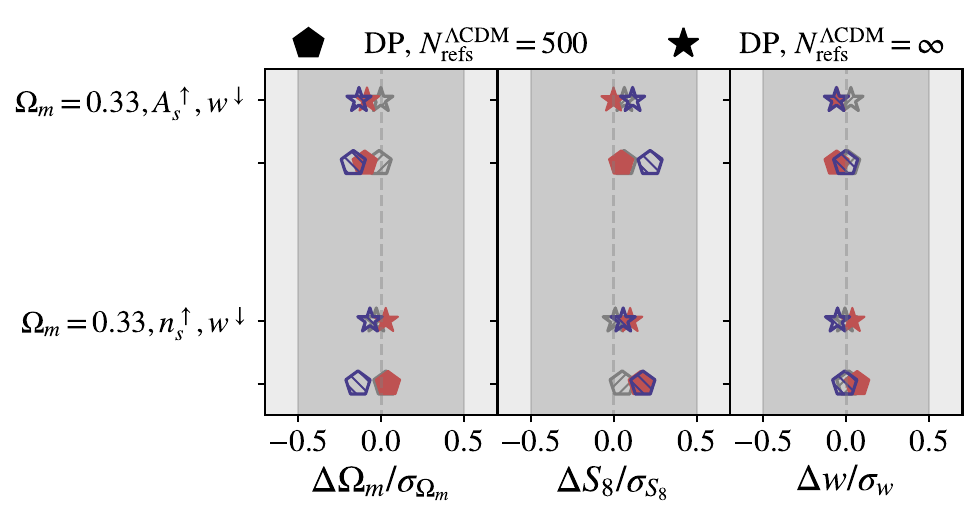}
    \caption{1D biases between \textsc{cola} emulators and \textsc{ee2} with fiducial data vectors generated with \textsc{bacco}. These test the robustness of the $N_{\rm refs}^{\Lambda \rm CDM}=500$ and $N_{\rm refs}^{\Lambda \rm CDM}=\infty$ results presented in Fig.~\ref{fig:inf_refs}. Note that the fiducial $\Omega_m$ value differs from past tests in order to bring the cosmology inside the \textsc{bacco} parameter space. }
    \label{fig:1d_bias_baccodv}
\end{figure}

Fig.~\ref{fig:1d_bias_baccodv} shows the 1D bias measures of Eq.~\ref{eq:1d_fob} between \textsc{ee2} and \textsc{cola} emulators. Once again, we confirm that COLA emulation under the $N_{\rm refs}^{\Lambda \rm CDM}=500$ and $N_{\rm refs}^{\Lambda \rm CDM}=\infty$ approaches produced low bias relative to \textsc{ee2}, now with a different prescription for the fiducial data. For Cutoffs 1, 2, and 3 the worst FoB values produced by the $N_{\rm refs}^{\Lambda \rm CDM}=500$ emulator were 0.09, 0.23, and 0.50 respectively. Similarly the highest FoB values produced using the $N_{\rm refs}^{\Lambda \rm CDM}=\infty$ approach were 0.10, 0.16, and 0.38 for the three scale cuts, either meeting or near the target value of 0.3.

\begin{table}[t]
\centering
\setlength{\tabcolsep}{6pt}
\renewcommand{\arraystretch}{1.6}
\begin{tabular}{ |c |c|c| }
\hline
\rule{0pt}{15pt} \backslashbox{Cosmo.}{Emul.} &  \makecell{DP, 500 \\ ($\Lambda$CDM)} & \makecell{DP, $\infty$ \\ ($\Lambda$CDM)}  \\\hline\hline
$\Omega_m=0.33$, $A_s^\uparrow$, $w^\downarrow$, C1   & $95.0\%$  & $98.0\%$    \\ \hline
$\Omega_m=0.33$, $A_s^\uparrow$, $w^\downarrow$, C2   & $77.8\%$      & $87.5\%$     \\ \hline
$\Omega_m=0.33$, $A_s^\uparrow$, $w^\downarrow$, C3   & $43.0\%$      & $52.0\%$     \\ \hline\hline
$\Omega_m=0.33$, $n_s^\uparrow$, $w^\downarrow$, C1   & $94.0\%$     & $96.6\%$      \\ \hline
$\Omega_m=0.33$, $n_s^\uparrow$, $w^\downarrow$, C2   & $74.3\%$      & $87.1\%$       \\ \hline
$\Omega_m=0.33$, $n_s^\uparrow$, $w^\downarrow$, C3   & $50.3\%$      & $57.8\%$       \\ \hline
\end{tabular}
\caption{Percentages of MCMC accepted points with the difference in $\chi^2$ between the COLA emulators and \textsc{ee2} falling in the desired range of $|\Delta \chi^2| <1$, when the fiducial data vectors are generated with the \textsc{bacco} emulator.}
\label{tab:bacco_fid_chi2}
\end{table}

\section{Mitigating Biases Using Principal Component Analysis}
\label{app:pca}
As a final technique to mitigate the biases in the COLA analyses, we try a procedure inspired by baryonic effect mitigation using Principal Component Analysis \cite{baryonic_pcas, baryon_pca, des_baryon_pca}. As a high number of \textsc{ee2} $\Lambda$CDM references proved an effective approach to reducing the bias of COLA, we consider whether differences in the $\Lambda$CDM data vectors between COLA and \textsc{ee2} can effectively be marginalized over in extended models. We start by generating an LHS of 30 $\Lambda$CDM cosmologies in a space shrunken symmetrically by $30\%$ total from the \textsc{ee2} limits in each parameter, in order to lessen the variability of the data vectors. 

For each cosmology, we compute two cosmic shear data vectors $\mathbf{d}$, one using a COLA emulator for nonlinear corrections and another using \textsc{ee2}. We then generate a difference matrix $\Delta$ such that each column is the difference of the COLA and \textsc{ee2} data vectors for one cosmology. The shape of $\Delta$ is $N_\mathrm{DV} \times 30$, where $N_\mathrm{DV}$ is the number of elements in the data vector. Schematically, this reads:

\begin{equation}
\Delta = 
\left[
\begin{array}{ccc}
\vline &  & \vline \\
(\mathbf{d}_1^\textsc{ee2} - \mathbf{d}_1^\mathrm{COLA}) & ... & (\mathbf{d}_{30}^\textsc{ee2} - \mathbf{d}_{30}^\mathrm{COLA}) \\
\vline &  & \vline \\
\end{array}
\right]_{N_\mathrm{DV} \times 30}.
\end{equation}
Performing a Cholesky decomposition on the survey covariance matrix $C = LL^T$, we then weight the difference matrix by $\Delta_{L^{-1}} = L^{-1}\Delta$. We then perform a Singular Value Decomposition on $\Delta_{L^{-1}}$, such that $\Delta_{L^{-1}} = U \Sigma V$, where $U$ and $V$ are square orthogonal matrices of shapes $N_\mathrm{DV} \times N_\mathrm{DV}$ and $30 \times 30$, respectively. $\Sigma$ is a matrix of shape $N_\mathrm{DV} \times 30$ whose first $30$ rows form a diagonal matrix with the singular values, and the other elements are zero. The first $30$ columns of $U$ are eigenvectors of $\Delta_{L^{-1}}$, denoted by $\mathbf{e}_i$:

\begin{equation}
U = 
\left[
\begin{array}{cccc}
\vline &  & \vline & ... \\
\mathbf{e}_1 & ... & \mathbf{e}_{30} & ...\\
\vline &  & \vline & ... \\
\end{array}
\right]_{N_\mathrm{DV} \times N_\mathrm{DV}}.
\end{equation}

By reweighting these vectors with $L$, they can be used to describe the differences between COLA and \textsc{ee2} data vectors in the $\Lambda$CDM model. The Cholesky and Singular Value decompositions were performed using \textsc{scipy} \cite{2020SciPy-NMeth}. To incorporate the principal components in the analysis, at each step of the MCMC, we correct the final data vector as:

\begin{equation}
    \mathbf{d} = \mathbf{d}^\text{No PC} + L\sum_{i = 1}^{n_\mathrm{PC}}Q_i \mathbf{e}_i,
\end{equation}
where $Q_i$ is a parameter sampled in the MCMC with large flat priors $[-100, 100]$, and $n_\mathrm{PC}$ is the number of principal components chosen to correct the data vector. By marginalizing the cosmological parameter constraints over $Q_i$, we expect to mitigate the differences between COLA and \textsc{ee2} results at the cost of increasing the error bars. We remark that this method only relies on $\Lambda$CDM predictions of the matter power spectrum, and we want to test whether this procedure can mitigate biases in extended models, in our case wCDM.

\begin{figure}[h]
    \centering
    \includegraphics[width=0.9\columnwidth]{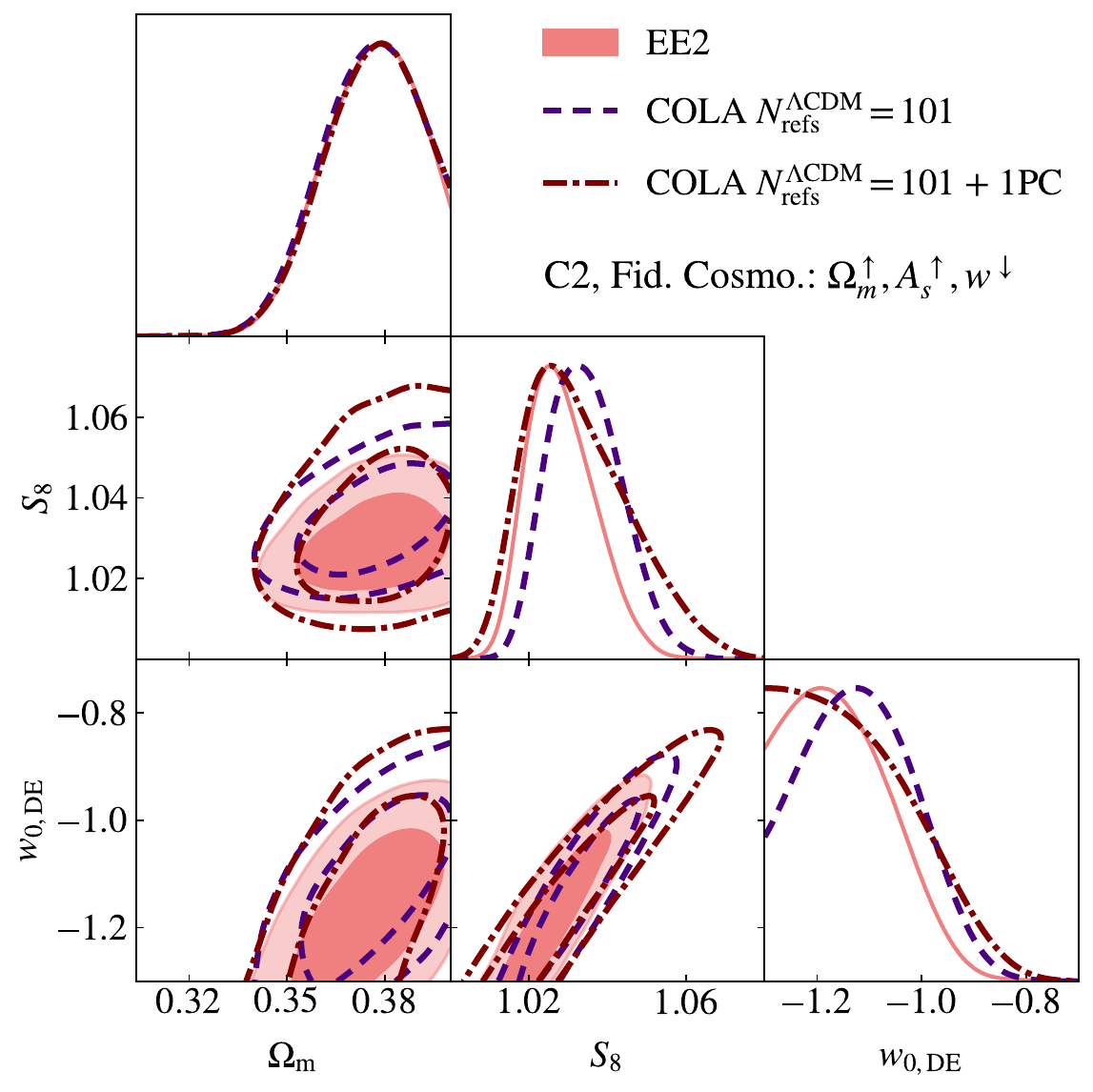}
    \includegraphics[width=0.9\columnwidth]{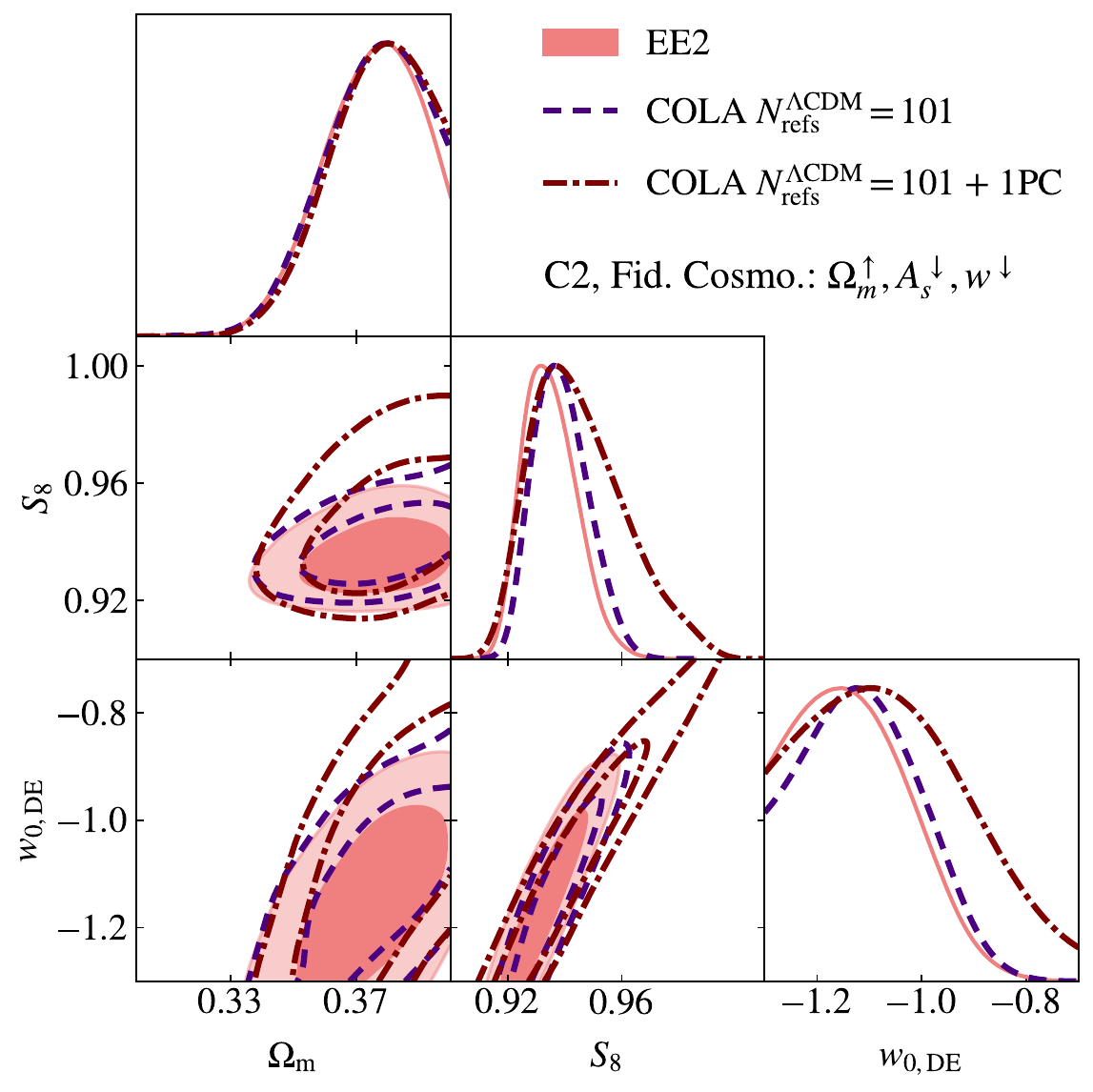}
    \caption{ Confidence contours (68\% and 95\%) for $\Omega_m$, $S_8$ and $w$, in three cases: \textsc{ee2} (pink filled contours), the $N_{\rm refs}^{\Lambda\rm CDM}=101$ \emph{default-precision} COLA emulator (blue dashed lines), and the same COLA emulator when marginalizing over 1 PC is employed (maroon dotted-dashed lines). The top panel shows results using the fiducial cosmology $(\Omega_m^\uparrow, A_s^\uparrow, w^\downarrow)$. We observe that the 2D contours from the COLA chain employing marginalization, completely enclose the \textsc{ee2} contours and spread the 1D distributions in the direction of the \textsc{ee2} distributions, relative to the COLA chains without PC marginalization. However, in the bottom panel where the fiducial cosmology $(\Omega_m^\uparrow, A_s^\downarrow, w^\downarrow)$ is used, the 1D distributions spread in the direction opposite of the \textsc{ee2} constraints. }
    \label{fig:pcs_triangle}
\end{figure}

\begin{figure}[h]
    \centering
    \includegraphics[width=\columnwidth]{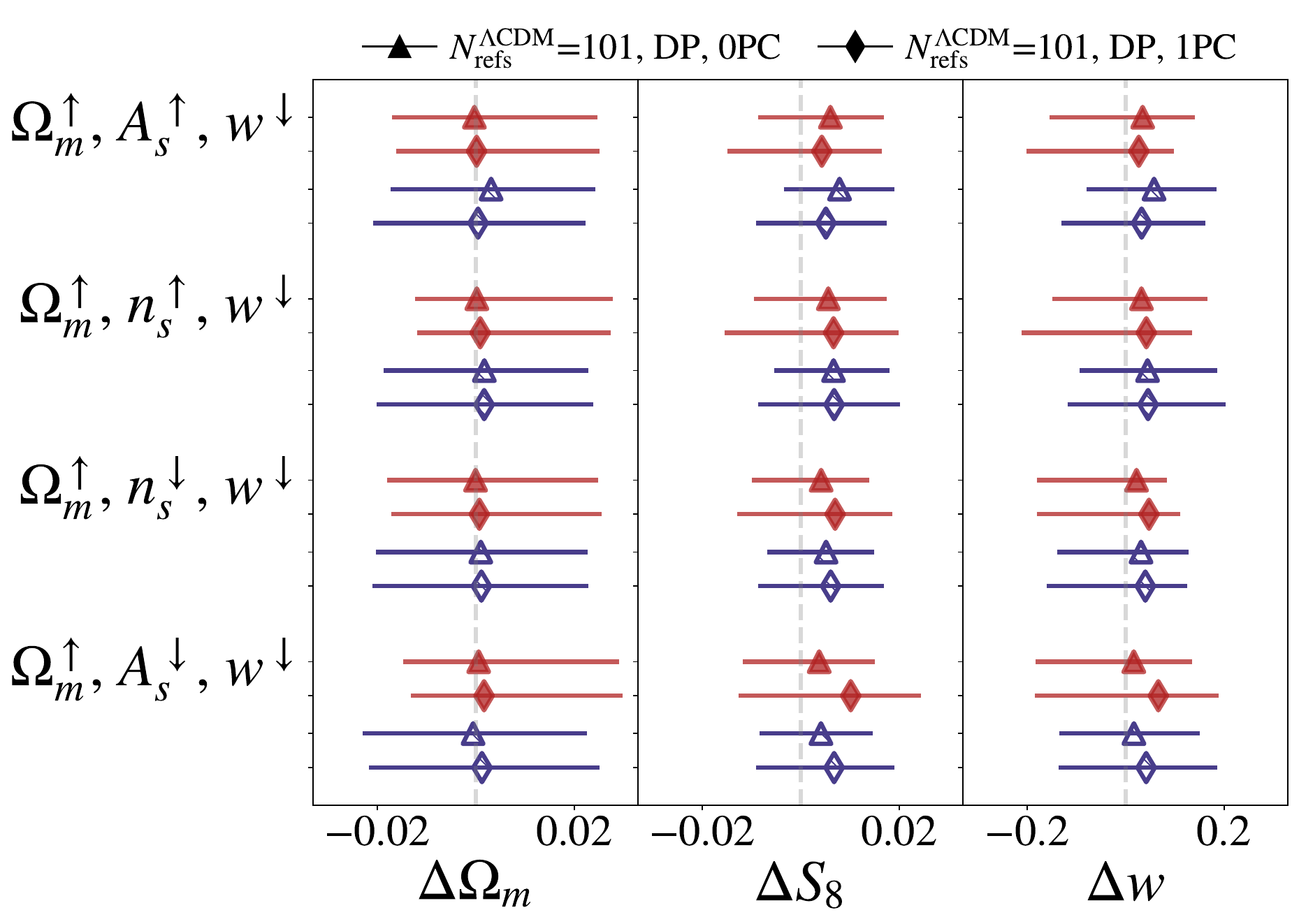}
    \caption{1D differences between COLA emulators and \textsc{ee2} for fiducial cosmologies outside of the $\Lambda$CDM region, for the two most aggressive scale cuts. We test the effect of marginalizing over a PC measuring the difference between the COLA and \textsc{ee2} data vector, using our \emph{default-precision} emulator with $N_{\rm refs}^{\Lambda \rm CDM}=101$.}
    \label{fig:w_fiducials_pcs}
\end{figure}

In Fig.~\ref{fig:pcs_triangle} we show 2D contours when marginalizing over PCs is employed and when it is not. The $N_{\rm refs}^{\Lambda\rm CDM}=101$ \emph{default-precision} emulator was used for these tests. We include two fiducial cosmologies using Cutoff 2, showing mixed results between the two cases. While marginalizing over 1 PC broadened the posterior, which can alleviate tension, the posterior did not always move in the correct direction so as to lessen the difference in the means between the COLA and \textsc{ee2} constraints. Fig.~\ref{fig:w_fiducials_pcs} shows a more complete assessment, using the 4 fiducial cosmologies with $w=w^{\downarrow}$, and the two more aggressive scale cuts. Here we plot the difference in the means $\Delta \theta = \theta_{\rm COLA} - \theta_{\textsc{ee2}}$ and the error bars separately $\sigma_{\theta} = \sqrt{\sigma_{\theta,\textsc{ee2}}^2 + \sigma_{\theta,\mathrm{COLA}}^2}$, to show the effect of PC marginalization on $\sigma_\theta$.

The results in Fig.~\ref{fig:w_fiducials_pcs} are mixed, showing a broadening of error bars but failing to consistently reduce the difference in means between the COLA emulators and \textsc{ee2}. Hence, we did not extensively test marginalizing over a larger number of PCs. This approach can likely be improved by adjusting the principal components for each cosmology sampled in the chain.

\label{lastpage}

\end{document}